%% file: ANA-SCTD-2019-01-PAPER.tex
\pdfoutput=1

\documentclass[cernpreprint,atlasdraft=true,texlive=2016,UKenglish, texmf]{atlasdoc}
 
\usepackage[subfig,block=none]{atlaspackage}
\usepackage{atlasbiblatex}
\usepackage{atlasphysics}
\usepackage{rotating}
 
\graphicspath{{logos/}{figures/}}
 
\usepackage{ANA-SCTD-2019-01-PAPER-defs}
 
\AtlasTitle{Operation and performance of the ATLAS semiconductor tracker in LHC Run~2}
 
\PreprintIdNumber{CERN-EP-2021-178}
 
\AtlasAbstract{
The semiconductor tracker (SCT) is one of the tracking systems for charged particles in the ATLAS detector. It consists of 4088 silicon strip sensor modules.
During Run~2 (2015--2018) the Large Hadron Collider delivered an integrated luminosity of 156~fb$^{-1}$ to the ATLAS experiment at a centre-of-mass proton-proton collision energy of 13~TeV. The instantaneous luminosity and pile-up conditions were far in excess of those assumed in the original design of the SCT detector.
Due to improvements to the data acquisition system, the SCT operated stably throughout Run~2.
It was available for 99.9\% of the integrated luminosity and achieved a data-quality efficiency of 99.85\%.
Detailed studies have been made of the leakage current in SCT modules and the evolution of the full depletion voltage, which are used to study the impact of radiation damage to the modules.
}
 
\author{The ATLAS Collaboration}
 
\AtlasRefCode{SCTD-2019-01}
 
\AtlasJournalRef{JINST 17 (2022) P01013}
\AtlasDOI{10.1088/1748-0221/17/01/P01013}

\hypersetup{pdftitle={ATLAS document},pdfauthor={The ATLAS Collaboration}}
 
\begin{document}
 
\maketitle
 
\tableofcontents

\section{Introduction}
\label{sec:intro}
 
The ATLAS experiment~\cite{PERF-2007-01} has been collecting data to study the interactions of elementary particles since 2009, using proton--proton ($pp$) collisions provided by the Large Hadron Collider (LHC)~\cite{Evans:2008zzb} at CERN. The LHC produces the highest-energy proton beams in the world, providing a unique opportunity to search for physics beyond the Standard Model (SM), as well as to precisely measure SM processes at very high energy.
The first operational period in 2009--2012, called Run~1, yielded physics data at centre-of-mass energies ($\sqrt{s}$) of 7 and 8~\TeV.
After the first long shutdown (LS1) in 2013--2014, dedicated to upgrading the accelerator and the detectors, Run-2 operation began in 2015 at $\sqrt{s} = 13$~\TeV\ and continued until 2018.
The total integrated luminosity of $pp$ collisions delivered to the ATLAS experiment by the LHC during Run~2 corresponds to 156~\ifb.
In addition to $pp$ collisions, special physics data-taking runs for heavy-ion physics were performed.
The second long shutdown (LS2) in 2019--2021 is for machine and detector upgrades towards Run-3 operation, which is scheduled to start in 2022.
 
The ATLAS detector is a multipurpose detector composed of the inner detector (ID), the calorimeters and the muon spectrometer.
The ID is located in an axial magnetic field of 2~T generated by a superconducting solenoid, and performs measurements of charged-particle trajectories and determines their momenta.
The electromagnetic and hadronic calorimeters are outside the solenoid and measure the energies of electrons, photons and hadrons.
Good hermeticity of the calorimetry is ensured by the forward calorimeters, enabling a precise measurement of the missing transverse momentum for each event.
The muon spectrometer is the outermost part of the detector. It provides muon identification and momentum measurement using toroidal magnets, as well as a muon-based trigger.
 
The ID is divided into three subsystems: the pixel detector (Pixel), the semiconductor tracker (SCT) and the transition radiation tracker (TRT).
The Pixel provides a four-layer measurement of charged-particle tracks, where the innermost layer is the insertable B-layer (IBL), which was installed during LS1~\cite{ATLAS-TDR-19, PIX-2018-001}.
It is located closest to the interaction point (IP) and hence it has a great impact on the reconstruction of primary and secondary decay vertices.
The SCT is a precise silicon microstrip detector, which extends the tracking volume to radial distances of $299 < r < 560$~mm.\footnote{
ATLAS uses a right-handed coordinate system with its origin at the nominal interaction point in the centre of the detector and the $z$-axis along the beam-pipe. The $x$-axis points from the interaction point to the centre of the LHC ring and the $y$-axis points upwards. Cylindrical coordinates ($r$,$\phi$) are used in the transverse plane, $\phi$ being the azimuthal angle around the beam-pipe. The pseudorapidity $\eta$ is defined in terms of the polar angle $\theta$ as $\eta = -\ln{\tan{(\theta/2)}}$.}
It plays an important role in measuring the transverse momenta of charged particles.
The outermost ID system is the TRT, consisting of 50 layers of drift tubes.
It is used for charged-particle tracking and for electron identification, utilising transition radiation.
 
\begin{figure}[t]
\centering
\includegraphics[width=0.9\textwidth]{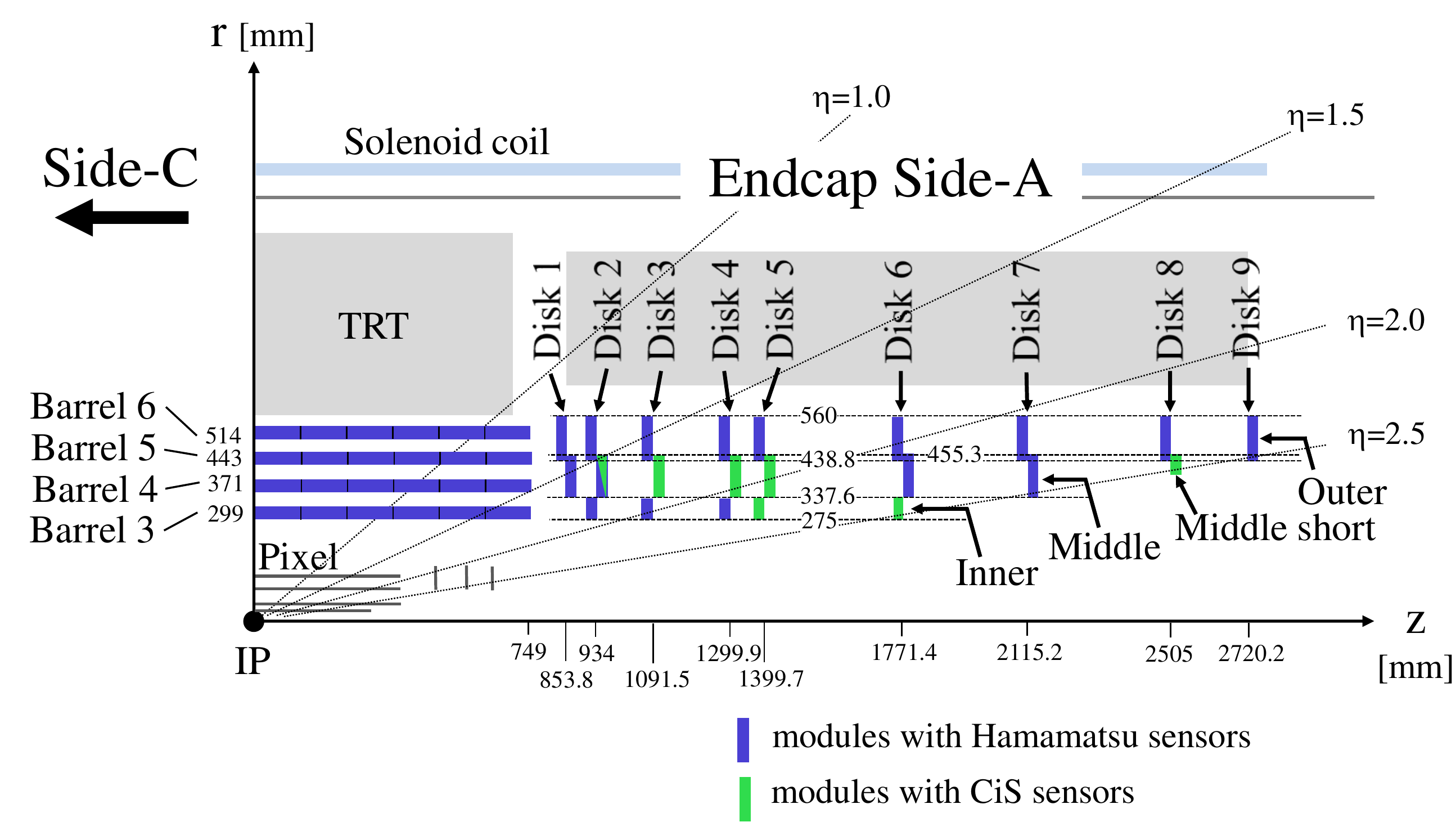}
\caption{
Schematic view of one quadrant of the SCT. The numbering schemes for the barrel layers and endcap disks are shown, together with the radial and longitudinal coordinates. The disks comprise up to three rings of modules, referred to as inner, middle and outer, at increasing radii from the beam-pipe.
The detector is symmetric between Side-A (positive $z$) and Side-C (negative $z$).
}
\label{fig:SCT_quadrant}
\end{figure}
 
As shown in Figure~\ref{fig:SCT_quadrant}, the SCT consists of four layers of silicon-strip sensors in the barrel (numbered 3 to 6), and nine disks (1 to 9) in each of the endcaps.
Each barrel layer consists of 12 rings of modules along the $z$-direction, with a total of 2112 barrel modules~\cite{NIMA-568-642-2006}.
There are 988 modules per endcap~\cite{NIMA-575-353-2007}, with each disk consisting of up to three rings, referred to as inner, middle and outer, at increasing radii from the beam-pipe.
Each barrel module consists of four rectangular 285-$\mu$m-thick $p^+$-on-$n$ silicon sensors~\cite{NIMA-578-98-2007} with strips of pitch 80~$\mu$m.
The sensors have 770 strips of length $\sim$6~cm, of which 768 are read out via wire bonds.
The unbonded outermost strips are present for field uniformity at the sensor edges.
Two sensors on each side of a module are wire-bonded together to form 12-cm-long strips.
A second pair of identical sensors is glued back-to-back with the first pair at a stereo angle of 40~mrad.
Each module also has two arrays of six ABCD chips~\cite{NIMA-552-292-2005}, where each chip is wire-bonded to 128 strips, in order to amplify, shape and discriminate pulses from the charges collected on the strips.
The ABCD chips are placed on hybrid boards, which are bridged over the sensors.
A binary threshold of around 1~fC in the detected charge is used to define a `hit' in the SCT.
A barrel module has dimensions of $6\,\text{cm} \times 12\,\text{cm}$. Most of the endcap modules have similar dimensions, but have a trapezoidal shape, except for the `short' endcap modules on the inner rings of disks~2--6 and also on the middle ring of disk~8, which consist of just one sensor on each side and therefore have only half the length.
 
The sensors were provided by two manufacturers: Hamamatsu Photonics (HPK)\footnote{Hamamatsu Photonics K.K., 1126-1, Ichino-cho, Higashi-ku, Hamamatsu, Shizuoka, 435-8558, Japan.} and CiS\footnote{CiS Forschungsinstitut f\"ur Mikrosensorik GmbH, Konrad-Zuse-Strasse 14, 99099 Erfurt, Germany.}.
Most of the sensors have a crystal lattice orientation (Miller indices) of <111>, while 92 modules in the barrel, provided by HPK, have <100>.
The difference in crystal orientation results in slightly different performance, such as different noise characteristics; this is discussed in detail in Section~\ref{sec:performance}.
The silicon wafers produced by CiS to manufacture short endcap modules were oxygenated to increase the radiation hardness of these innermost sensors~\cite{NIMA-578-98-2007}.
 
The SCT is cooled to about $0 \degr$C by an evaporative cooling system using C$_3$F$_8$ fluid~\cite{JINST-3-P07003-2008}.
A heater pad system is intended to provide a thermal barrier between the SCT and the TRT, because the TRT operates at room temperature.
However, it was only partially operative in the barrel region; hence barrel layer~6 was operated 6$\degC$ warmer than the rest of the SCT to prevent mechanical stress on the TRT.
The cooling system was originally driven by compressors; these were replaced by a thermosiphon system~\cite{IJCRE-13-511-2015} in 2018.
The SCT position was monitored using frequency-scanning interferometry in Run~1~\cite{ATL-INDET-PROC-2010-037}.
In Run~2, the SCT alignment was performed fully offline using reconstructed tracks~\cite{IDTR-2019-05}.
 
The SCT has been fully operational since the commissioning period using cosmic-ray data in 2008~\cite{IDET-2010-01}.
It showed excellent performance during LHC Run~1, as reported in Ref.~\cite{IDET-2013-01}.
In Run~2, the experimental environment became more challenging for the SCT, because an instantaneous luminosity up to $2.1 \times 10^{34}~\mathrm{cm}^{-2}\mathrm{s}^{-1}$ was achieved, corresponding to about twice the design value.
Consequently $\mu$, the average number of $pp$ interactions per bunch crossing (pile-up), increased and the SCT was exposed to a high density of charged particles, as shown in Figure~\ref{fig:SCT_eventDisplay}.
This affected the SCT data transmission rate and the SCT data quality, as discussed in Section~\ref{sec:operation}.
 
Radiation damage was also a key feature of SCT operation in Run~2.
The radiation received by the SCT modules until the end of Run~2 is estimated to be equivalent to a fluence of up to $5.6 \times 10^{13}$ 1~\MeV\ neutrons~\cite{Baranov:2005ewa,CERN-2021-001}.
The operational strategy for the SCT therefore needed to be modified to account for the changes in sensor properties caused by radiation damage.
At the same time, this provided valuable data for studies of radiation damage to silicon sensors in a hadron collider experiment; in particular, the SCT uses $p^+$-on-$n$ type silicon sensors, for which the $n$-bulk of all sensors was expected to invert into quasi-$p$-type (type inversion) during the second half of Run~2.
 
Several improvements to the SCT system are discussed in this paper. The operation of the SCT in Run~2 is described in Section~\ref{sec:operation}, and its performance is assessed in Section~\ref{sec:performance}. The effects of radiation on the detector are discussed in Section~\ref{sec:radiation}.
 
\begin{figure}[t]
\centering
\includegraphics[width=0.5\textwidth]{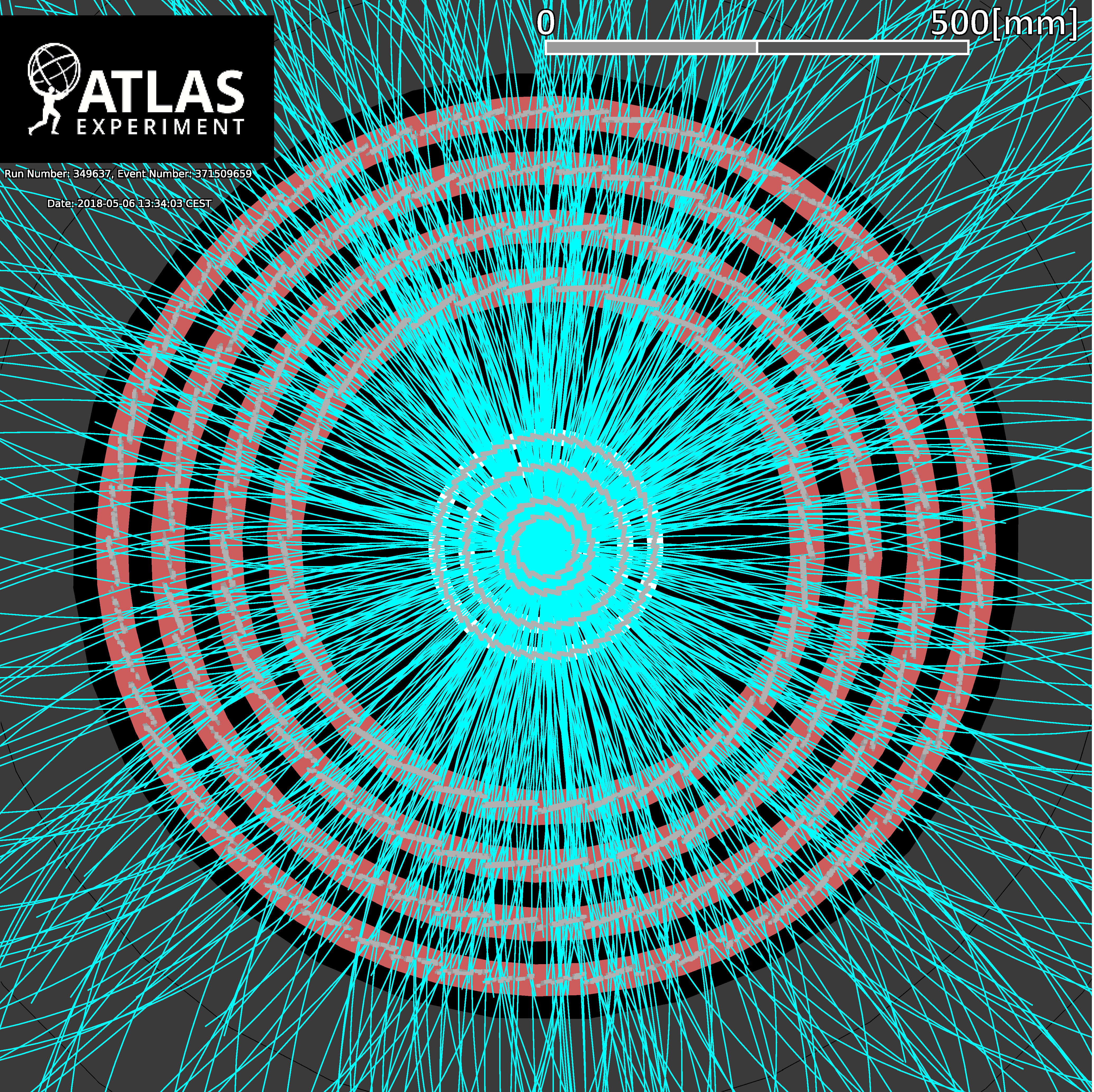}
\caption{
Event display projected onto the $r$--$\phi$ plane, taken by an unbiased random trigger in a typical $pp$ collision run at $\sqrt{s} = 13$~\TeV\ in May 2018.
This event has approximately 60 $pp$ interactions, which is nearly the maximum experienced in Run~2.
The outer four circles correspond to the four barrel layers of the SCT, while the inner four circles are the Pixel layers.
The lines emerging almost radially from the centre (interaction point) represent reconstructed tracks of charged particles with transverse momentum greater than 0.5~\GeV.
}
\label{fig:SCT_eventDisplay}
\end{figure}
 
 
\section{Operation of the SCT in LHC Run 2}
\label{sec:operation}
 
\label{sec:operation_overview}
The typical cycle of operation of the SCT in Run~2 followed the same procedure that had been established in Run~1.
During periods of beam injection, energy ramping and optimisation, the SCT was put into the standby state, where the low voltage (LV) supplied to the ABCD chips was fully maintained, but the high voltage (HV) to the sensors was reduced to 50~V.
This is sufficiently low so as to avoid electrical breakdown if there was any significant beam loss, which could cause many particles to enter the SCT simultaneously.
Once stable beam conditions were declared, the HV was automatically raised back to the operational voltage (typically 150~V), provided that the LHC collimators were at their predefined positions for physics data-taking, the background rates measured by the beam conditions monitor~\cite{JINST-3-P02004-2008}, the beam loss monitor and the forward detectors were low enough, and the hit occupancy of the SCT was as expected for the instantaneous luminosity.
This `warm start' procedure took about 30 seconds before the SCT reached its operational HV state.
When the LHC beam was dumped, the HV was immediately ramped back down to its standby state.
An operator in the ATLAS control room was responsible for monitoring the SCT data acquisition (DAQ) system, detector control system (DCS) and data quality (DQ).
In case of problems, the operator attempted to recover normal operation by following predefined procedures or by calling an SCT expert if necessary.
 
The experimental conditions in Run~2 were challenging due to high pile-up, resulting in up to $\sim$2\% hit occupancy in SCT modules during periods of operation at the maximum LHC luminosity. This resulted in a very high load on the DAQ system.
There was significant radiation damage to the SCT sensors, which required modification to the operational strategy.
Nevertheless, the SCT was available for 99.9\% of the integrated luminosity during Run 2.
In terms of the DQ efficiency, which is the fraction of data usable for physics analyses relative to the recorded integrated luminosity, it achieved 99.85\%.
This was due to several improvements, in particular to the DAQ system in 2015--16, which are discussed in more detail in the following sections.
As shown in Table~\ref{tab:DisabledElements}, 98.6\% of the SCT elements remained active until the end of Run~2.
During LS1, two of the modules that had been disabled during Run 1 were recovered.
An additional 14 modules were disabled in Run 2.
For ten of these modules, this was attributed to HV-related problems such as trips of the HV and unstable HV currents.
The remaining four modules were disabled due to LV-related problems or other problematic on-detector components.
 
\begin{table}[t]
\centering
\caption{Numbers of disabled SCT elements as of 11th December 2018, after the end of Run~2. The values in parentheses are those from the end of Run 1~\cite{IDET-2013-01}. The numbers of chips (strips) exclude those in disabled modules (disabled modules and chips).}
\begin{tabular}{l|rrrrrr|rr}
\hline
Component & \multicolumn{2}{c}{Barrel} & \multicolumn{2}{c}{Endcaps} & \multicolumn{2}{c|}{SCT} & \multicolumn{2}{c}{Fraction [\%]}\\
\hline
Modules &    12 &   (11) &    30 &   (19) &    42 &    (30) & \tablenum{1.0}  & (0.73)\\
Chips   &    55 &   (38) &    22 &   (17) &    77 &    (55) & \tablenum{0.16} & (0.11)\\
Strips  &  6398 & (4111) &  9505 & (7252) & 15903 & (11363) & \tablenum{0.25} & (0.18)\\
\hline
\end{tabular}
\label{tab:DisabledElements}
\end{table}
 
The operational HV for most sensors was kept at 150~V until 2017.
In 2018, a small drop in hit efficiency of $\sim$0.5\% was observed. This was related to an increase in the full depletion voltage ($\vfd$), due to radiation damage.
Based on HV scan data (discussed in detail in Section~\ref{sec:HitEfficiency_gen}), the operational HV of modules located in regions of high radiation was increased to 200--250~V.
In Run~1, it was reported that 27\% of the modules with CiS-manufactured sensors exhibited anomalous fluctuations of up to $\sim$200~$\mu$A in their leakage current during $pp$ collisions~\cite{IDET-2013-01}, and consequently the operational HV of these modules was reduced by 30--40~V to suppress these fluctuations.
This condition was retained at the start of Run 2.
In 2018, however, their behaviour was found to be more stable, since the nominal leakage current of $\sim$1000~$\mu$A exceeded these fluctuations.
Hence, the operational HV of these modules was set back to the nominal value in order to recover their hit efficiency.
 
To keep the temperature of the SCT close to $0\degr$C, the evaporative cooling system was operated stably at $-20\degr$C to $-14\degr$C, except for barrel layer~6, which was set to $-10\degr$C (see Section~\ref{sec:intro}).
The cooling temperatures set depended on the location within the SCT, because of variations in the hydraulic impedance among the cooling pipes, and the effects of gravity, which resulted in higher fluid concentrations in the lower part of the detector.
These temperatures were maintained throughout most of Run~2, except for two weeks during each winter shutdown period, when the evaporative cooling system was turned off for maintenance work and the modules warmed to room temperature, and also part of 2015.
During that time, early in Run~2, the temperature of the evaporative cooling system was set to $-10\degr$C, because increased humidity within the ID volume resulted in a higher dew point around the SCT.
The cause of this was identified as an over-pressure safety valve to the ID volume, which was unexpectedly open.
This problem was fixed in October 2015, after which the temperature was restored to the nominal value.
This small change in temperature had no impact on detector operation.
 
\subsection{Data acquisition}
 
The DAQ system of the SCT uses a binary readout architecture.
Charge created in the sensor is immediately digitised and stored in a 132-cell-deep pipeline within the ABCD chip.
On receipt of a trigger, the column of binary data associated with the bunch crossing is read out and transmitted to the off-detector DAQ system.
 
Figure~\ref{fig:DAQSchematic} shows a schematic diagram of the DAQ system~\cite{IDET-2013-01, JINST-3-P01003-2008}.
It comprises 128 Readout Driver (ROD) and Back of Crate (BOC) cards.
The BOC provides the optical interface between a ROD and up to 48 SCT modules; for each module the BOC transmits the clock and trigger via a single command stream and receives data back via two optical links (FE-links) from the six ABCD chips on each side of the module.
For each level-1 (L1) trigger, the ROD processes the incoming data from up to 96 FE-links and combines those data into a single event fragment.
The event fragment is then transmitted at the level-1 trigger rate on an optical link (S-link)~\cite{JINST-2-P09003-2007} to the global DAQ system of ATLAS, where the data are buffered until a higher-level trigger decision is made.
The commands and triggers, as well as a 40~MHz clock, are distributed by the trigger, timing and control (TTC) interface module.
 
\begin{figure}[t]
\centering
\includegraphics[width=0.8\linewidth]{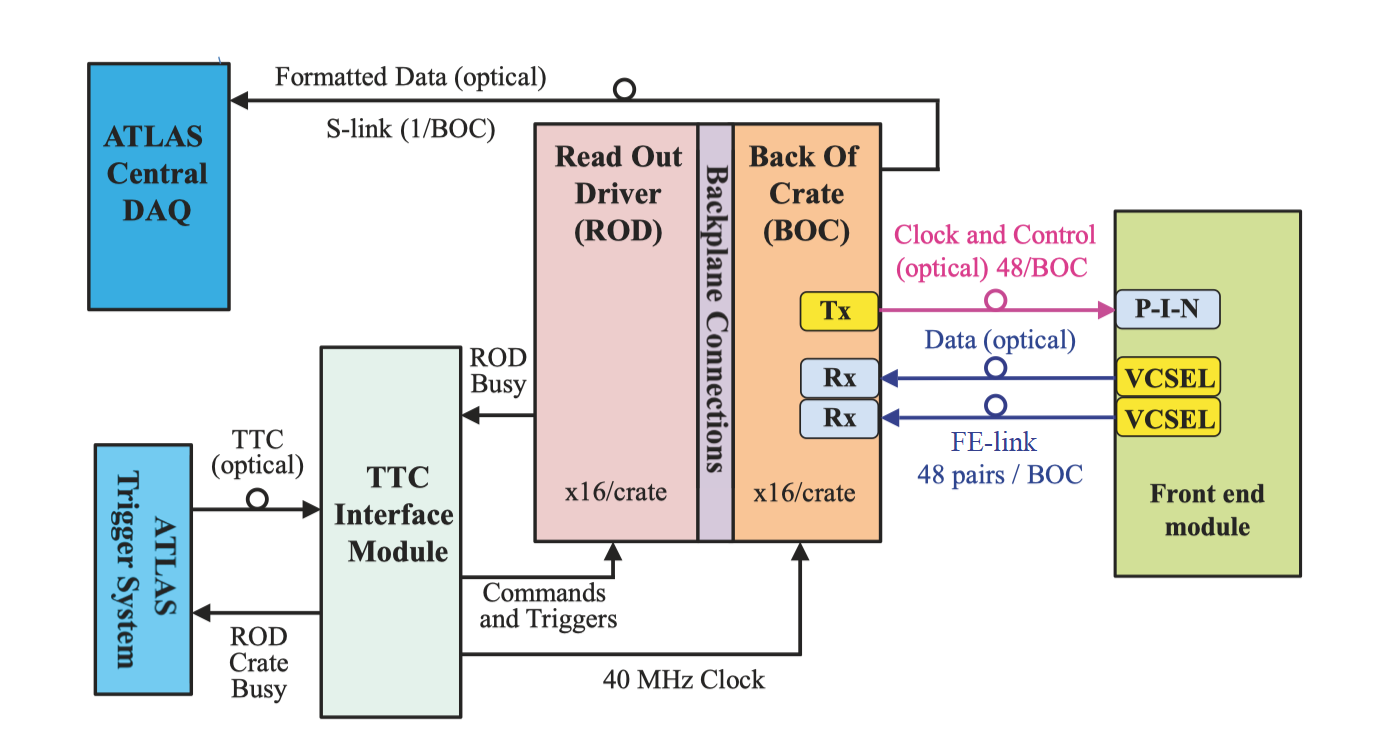}
\caption{Schematic diagram of the SCT DAQ~\cite{IDET-2013-01}. The front-end module is located on the detector, while the ROD, BOC and TTC Interface Module are located in the underground cavern outside the detector.}
\label{fig:DAQSchematic}
\end{figure}
 
Both the transmission (TX) and receiver (RX) optical paths incorporate redundancy mechanisms which can be used in the case of failure of individual links~\cite{JINST-2-P09003-2007}.
The TX redundancy allows the module to be configured to use the trigger command stream from the adjacent module; the RX redundancy allows the module to transmit data from all 12 ABCD chips using one of the two data links.
The use of optical link redundancy at the end of Run~2 is shown in Table~\ref{tab:RXandTX}.
Although the total number of links using redundancy mode increased during Run~2, the fractions are still only 1.3\% (1.9\%) for the TX (RX).
 
\begin{table}[t]
\centering
\caption{Numbers of optical links configured using the redundancy mode at the end of Run~2. The numbers in parentheses are the status at the end of Run~1~\cite{IDET-2013-01}.}
\begin{tabular}{l|rrrrrr|rr}
\hline
Link type & \multicolumn{2}{c}{Barrel} & \multicolumn{2}{c}{Endcaps} & \multicolumn{2}{c|}{Total} & \multicolumn{2}{c}{Fraction}\\
\hline
TX & 36 &  (9) & 19 &  (5) &  55 &  (14) & 1.3\% & (0.3\%)\\
RX & 53 & (41) & 99 & (91) & 152 & (132) & 1.9\% & (1.6\%)\\
\hline
\end{tabular}
\label{tab:RXandTX}
\end{table}
 
In the following subsections, specific DAQ developments during Run~2 are discussed in more detail.
 
\subsubsection{Optical data transmission}
 
The dominant failure mechanism for the DAQ system during Run~1 was the failure of individual channels within the 12-channel array of the vertical-cavity surface-emitting laser (VCSEL) for the TX~\cite{JINST-2-P09003-2007}.
The loss of a channel meant that the module no longer received the clock and trigger signals, resulting in the loss of data from that module.
Although the TX redundancy could be applied to re-enable the module, failures of two adjacent channels resulted in the loss of data from one module until that TX plug-in was physically replaced.
The peak rate of channel deaths in Run~1 was as high as 10--12 per week, and TX redundancy was used for up to 220 modules.
 
A number of TX replacement campaigns were carried out; failures in the original TX installation were attributed to inadequate precautions against electrostatic discharge by the manufacturer, and then in the second installation to the ingress of humidity due to a non-hermetic seal.
A third installation, carried out early in 2012, used a new VCSEL product manufactured with a complete hermetic seal, leading to a dramatic reduction in the channel death rate.
Nonetheless, a small rate of deaths continued ($\sim$5--10 per year) due to a coefficient of thermal expansion mismatch between the epoxy and the surface of the VCSEL array, leading to cracks in the VCSEL and the subsequent ingress of humidity.
 
During LS1, a final batch was produced incorporating the commercially available VCSEL array (10 Gbps LightABLE surface-mount parallel fibre-optic engine, Reflex Photonics), using a re-engineered TX plug-in to match the LightABLE to the BOC.
Three of the eight ROD crates were fully populated with LightABLEs, while the BOCs in the remaining five crates continued to be equipped with the original TX, and they operated throughout Run 2.
The LightABLEs have proved to be very reliable; only two channel deaths occurred, of which one was infant mortality.
In the remaining ROD crates, failures were only about 2--3 per year, which is easily compensated for by the TX redundancy mechanism.
It is envisaged that this configuration will be maintained in Run~3, as the very low death rate does not justify the significant effort and risk involved in changing the TXs in the remaining five crates.
 
\subsubsection{Developments for operation with high pile-up}
\label{sec:DAQ:HighPileup}
 
The SCT was designed to operate with 0.2\%--0.5\% occupancy in its 6.3~million sampled strips at an LHC pile-up rate of up to $\mu\sim23$.
The fundamental bottlenecks that restrict data-taking with increased occupancy, arising from higher pile-up, are the bandwidth limitations of the FE-links (which transmit bit streams at 40~Mbps) and the S-links (which transmit 32-bit words at 40~MHz, a throughput of 1.28~Gbps).
For the FE-links, exceeding the bandwidth leads to a loss of data from the corresponding module.
As shown in Figure~\ref{fig:ABCDErrors}, the fraction of errors issued by the ABCD chips increased for high pile-up and trigger rates, in particular for modules relying on RX redundancy.
If data throughput on the S-link exceeds $\sim$90\% of the bandwidth, the ROD starts to receive back-pressure from the S-link and asserts a busy flag (BUSY) to suspend the entire data-taking of the global DAQ system~\cite{JINST-3-P01003-2008} via the TTC interface module.
Extrapolations of the occupancy with increasing pile-up during Run~1 suggested that data flow in the FE-links and S-links would exceed bandwidth limitations at pile-up rates of $\mu \sim 87$ and $\mu \sim 33$ respectively, at a level-1 trigger rate of 100~kHz.
These projections assumed optimum on-chip data compression: only hits from the in-time bunch crossing that exceed a threshold were read out, while hits from the preceding bunch crossing were simultaneously vetoed (the so-called `01X' mode). In addition, the standard data-packing scheme for the ROD was used, which included readout of the hits from each strip over three consecutive bunch crossings.
In any case, FE-links using the RX redundancy would reach the bandwidth limitation at a much lower pile-up level, because all data would then be read out through only one of the data links.
 
\begin{figure}
\centering
\includegraphics[width=0.6\linewidth]{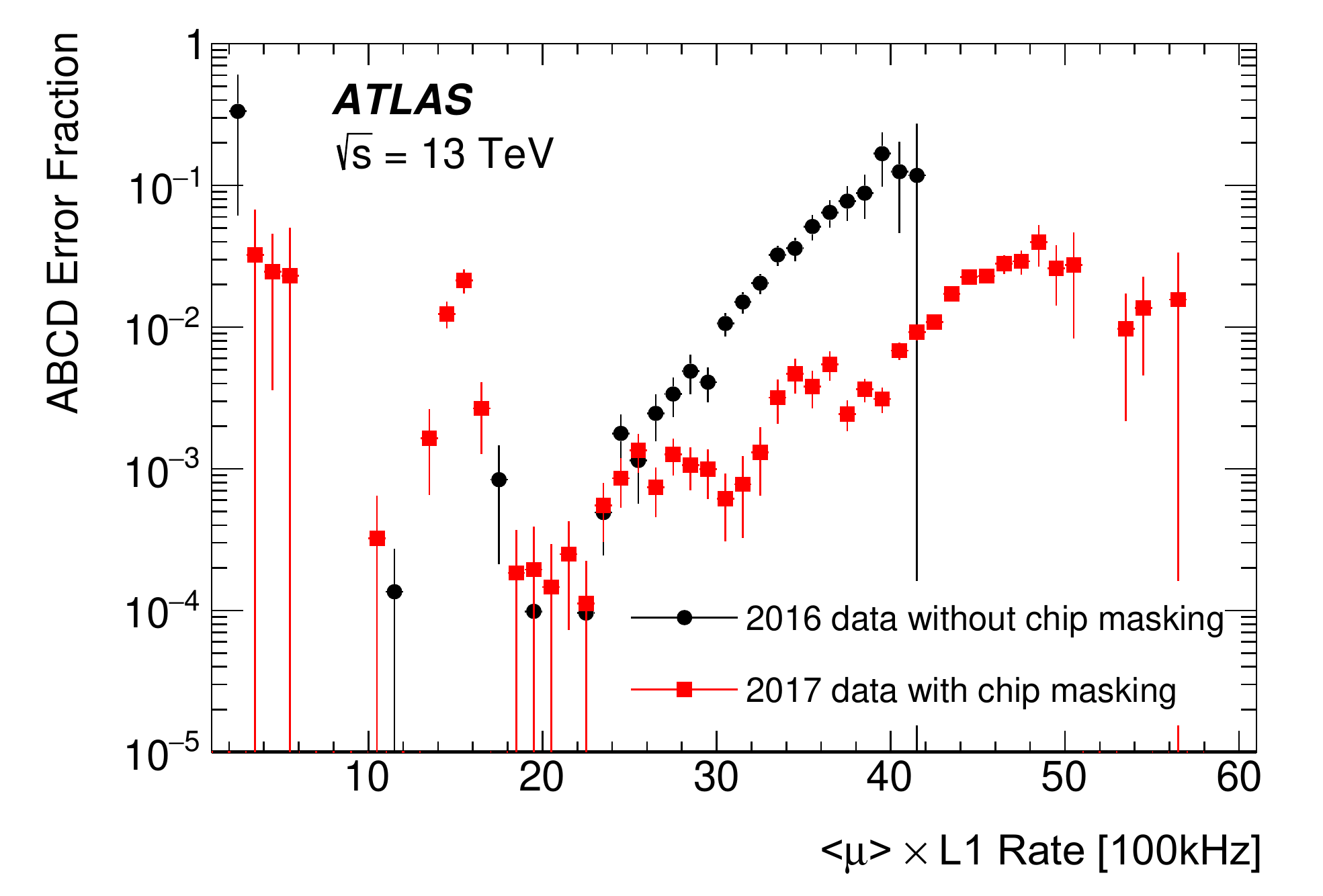}
\caption{Fraction of ABCD chip errors as a function of average pile-up, $\mu$, times the level-1 trigger rate for a module using RX redundancy. The black dots are from 2016, before introduction of the chip-masking mechanism, while the red squares are from 2017 with chip masking. Changes to the set of masked chips are visible as steps in the ABCD error fraction around $\mu$ $\times$ L1 rate $\sim 30$ and 40. The significant peak at $\sim 15$ is due to a chip that was transiently very noisy, while the peak at $\sim 4$ is due to an instantaneously low level-1 trigger rate with $\mu \sim 55$.}
\label{fig:ABCDErrors}
\end{figure}
 
It was anticipated late in Run~1 that the LHC would provide higher luminosity, causing more pile-up.
Furthermore, it was discovered early in Run~2 that the SCT hit occupancy had already increased significantly compared to Run~1; this was attributed to secondary interactions with material in the newly installed IBL services.
In order to achieve a comfortable margin of safety for operation with pile-up at $\mu \sim 60$, the four mitigation steps listed in Table~\ref{tab:DAQMigrationSteps}, and described below, were developed.
 
\begin{table}[t]
\centering
\caption{Mitigation steps deployed to address SCT bandwidth limitations.}
\begin{tabular}{llll}
\hline
Action & Type & Date & Comments\\
\hline
Expanded DAQ & Hardware & 2014 (LS1) & New ROD and BOC production\\
FE-link remapping & Hardware & 2015 & Reshuffle of data links to BOCs\\
Data compression & Firmware & 2016 & Improved data compression on the ROD\\
Chip masking & Software & 2017 & Implementation of on-the-fly chip masking\\
\hline
\end{tabular}
\label{tab:DAQMigrationSteps}
\end{table}
 
\begin{figure}[t]
\centering
\includegraphics[width=0.6\linewidth]{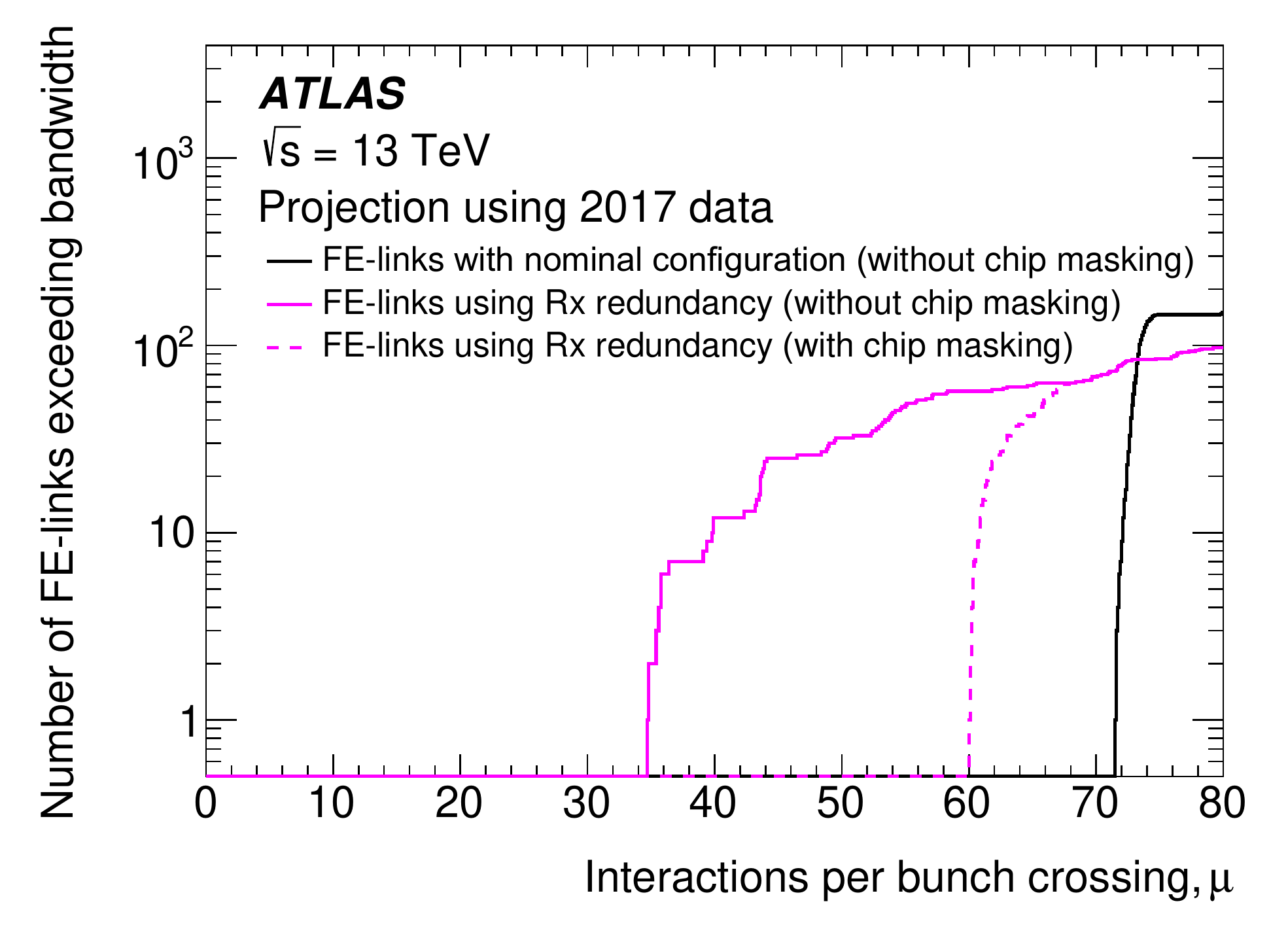}
\caption{
Number of FE-links exceeding the bandwidth limitation as a function of $\mu$, at a level-1 trigger rate of 100~kHz.
In the nominal configuration (black curve at right), signals are read out by two FE-links and therefore one FE-link transmits signals from up to six ABCD chips.
For the FE-links using RX redundancy, where one FE-link transmits signals from up to 12 ABCD chips on a module, the dashed middle (solid left) curve corresponds to the case with chip masking enabled (disabled).
}
\label{fig:BandwidthWithDAQMigration}
\end{figure}
 
\begin{description}
\item[Expanded DAQ:] To address the pile-up limit imposed by the S-links, the numbers of RODs and BOCs, and consequently the number of S-links, were increased from 90 to 128 during LS1, thereby reducing the number of modules serviced by each ROD from 48 to no more than 36.
 
\item[FE-link remapping:] The mapping of the FE-links to the BOCs was re-optimised in order to ensure a more uniform data load on each ROD.
This resulted in a flatter occupancy distribution across the S-links, removing spikes from the highest-occupancy S-links and thereby improving the stability of data-taking.
 
\item[Data compression:] A new highly efficient data-packing scheme (the `supercondensed mode') was implemented on the RODs routinely from 2016 onwards.
In this mode, clusters of up to 16 strips are packed into a single 16-bit word, resulting in a typical data size reduction of $\sim$25\% during $pp$ collisions compared to the expanded mode.
This data size reduction is achieved at the expense of some error flags which were superfluous to efficient data-taking.
It significantly reduces the data size used to encode large clusters, which otherwise unduly bias the distribution.
 
\item[Chip masking:] This mechanism was developed in order to mask the noisiest chips during operation with the highest level of LHC pile-up.
Lists of chips with the highest noise were prepared so that various chips could be disabled when different pile-up levels were exceeded.
The value of $\mu$ was continuously monitored by the LHC, and published within the ATLAS DAQ framework.
This allowed reductions in the volume of hit data, so as to remain below the bandwidth limitations of the data links at the cost of a negligible loss of data.
The mechanism worked `on the fly', gradually reducing the number of masked chips to zero as the LHC luminosity, and hence the rate of pile-up, decreased during a data-taking run, effectively suppressing the ABCD chip errors, as shown in Figure~\ref{fig:ABCDErrors}.
The principle of chip masking has been demonstrated successfully, leading to all FE-links being operable within bandwidth constraints up to $\mu=60$, as shown in Figure~\ref{fig:BandwidthWithDAQMigration}.
This concept was first developed to cope with high-pile-up conditions in 2017. At this time the number of LHC bunches was reduced in order to avoid unexpected beam dumps, which had been occurring frequently during 2017, and consequently the average $\mu$ was increased to maintain the luminosity.
However, this situation was mitigated in 2018, leading to a return to nominal pile-up conditions for the remainder of Run~2.
Chip masking was therefore not deployed routinely; however, this capability is expected to be a valuable asset in efforts to maintain tracking efficiency if the level of pile-up increases significantly in Run~3.
 
\end{description}
 
The impact of the different mitigation steps related to S-links is further illustrated in Figure~\ref{fig:SlinkOccupancies}, using extrapolations of measured data occupancy as a function of pile-up.
When using the configuration from 2016, all S-links are seen to operate below a bandwidth occupancy of 90\%, which shows that the bandwidth limitation problem was successfully solved.
 
\begin{figure}[t]
\centering
\includegraphics[width=0.6\linewidth]{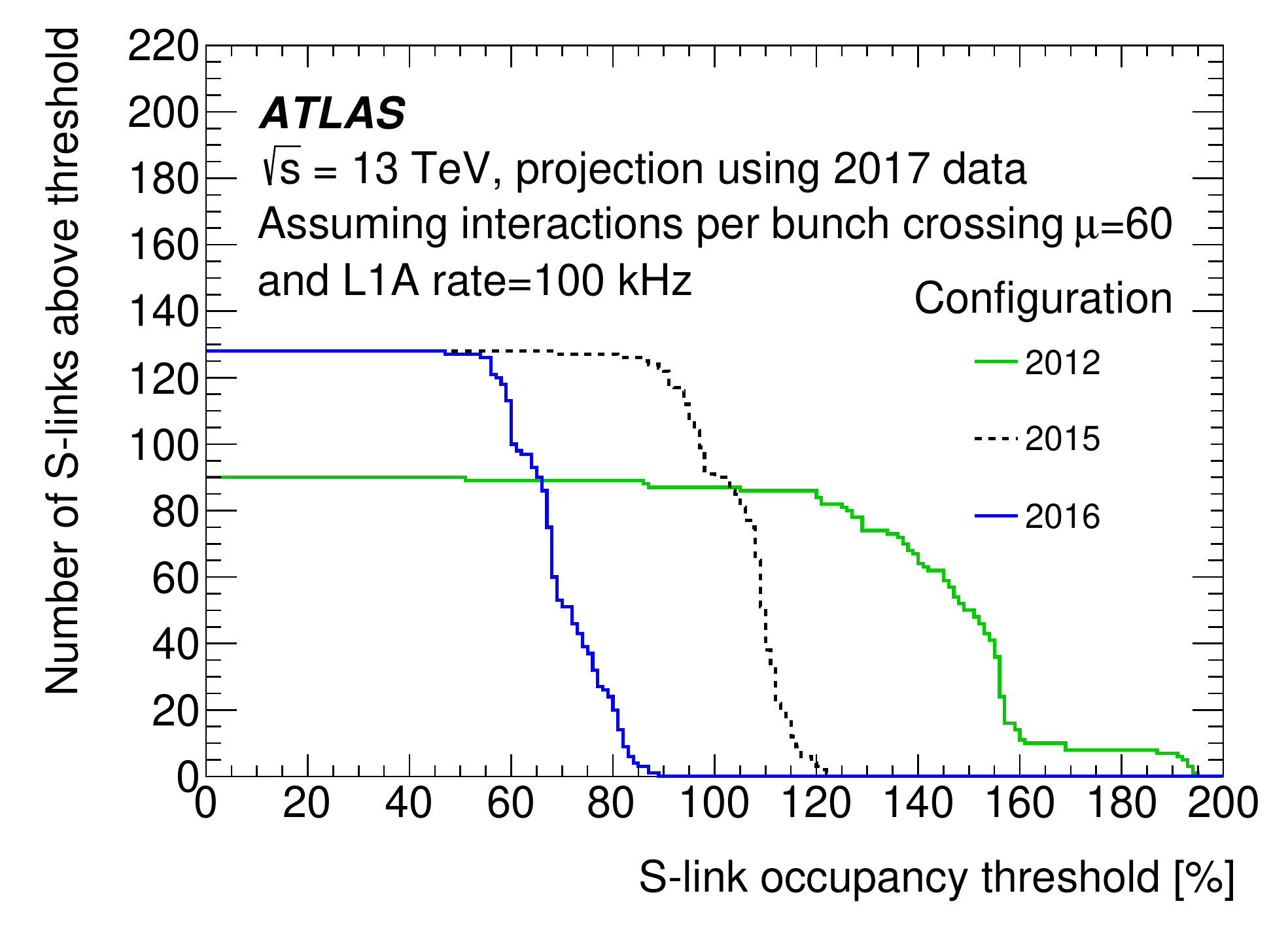}
\caption{Number of S-links operating above the bandwidth occupancy threshold specified on the $x$-axis.
During operation, once the S-link occupancy exceeds 90\% of its bandwidth, the corresponding ROD starts to fall into the busy state, so that it becomes impossible to transfer data correctly.
The bandwidth occupancy in 2012 (green line, right), 2015 (dashed black, middle) and 2016 (blue, left) is simulated using the extreme condition of a level-1 trigger rate of 100~kHz with $\mu = 60$.
The line labelled 2012 shows the 90-ROD system used in Run~1, where many S-links were not able to handle the high data rate.
Several S-links for endcap modules showed significantly higher occupancy than the other S-links, due to an imbalanced distribution of data rate per ROD.
These features were significantly improved in the 2015 configuration with 128 RODs.
In 2016 the supercondensed mode was introduced and the FE-link remapping was performed, in order to distribute the S-link occupancy more uniformly.
}
\label{fig:SlinkOccupancies}
\end{figure}
 
\begin{figure}[t]
\centering
\includegraphics[width=0.6\linewidth]{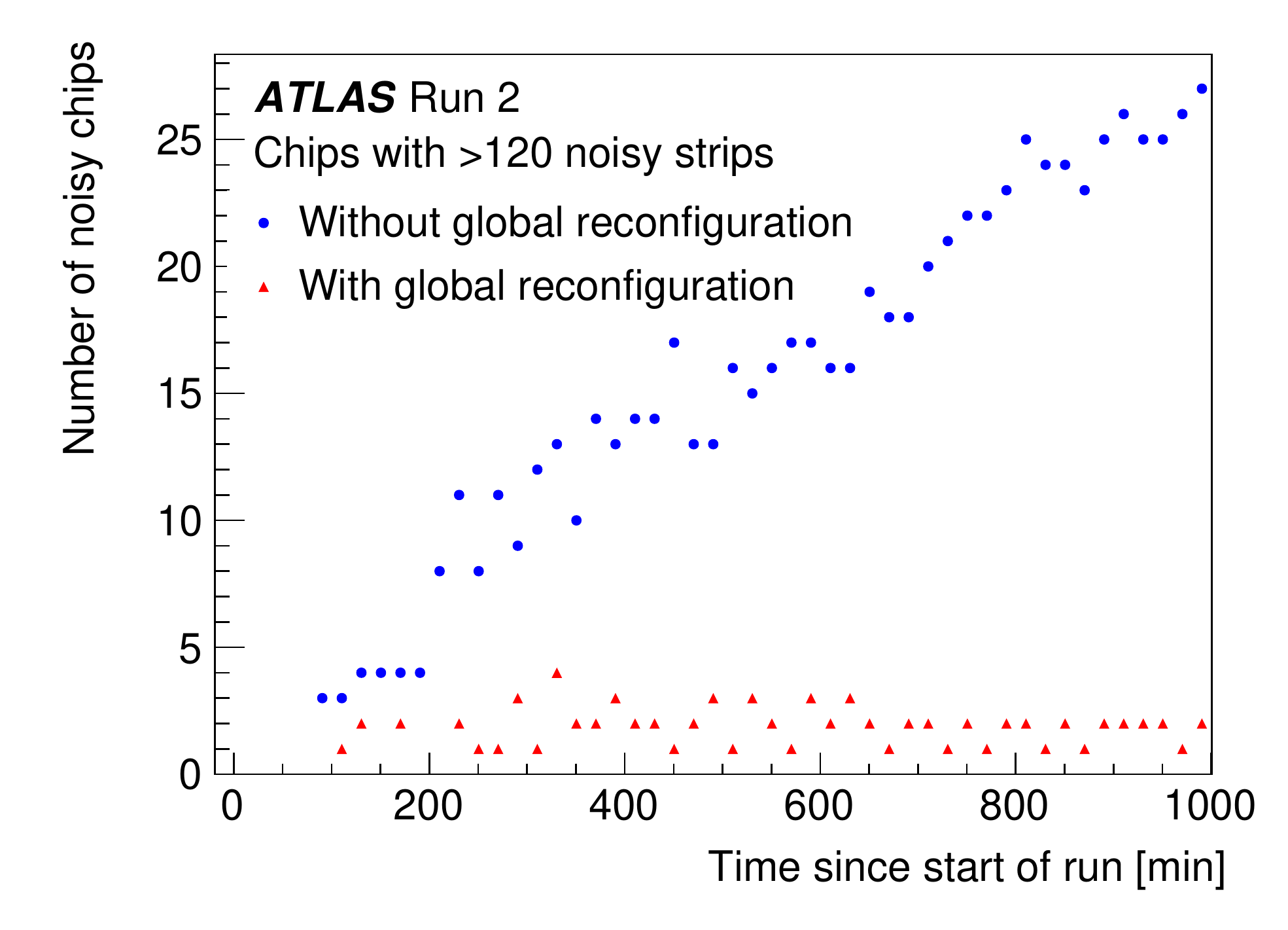}
\caption{Evolution of the number of noisy chips in a run with and without global reconfiguration activated, for two typical runs in Run~2.}
\label{fig:GRImpact}
\end{figure}
 
\subsubsection{Online module recovery}
\label{sec:DAQ:OnlineModuleRecovery}
 
Due to intense radiation from the $pp$ collisions, a charged particle occasionally passed through front-end electronics on the detector, causing an instantaneous error.
Energy deposited in the $p$-$i$-$n$ diodes in the modules, which receive the clock and trigger signals, resulted in desynchronisation of the module relative to the global ATLAS DAQ system.
An automated module recovery mechanism monitored desynchronisation errors in data buffered downstream of the S-link and reconfigured desynchronised modules.
This entire process, from detection of the desynchronisation error to the resumption of synchronised data arriving from the module, took 20--30~seconds.
In high-pile-up conditions ($\mu \sim 60$), module recovery was typically activated a few times per minute.
 
Single-event upsets (SEUs) within the internal threshold register of an ABCD chip caused the chip to become noisy or inefficient.
If this happened on the configuration register, the readout sequence of the ABCD chip was interrupted.
The rate of SEUs was studied during Run-1 operation~\cite{IDET-2013-01}.
As expected, a higher SEU rate was observed as the hit occupancy increased.
This could be understood approximately from earlier studies using test-beam data~\cite{NIMA-515-415-2003}, and indicated that errors induced by the SEUs would become more important in higher-luminosity operation in Run~2.
Since it is difficult to identify these problematic chips online, a system was implemented to apply a global reconfiguration to the entire SCT DAQ system, instead of using a targeted module recovery mechanism.
This process was activated every $\sim$90~minutes and took $\sim$1.2~seconds, so the loss of data is expected to be negligible.
Figure~\ref{fig:GRImpact} shows the number of noisy chips (defined as at least 120 channels with continuous high noise) as a function of time since the beginning of the run.
Typically, the number of noisy chips was reduced to less than five at any time.
 
\subsubsection{Handling ROD-busy assertions}
\label{sec:DAQ:onRODprocessing}
 
Towards the end of Run~1, the SCT started to have an average data-taking inefficiency of $\sim$0.9\%, caused by an increase in the number of permanent busy assertions coming from the RODs.
The impact on the ATLAS data-taking efficiency was limited by the automatic exclusion of the ROD from data-taking within about 0.2~seconds of the onset of BUSY.
However, the re-integration of the ROD into data-taking mode imposed about 7~seconds of BUSY, as the ROD must be masked from incoming triggers during its reconfiguration.
The overall time that the ROD was excluded from data-taking, from the initial ROD removal to its full re-integration, was $\sim$20~seconds, including the $\sim$7.2~seconds of dead time to ATLAS.
The permanent ROD-busy signals were induced by unrecognised illegal incoming bit patterns on the FE-links.
The impact on the DQ of taking data while the ROD was excluded was evaluated offline (discussed in Section~\ref{sec:DQmonitoring}).
 
This problem was significantly mitigated early in Run~2 by ROD firmware improvements, as well as by imposing a limit on the bit-stream length for incoming events on the FE-links.
Towards the end of Run~2, a ROD busy occurred once per day in typical $pp$ run conditions, and the average SCT dead time was typically less than 0.1\%.
 
\subsection{Detector calibration}
\label{sec:calibration}
 
Routine calibration of the SCT was performed to keep the hit efficiency above 99\% and the noise occupancy below $5 \times 10^{-4}$.
The basic procedure is described here, while more details can be found in Refs.~\cite{IDET-2013-01, JINST-3-P01003-2008}.
Optical calibrations were typically run first, to establish reliable communication for sending and receiving data via the data links.
Digital tests were then performed to exercise and verify the functionality of the front-end chips, the mask registers and the pipeline buffers.
Analogue optimisation was the main part of the calibration procedure, consisting of the trim range scan and the response curve test.
 
The trim range scan found the best configuration of the digital-to-analogue converters (DACs)\footnote{There are two types of DACs implemented in each ABCD chip. An 8-bit DAC changes the threshold setting for all 128 strips belonging to the chip. Variations in individual strips are compensated by a 4-bit DAC. Details are fully described in Refs.~\cite{IDET-2013-01, JINST-3-P01003-2008}.} to give a threshold equivalent to 1~fC for each ABCD chip.
This was achieved by injecting a fixed 1~fC test charge into the chips and measuring the variation in hit occupancy while changing the threshold. The occupancy was expected to be 100\% when the threshold was sufficiently low and 0\% when the threshold was high. The optimal setting was found in the DAC configuration where the hit occupancy was 50\% (vt50).
This measurement also provided an estimate of the noise, from the Gaussian spread of the hit occupancy as a function of the threshold, which was modelled by an error function.
In the response curve test, a similar threshold scan was performed using ten different values for the charge injected by the ABCD chips.
The gain was estimated from the slope of the curve relating the injected charge to vt50, and the input equivalent noise charge (ENC) was calculated from the output noise divided by the gain.
 
The calibration procedure also provided a list of problematic or unusable strips, which was uploaded to the conditions database to be used in the offline data reconstruction.
Another important calibration was the noise occupancy test, in which the noise occupancy was measured for ten different threshold settings between 0.3~fC and 1.0~fC. This test was done occasionally to understand the noise level in terms of the noise occupancy.
Results from the noise occupancy test as well as the response curve test are discussed in Section~\ref{sec:Noise}.
 
The calibration procedure was typically performed every few weeks.
Since the HV needed to be set to the nominal values during calibration, to ensure operational detector response, it could only be done without beam circulating in the LHC, for safety reasons in case of unexpected beam losses.
The times available for calibration were dedicated calibration periods between LHC fills, and longer intervals in LHC operations to allow technical interventions.
Due to improvements, the time between LHC fills was gradually reduced to about 1.5~hours.
This allowed more physics data to be provided in the later years of Run~2, but the calibration time was then significantly limited.
To speed up the calibration, the number of scan points for the trim range scan was reduced by $\sim$50\%, for example, while still retaining reliability and robustness.
The total time for the full SCT calibration was reduced to 1.5~hours.
This allowed a calibration frequency of at least once per month, and the fraction of noisy strips was kept below 0.3\%, as discussed in detail in Section~\ref{sec:OfflineCalibration}.
 
\subsection{Data quality}
 
\subsubsection{Data quality monitoring and assessment}
\label{sec:DQmonitoring}
 
To assure high-quality data for physics analyses, it is essential to monitor the DQ during data-taking.
Evaluation of the SCT DQ was based on the ATLAS experiment's common DQ framework, which is summarised in Ref.~\cite{ATLAS-DAPR-2018-01-001}.
Several important parameters were monitored using histograms produced during the fast processing of sampled events~\cite{JPhysConf-219-032037-2010}.
These included information about the SCT performance, such as hit efficiency and occupancy, and also information from reconstructed tracks, such as the number of clusters associated with a track.
These histograms were automatically produced and evaluated by the online data-quality monitoring framework~\cite{JPhysConf-219-022035-2010}, which would warn the DQ monitoring operators when a potential problem was identified, based on predefined criteria.
 
When the data had been fully reconstructed, the DQ was thoroughly assessed offline.
When a problem was found, a `defect' flag was set for the corresponding luminosity block. This is a period of time (typically 1~minute), during which the instantaneous luminosity, the detector and trigger configurations, and the other important run conditions are considered to be constant.
The defect flags are stored in a database~\cite{EPJC-72-1960-2012}.
Several different defect flags were prepared, and classified as being either `intolerable' or `tolerable', depending on the severity of the problem and its impact on the data quality.
The SCT-specific DQ flags were mostly the same as those used in Run~1~\cite{IDET-2013-01}.
 
Data flagged with an intolerable defect are excluded from physics analyses. For the SCT, three major types of intolerable defect were defined:
\begin{itemize}
\item Significant acceptance loss for the tracking, because multiple RODs were excluded from the DAQ.
\item One or more DAQ crates (each holding about 12.5\% of the modules) were excluded from the DAQ.
\item The SCT HV was not in the `ready' state: the HV was less than the operational setting, which was typically 150--250~V depending on module location.
\end{itemize}
 
\begin{table}[t]
\centering
\caption{SCT DQ defect summary in Run 2.}
\begin{tabular}{l|cccc|c}
\hline
Year & 2015 & 2016 & 2017 & 2018 & Total\\
Luminosity recorded by ATLAS [fb$^{-1}$] & 3.6 & 35.5 & 46.5 & 60.2 & 145.8\\
\hline
Defect description & \multicolumn{5}{c}{Fraction of luminosity affected [\%]}\\
\hline
\multicolumn{6}{c}{Intolerable defects}\\
\hline
Major acceptance loss due to ROD exclusion  & \tablenum{0.22} & \tablenum{0}    & \tablenum{0.01} & \tablenum{0}    & \tablenum{0.01}\\
DAQ crate(s) excluded from readout          & \tablenum{0}    & \tablenum{0.01} & \tablenum{0.01} & \tablenum{0}    & \tablenum{0.01}\\
SCT HV not ready                            & \tablenum{0.01} & \tablenum{0.10} & \tablenum{0.04} & \tablenum{0.01} & \tablenum{0.04}\\
SCT simulation data flag activated          & \tablenum{0}    & \tablenum{0}    & \tablenum{0}    & \tablenum{0.21} & \tablenum{0.09}\\
\hline
\multicolumn{6}{c}{Tolerable defects}\\
\hline
$>40$ modules noisy                         & \tablenum{6.0}  & \tablenum{13}   & \tablenum{20}   & \tablenum{51}   & \tablenum{31}\\
Minor acceptance loss due to ROD exclusion  & \tablenum{14}   & \tablenum{5.1}  & \tablenum{0.68} & \tablenum{0.55} & \tablenum{2.0}\\
Low efficiency                              & \tablenum{4.3}  & \tablenum{0}    & \tablenum{1.4}  & \tablenum{1.6}  & \tablenum{1.2}\\
$>40$ modules with link errors              & \tablenum{2.4}  & \tablenum{1.6}  & \tablenum{0.34} & \tablenum{0}    & \tablenum{0.57}\\
SCT not at standard HV                      & \tablenum{0.58} & \tablenum{0.08} & \tablenum{0.08} & \tablenum{1.2}  & \tablenum{0.54}\\
Trip of one power supply crate              & \tablenum{0}    & \tablenum{0}    & \tablenum{0}    & \tablenum{0.06} & \tablenum{0.03}\\
\hline
\end{tabular}
\label{table:DQ_summary}
\end{table}
 
The fraction and evolution of intolerable SCT defects during Run~2 is summarised in Table~\ref{table:DQ_summary}.
In 2015, the most significant DQ inefficiency was caused by multiple ROD failures.
This was because the trigger rate was significantly higher than that in Run~1.
This problem became less severe in later years, due to the ROD firmware improvements described in Section~\ref{sec:DAQ:onRODprocessing}.
From 2016 onwards, high DQ efficiency was maintained.
Manual adjustments to the SCT were sometimes necessary, for example when unexpected changes in beam conditions occurred. These led to inevitable small delays in turning the SCT back on, resulting in some data being recorded with intolerable defects because the SCT voltage was not ready.
In 2018, a one-time incident resulted in significant DQ loss. An SCT simulation-mode data flag was unintentionally activated during a physics run.
This inhibited further SCT data reconstruction, and consequently no track was provided to the high-level trigger.
Software changes have been implemented to prevent this from happening again.
Through the whole of Run~2, the DQ inefficiency due to SCT intolerable defects was 0.15\%, corresponding to 217~pb$^{-1}$.
 
The tolerable SCT defect affecting the most integrated luminosity originated from having more than 40 noisy modules, where a noisy module is defined by the average noise occupancy being greater than 0.15\%. The frequency of this defect increased in the later years of Run~2. Other tolerable defects were:
\begin{itemize}
\item Minor acceptance loss for tracking, because a single ROD was excluded from the DAQ.
\item Hit efficiency less than 98\% on average, and less than 99\% when measured using only the first bunch crossing of a bunch train.
\item More than 40 modules with errors in the FE-link.
\item SCT being operated at non-standard HV values. This happened rarely, when the HV was scanned to study radiation damage to the sensors. The HV was kept sufficiently high that the impact on tracking is considered marginal.
\item At least one power supply crate tripped (where one crate holds $\sim$1.14\% of the modules).
\end{itemize}
This final defect flag was introduced in 2018, because of a few incidents related to a tripped power supply crate observed during physics runs. This was attributed to aging of the components. The degraded components will be replaced to mitigate this problem in Run~3.
 
\subsubsection{Noisy strip identification}
\label{sec:OfflineCalibration}
 
Noisy strips were searched for using a special data stream triggered on empty LHC bunch crossings, so that no collision hits should be present.
Events were triggered randomly at a rate of 2--10~Hz, giving sufficient data to determine the number of noisy strips run-by-run.
When a strip had a noise occupancy greater than 1.5\%, the strip was identified as being noisy and masked during offline data reconstruction.
 
Figure \ref{fig:NoStrVsDate} shows the fraction of noisy strips as a function of date, for both Run~1 and Run~2.
Typically, this number rose from about $10^{-4}$ to $3 \times 10^{-3}$, until a calibration was performed and the fraction of noisy strips dropped again.
This was due to resetting the thresholds, which drift as charge is accumulated in the chips.
On a shorter timescale, large run-by-run variations are observed.
The main reason for these fluctuations was bit flips in the threshold register, as discussed in Section~\ref{sec:DAQ:OnlineModuleRecovery}.
Although these strips were usually recovered by a global reconfiguration during the run, they are identified as noisy when the average noise occupancy in the run exceeds the threshold.
 
\begin{figure}[t!]
\centering
\includegraphics[width=0.8\columnwidth]{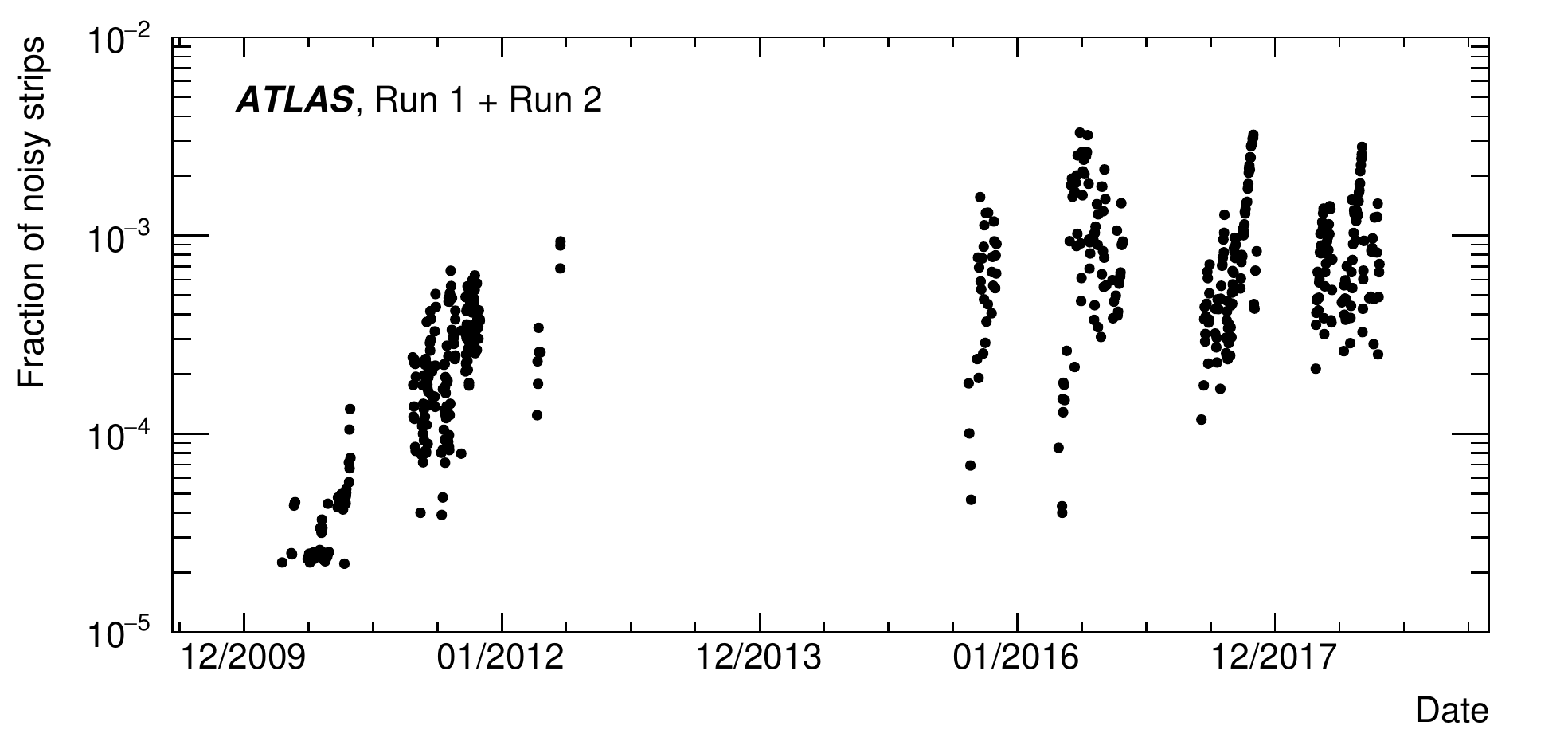}
\caption{Evolution of the fraction of noisy strips as a function of date, including both Run~1 and Run~2. Each point corresponds to one run.}
\label{fig:NoStrVsDate}
\end{figure}
 
 
\section{SCT performance}
\label{sec:performance}
 
The performance of the SCT had to be maintained in the harsh experimental environment of LHC Run~2, with high levels of pile-up and radiation.
To maximise the tracking efficiency for charged particles, the SCT is regularly calibrated to maintain a hit efficiency above $99\%$, while the noise occupancy remains below $5 \times 10^{-4}$. Hits found on contiguous SCT strips are grouped into a cluster, which forms the input to the tracking algorithm.
It is therefore important that the distribution of cluster sizes remains stable.
 
These key parameters are studied and discussed in this section, using the abundant data sample collected during Run~2.
Since the radiation levels depend on the location within the SCT, some performance results are shown separately for different regions.
For convenience, the parameter $\etaindex$ is introduced to specify the position along the $z$-axis of the 12 modules forming a stave of the SCT barrel: $\etaindex = -6, ..., -1, 1, ..., 6$, with increasing $z$-coordinate.
For the endcaps, the terms `inner ring', `middle ring' and `outer ring' specify the radial positions of the modules.
 
\subsection{Hit efficiency}
\label{sec:HitEfficiency_gen}
 
\subsubsection{Hit efficiency measurement}
\label{sec:HitEffic}
 
For a given SCT sensor, a hit efficiency for each track is defined by the equation
\begin{eqnarray}
\varepsilon &=& \frac{\nclu}{\nclu + \nhole}\ ,
\label{equation:efficiency}
\end{eqnarray}
where $\nclu$ and $\nhole$ denote, respectively, the number of clusters and the number of holes associated with the track.
If no cluster is found along the track in an SCT layer where one is expected, its absence is counted as a hole, unless there is a known reason for this absence (such as disabled modules, links or chips, or transient errors), in which case it is excluded from the hit efficiency calculation.
However, an exception is made for disabled or noisy strips, which can still be counted as a hole.
 
The method used to measure the hit efficiency is the same as that discussed in Ref.~\cite{IDET-2013-01}.
The calculation uses only well-reconstructed tracks in the ID.
Events with more than 500 tracks are rejected, because of the high probability of reconstructing fake tracks from incorrect combinations of clusters in such events.
Furthermore, the tracks used are required to satisfy the following:
\begin{itemize}
\item transverse momentum $\pt > 1$~GeV;
\item track fit quality identified as `good';\footnote{All tracks must have a fit quality satisfying $\chi^2 / N_\mathrm{dof} < 3$ ($N_\mathrm{dof}$ = number of degrees of freedom). Details are given in Ref.~\cite{SOFT-PUB-2007-007}.}
\item transverse impact parameter $|d_0|<10$~mm;
\item number of clusters on SCT sensors, excluding the sensor under consideration, $\nhitsct\ge 6$;
\item number of holes in Pixel detector, $\nholepix=0$;
\item number of clusters in Pixel IBL or B-layer (second innermost Pixel layer), $\nhitpix\ge1$;
\item incident angle with respect to SCT module surface, $|\phiin|<40\degr$.
\end{itemize}
In Run~2, the track-finding required the total number of clusters (holes) in the Pixel and SCT combined to be $\nhit^\mathrm{Si}\ge7$ ($\nhole^\mathrm{Si}\le2$).
To avoid introducing any bias into the hit efficiency measurement from these requirements, only tracks that would be reconstructed regardless of the state of the sensor under consideration are used.
Therefore, when excluding the sensor under consideration, $\nhit^\mathrm{Si}\ge7$ and $\nhole^\mathrm{Si}\le1$ are required.
Additionally, any cluster unassociated with the track, but found within a distance $\dcut$ around the corresponding track position, is included in $\nclu$ and excluded from $\nhole$ in Eq.~(\ref{equation:efficiency}).
The value of $\dcut$ is set to 200~$\mu$m unless stated otherwise.
 
\begin{figure}[t]
\centering
\subfloat[]{
\includegraphics[width=0.48\textwidth]{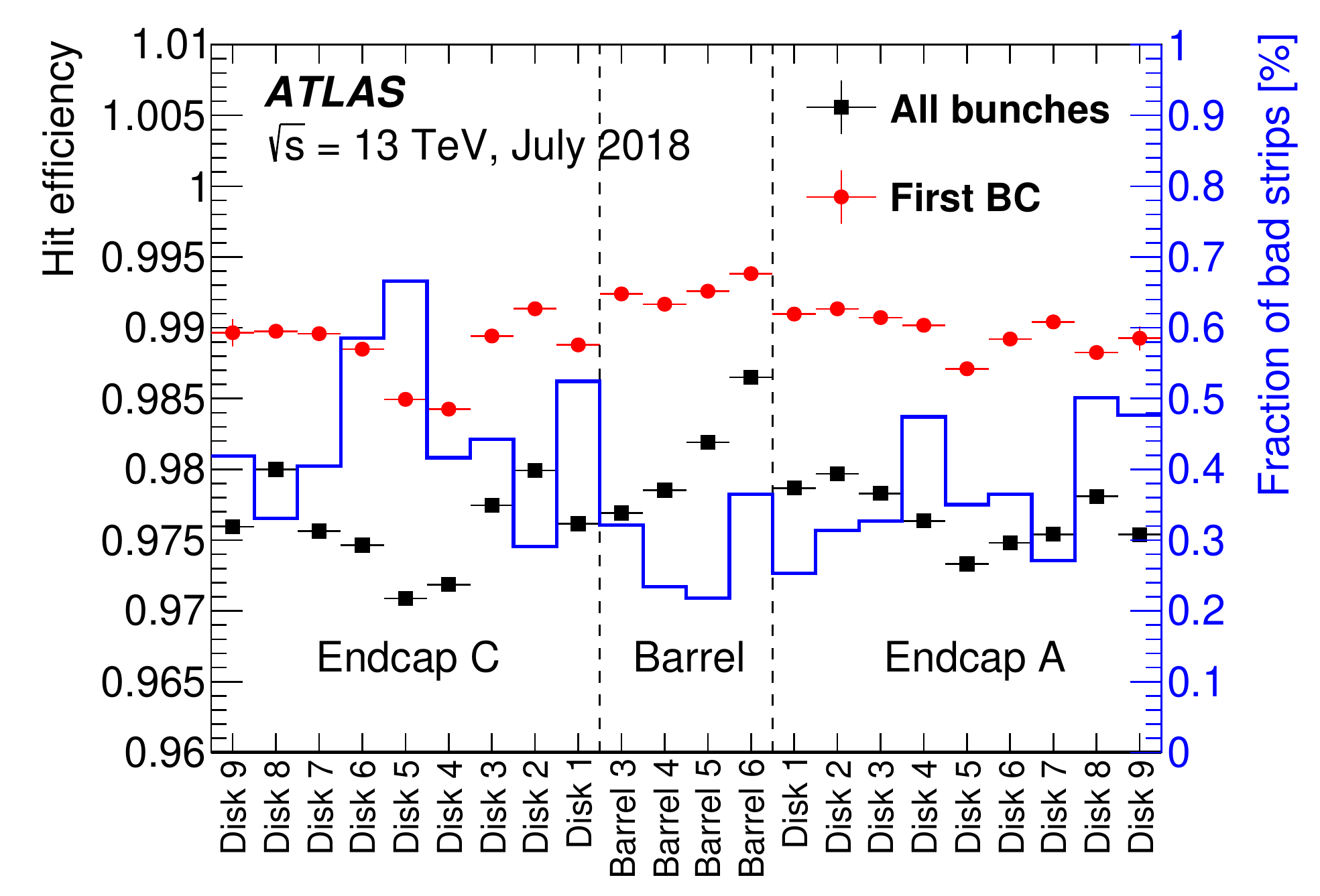}
\label{fig:HitEfficiencySummary:a}
}
\subfloat[]{
\includegraphics[width=0.48\textwidth]{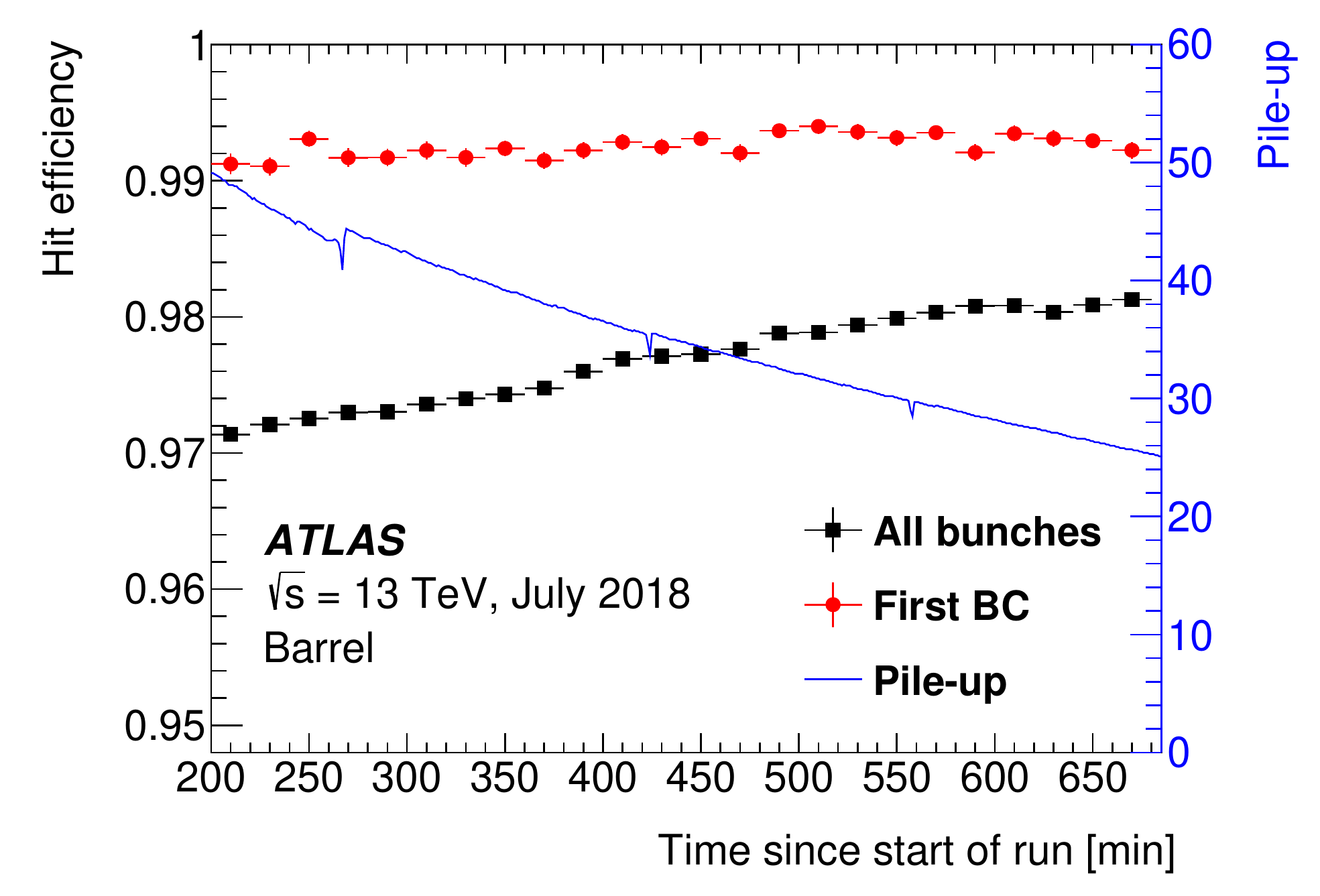}
\label{fig:HitEfficiencySummary:b}
}
\caption{
Hit efficiency measured in a typical $pp$ collision run in July 2018.
Values measured using all bunches (only first BC) are shown by the black squares (red dots).
\protect\subref{fig:HitEfficiencySummary:a} The hit efficiency averaged over each SCT barrel layer and endcap disk.
The fraction of bad strips is superimposed.
\protect\subref{fig:HitEfficiencySummary:b} Evolution of the hit efficiency from barrel modules in the run.
The blue line shows the evolution of pile-up in this run.
} \label{fig:HitEfficiencySummary}
\end{figure}
 
Figure~\ref{fig:HitEfficiencySummary}\subref{fig:HitEfficiencySummary:a} shows the hit efficiency for each barrel layer and each endcap disk measured in a $pp$ collision run in July 2018, with a pile-up value $\mu$ between 25 and 50.
The average hit efficiency measured using all bunches in a train ranges between 97\% and 98\%, which is slightly lower than the expected hit efficiency for the SCT sensors.
This is due to the `01X' readout mode, defined in Section~\ref{sec:DAQ:HighPileup}.
A real hit is lost when, by chance, the strip also has another hit in the preceding bunch.
If the hit efficiency is measured using only the first bunch crossing in each bunch train (called `first BC'), it is around 99\%.
The result using the first BC is therefore considered to be the intrinsic hit efficiency of the SCT sensor, while that determined using all bunches is also important because it impacts the tracking performance.
Thanks to the redundant design of the SCT, consisting of multiple barrel layers and endcap disks typically providing eight SCT clusters per track, this 1\% loss of hit efficiency has a negligible impact on the tracking efficiency.
The fraction of bad strips in each layer, including both the problematic strips identified by the calibration (Section~\ref{sec:calibration}) and the noisy strips (Section~\ref{sec:OfflineCalibration}), is superimposed on Figure~\ref{fig:HitEfficiencySummary}\subref{fig:HitEfficiencySummary:a}.
There tends to be anti-correlation between the hit efficiency and the fraction of bad strips, indicating that the non-uniform distribution of hit efficiency across the SCT regions can be partially attributed to the bad strips.
 
Figure~\ref{fig:HitEfficiencySummary}\subref{fig:HitEfficiencySummary:b} shows the hit efficiency measured using barrel modules as a function of time since the start of a run, with the evolution of the level of pile-up superimposed.
The intrinsic hit efficiency found using only the first BC is constant and remains above 99\%.
The difference between the intrinsic hit efficiency and the hit efficiency measured using all bunches is about 2\% at the beginning of the run, when $\mu = 50$, decreasing to about 1\% at the end of the run, when $\mu = 25$.
From this, the inefficiency induced by the 01X readout mode is seen to be proportional to the level of pile-up, as expected.
 
Systematic uncertainties in the hit efficiency are estimated as described in Ref.~\cite{IDET-2013-01}. All uncertainties are symmetrised by taking the larger of the upward and downward variations.
Uncertainties arising from inclusion of unassigned clusters within $\dcut$ are estimated by varying the value of $\dcut$ between 0 and $500~\mu$m.
The change in the measured efficiency when a cluster is also required on the other side of the sensor in the same module is taken to be another systematic uncertainty.
The intrinsic hit efficiencies using the first BC in a low-pile-up run are measured to be (99.00$\pm$0.53)$\%$ (endcap A), (99.26$\pm$0.21)$\%$ (barrel) and (98.83$\pm$0.51)$\%$ (endcap C).
These values are consistent with $99\%$ and the impact on the tracking efficiency is negligible.
 
\subsubsection{Interstrip hit efficiency}
 
A precise study of the hit efficiency as a function of position, for charged particles passing in between two strips of an SCT sensor (referred to as the interstrip hit efficiency), was performed using test beams~\cite{NIMA-538-384-2005} before the construction of the SCT.
A slight drop in the hit efficiency was observed in the centre of the gap between two strips, after irradiation and in the presence of a magnetic field.
A comparable feature can be investigated using data collected during Run~2 for the real SCT sensors, which have been irradiated during ten years of LHC operations.
Figure~\ref{fig:InterStripB3E1} shows this efficiency as a function of relative $x$, the distance of a track from one strip towards the next.
Position resolution results in an uncertainty of about 20~$\mu$m~\cite{IDET-2013-01} in the estimate of relative $x$ for each track.
For this measurement, only modules in barrel layer~3 with $|\etaindex| = 1$ are chosen, because they are located in the innermost layer, where high radiation fluence is expected, and they have tracks almost perpendicular to the sensors.
In order to avoid inefficiencies from other sources, only modules that are fully active and have HV greater than 150~V are used.
An incident angle of $\phiin = -5^\circ$ (with the direction defined in the same sense as for global $\phi$) was chosen because it is close to the Lorentz angle~\cite{IDET-2013-01}.
A small dip of 0.8\% is observed in the interstrip hit efficiency in the middle of the gap between two strips, qualitatively reproducing the result obtained from test-beam data.
 
\begin{figure}[t]
\centering
\includegraphics[width=0.6\textwidth]{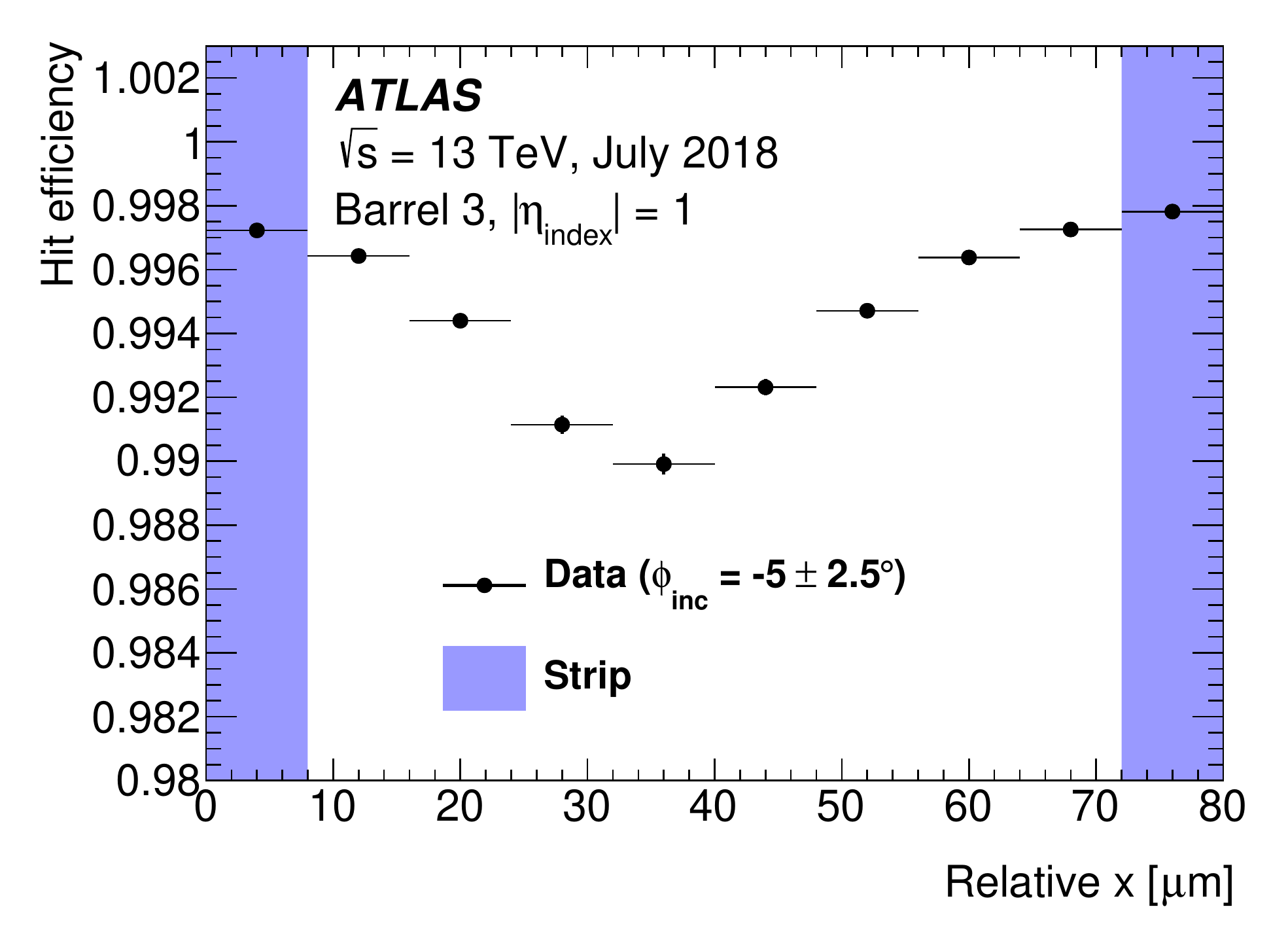}
\caption{
Interstrip hit efficiency as a function of relative $x$, the distance from one strip towards the next, with $\phiin = -5\degr$.
The measurement is performed using data from a low-pile-up run in July 2018.
Only modules with $|\etaindex| = 1$ are used.
All possible interstrip combinations are averaged.
The blue shaded regions near 0 and 80~$\mu$m correspond to the strip implant positions.
}
\label{fig:InterStripB3E1}
\end{figure}
 
\subsubsection{HV dependence of hit efficiency}
\label{sec:HitEffHV}
 
During the second half of Run~2, the $n$-type bulk of the silicon sensors was expected to invert into quasi-$p$-type.
For type-inverted sensors, the depletion region develops from the backplane towards the strips.
The operational HV therefore needs to be sufficiently higher than $\vfd$ to allow the depletion region to reach the strips and maximise the charge collection efficiency.
In order to confirm the optimal HV, as well as to study radiation damage effects, the HV dependence of the hit efficiency was occasionally measured during physics runs.
 
\begin{figure}[t]
\centering
\includegraphics[width=0.6\textwidth]{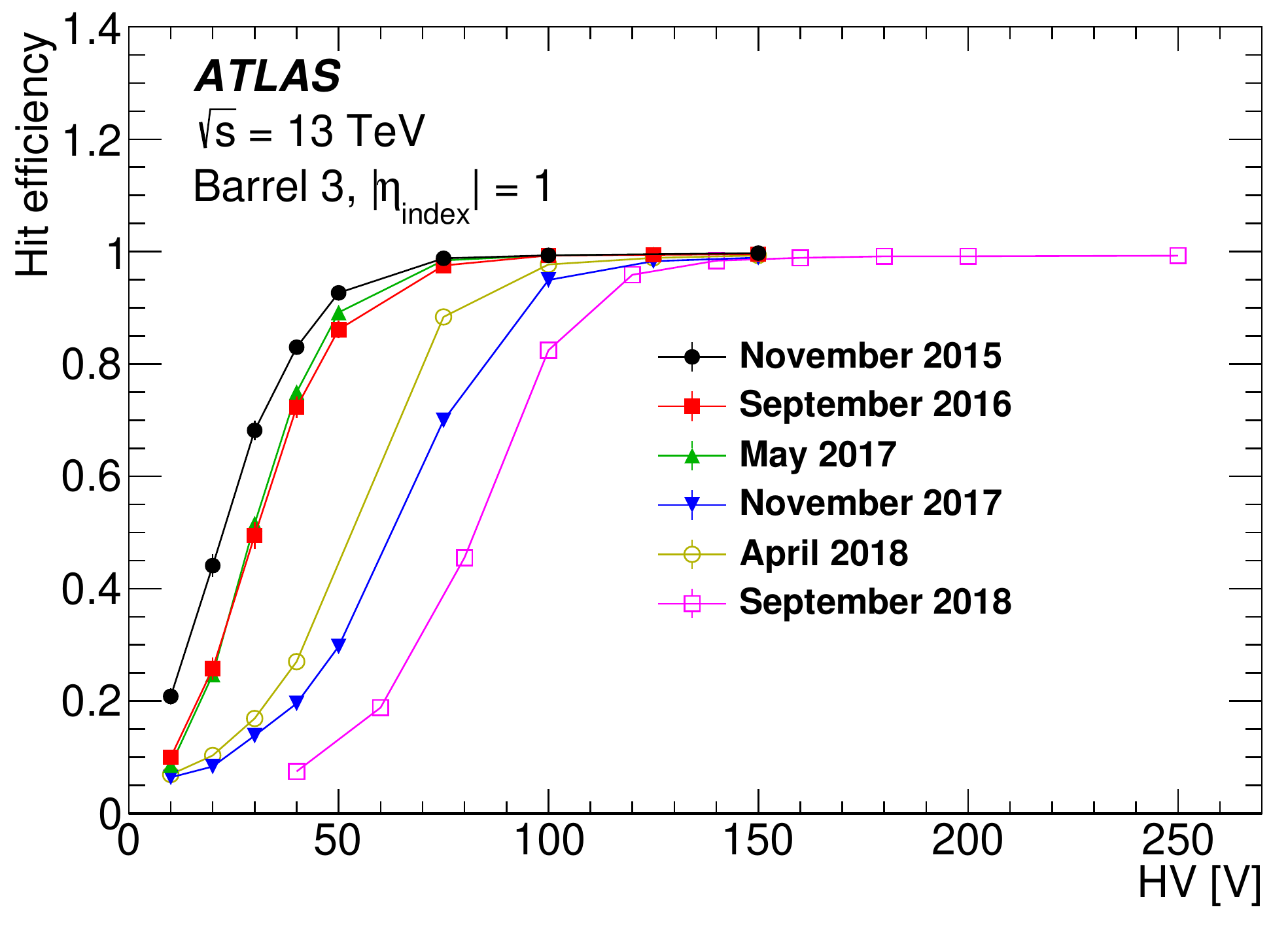}
\caption{Hit efficiency as a function of HV for modules with $|\etaindex| = 1$ in barrel layer 3, measured from November 2015 to September 2018.
}
\label{fig:HVdependence}
\end{figure}
 
Figure~\ref{fig:HVdependence} shows comparisons of hit efficiency as a function of HV for modules with $|\etaindex| = 1$ in barrel layer~3.
The data were collected throughout Run~2, between November 2015 and September 2018.
Until May 2017, the efficiency curves remained almost unchanged, rising sharply as the HV is increased.
In November 2017, however, the increase in efficiency with HV was slower, as expected from type inversion.
Between November 2017 and April 2018, this increase became slightly faster again, due to beneficial annealing which occurred during the year-end technical shutdown.
Around the end of Run~2 (September 2018), the start of the efficiency plateau was at around 140~V, close to the original operational HV of 150~V.
The hit efficiency was measured to be slightly lower in September 2018, and the operational HV was therefore increased to 250~V to recover the efficiency.
 
\subsection{Cluster width and Lorentz angle}
 
The cluster width is defined as the number of contiguous strips with hits that form a cluster.
Tracks used to measure the cluster width must satisfy the following criteria, matching those used in the previous study~\cite{IDET-2013-01}: $\pt > 0.5$~GeV, $|d_0| < 1$~mm, $\nhitsct \geq 8$ and $\nhitpix \geq 2$. Other track selection requirements are the same as those used in the hit efficiency measurement (see Section~\ref{sec:HitEffic}).
Figure~\ref{fig:oneDCluster} shows cluster widths measured in each of the four barrel layers at a HV of 150~V.
These distributions peak at 1 and have a long tail, with a mean width of about 1.2.
A cluster width of 2 is induced by charge sharing between two neighbouring strips.
The exponentially decreasing tail of the cluster width distribution up to ${\sim}10$ is induced by $\delta$-rays~\cite{IDET-2013-01}; for the long tail component greater than ${\sim}10$, contributions from multiple charged particles, where their hits are merged into one cluster, may be more prominent.
Clusters with zero width are due to tracks with a hole in the corresponding barrel layer.
 
\begin{figure}[t!]
\centering
\includegraphics[width=0.6\textwidth]{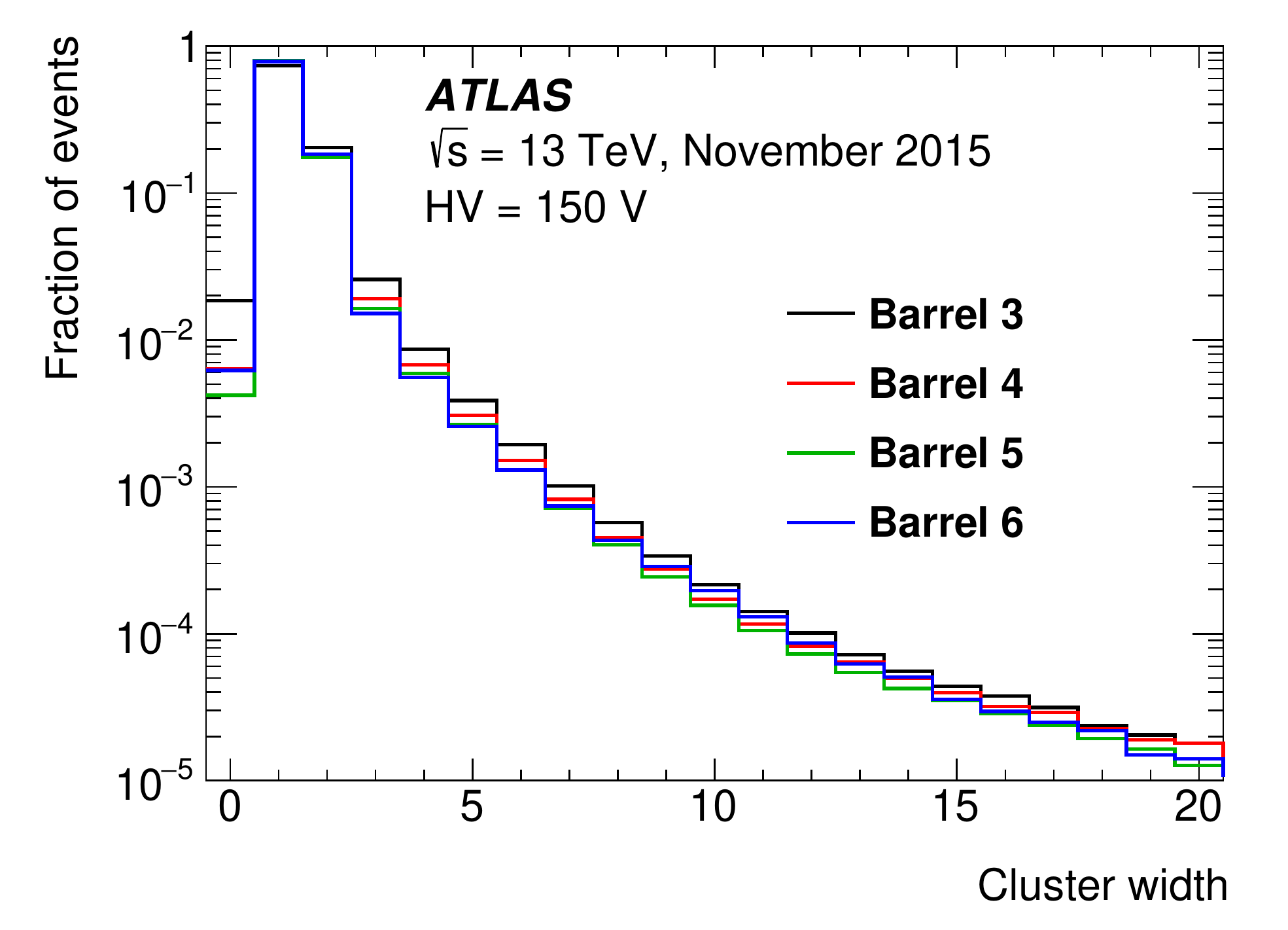}
\caption{Cluster width distributions in the four barrel layers, measured from a $pp$ collision run in November 2015 at a HV of 150~V.}
\label{fig:oneDCluster}
\end{figure}
 
\begin{figure}[t!]
\centering
\subfloat[]{
\includegraphics[width=0.45\textwidth]{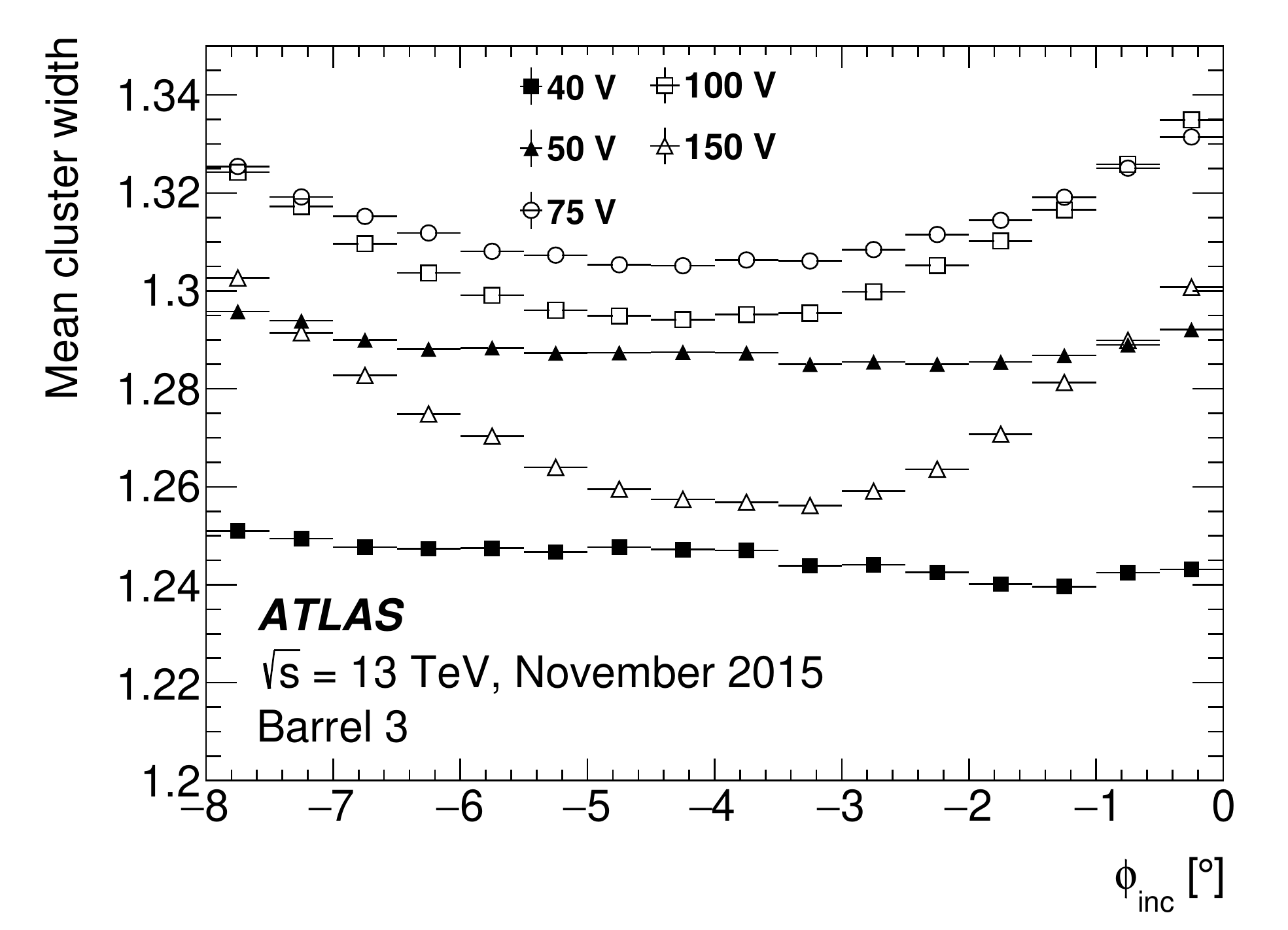}
\label{fig:summaryScanSCT:a}
}
\subfloat[]{
\includegraphics[width=0.45\textwidth]{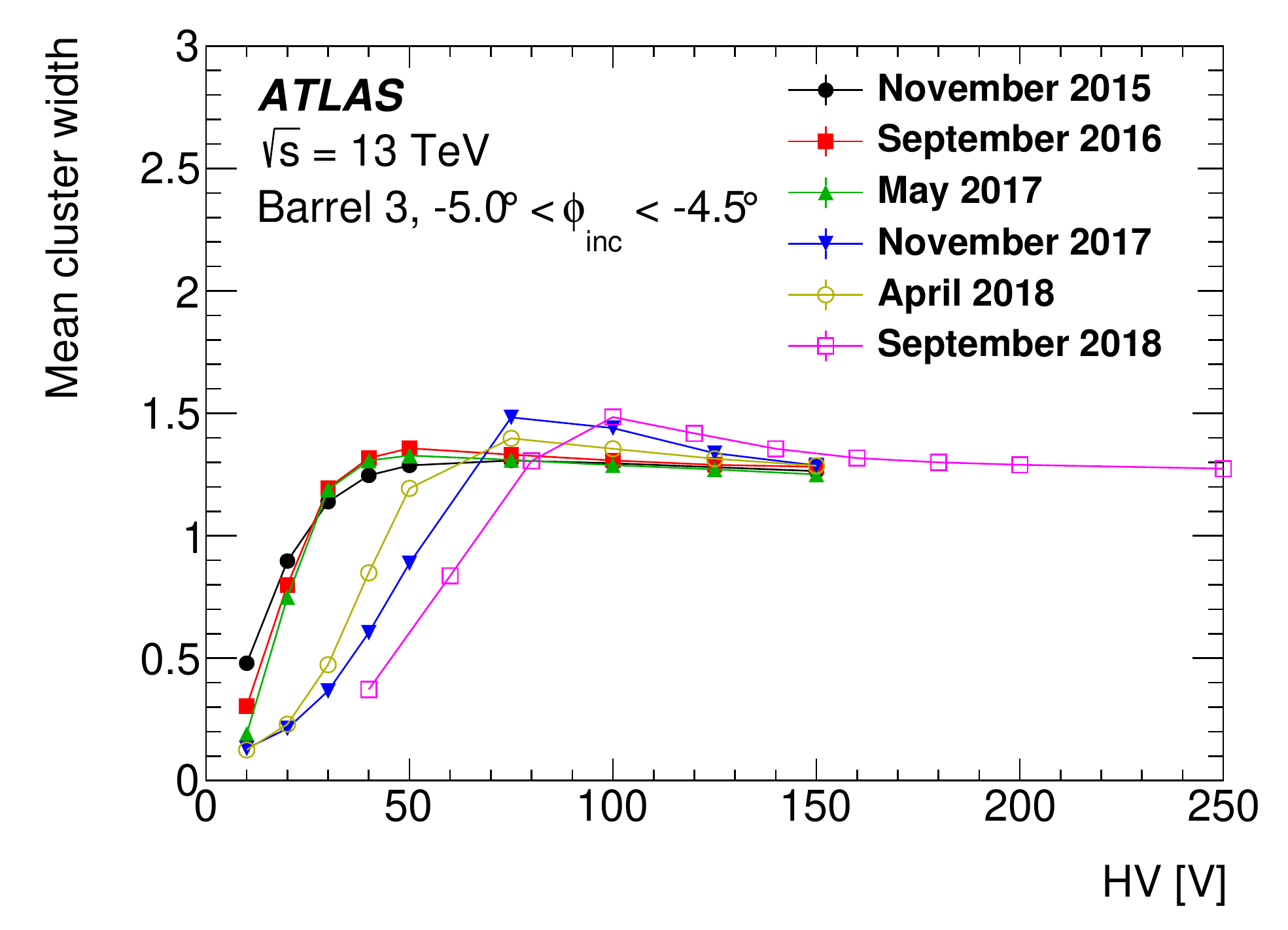}
\label{fig:summaryScanSCT:b}
}
\caption{
\protect\subref{fig:summaryScanSCT:a} Mean cluster width as a function of $\phiin$ for barrel layer~3 at five values of HV, measured in November 2015.
\protect\subref{fig:summaryScanSCT:b} Evolution of mean cluster width as a function of HV for barrel layer~3 during Run~2.
}
\label{fig:summaryScanSCT}
\end{figure}
 
The HV dependence of the mean cluster width has been investigated using HV scan data collected during physics data-taking runs.
The level of pile-up was constant to within 4\%--7\% during these HV scans.
Figure~\ref{fig:summaryScanSCT}\subref{fig:summaryScanSCT:a} shows distributions of mean cluster width as a function of $\phiin$ for barrel layer~3 at five different HV values.
The mean cluster width increases as the HV is increased up to 75~V, but then it decreases by a few percent as the HV is increased further to 150~V.
This behaviour suggests there is a trade-off between better charge collection at higher HV, increasing the cluster width, and faster charge collection, which decreases the cluster width by suppressing transverse charge diffusion.
To evaluate the HV dependence of the cluster width, only tracks with $\phiin = [-5.0\degr, -4.5\degr]$, close to the minimum in cluster width, are used.
Figure~\ref{fig:summaryScanSCT}\subref{fig:summaryScanSCT:b} shows the HV dependence of the mean cluster width measured in barrel layer~3, from data collected between 2015 and 2018 in Run~2.
Until May 2017, the cluster width increased rapidly as the HV was increased to around 50~V.
Since November 2017, however, the increase of the cluster width has become slower.
This is consistent with the observed change in the HV dependence of the hit efficiency, discussed in Section~\ref{sec:HitEffHV}, and appears to be induced by type inversion.
The peak in the mean cluster width around 70--100 V is more visible in the later curves; further investigation is needed to understand this feature.
 
The displacement from zero of the position of the minimum cluster width (MCW) in each of these distributions, $|\phiMCW|$, gives a good estimate of the Lorentz angle, $\thetaL$, resulting from the deviation of the drift-charge direction from the electric field vector, due to the Lorentz force acting on the charge carrier in the axial magnetic field.
The Lorentz angle in a magnetic field $\vec{B}$ is given by
\begin{eqnarray*}
\tan \thetaL &= \gamma_\mathrm{H} \mu_\mathrm{d} |\vec{B}|,
\end{eqnarray*}
where $\gamma_\mathrm{H}$ and $\mu_\mathrm{d}$ denote the Hall factor and the charge-carrier mobility, respectively.
The charge-carrier mobility is affected by the HV, the thickness of the depleted region and the temperature.
The Lorentz angle is important in the barrel modules, where the electric field from the HV is perpendicular to the magnetic field from the solenoid.
When the incident angle of a particle equals the Lorentz angle, all induced charge drifts along the particle direction, giving the minimum cluster width.
Only negatively charged tracks are selected, because the tilt of barrel SCT modules relative to the radial direction results in positively charged tracks populating only the lower end of the $\phiin$ distribution, below the minimum.
The track selection requirement is lowered to $\nhitsct \geq 7$, because the tighter requirement was found to bias the shape of the cluster width distributions.
 
\begin{figure}[t!]
\centering
\subfloat[]{
\includegraphics[width=0.45\textwidth]{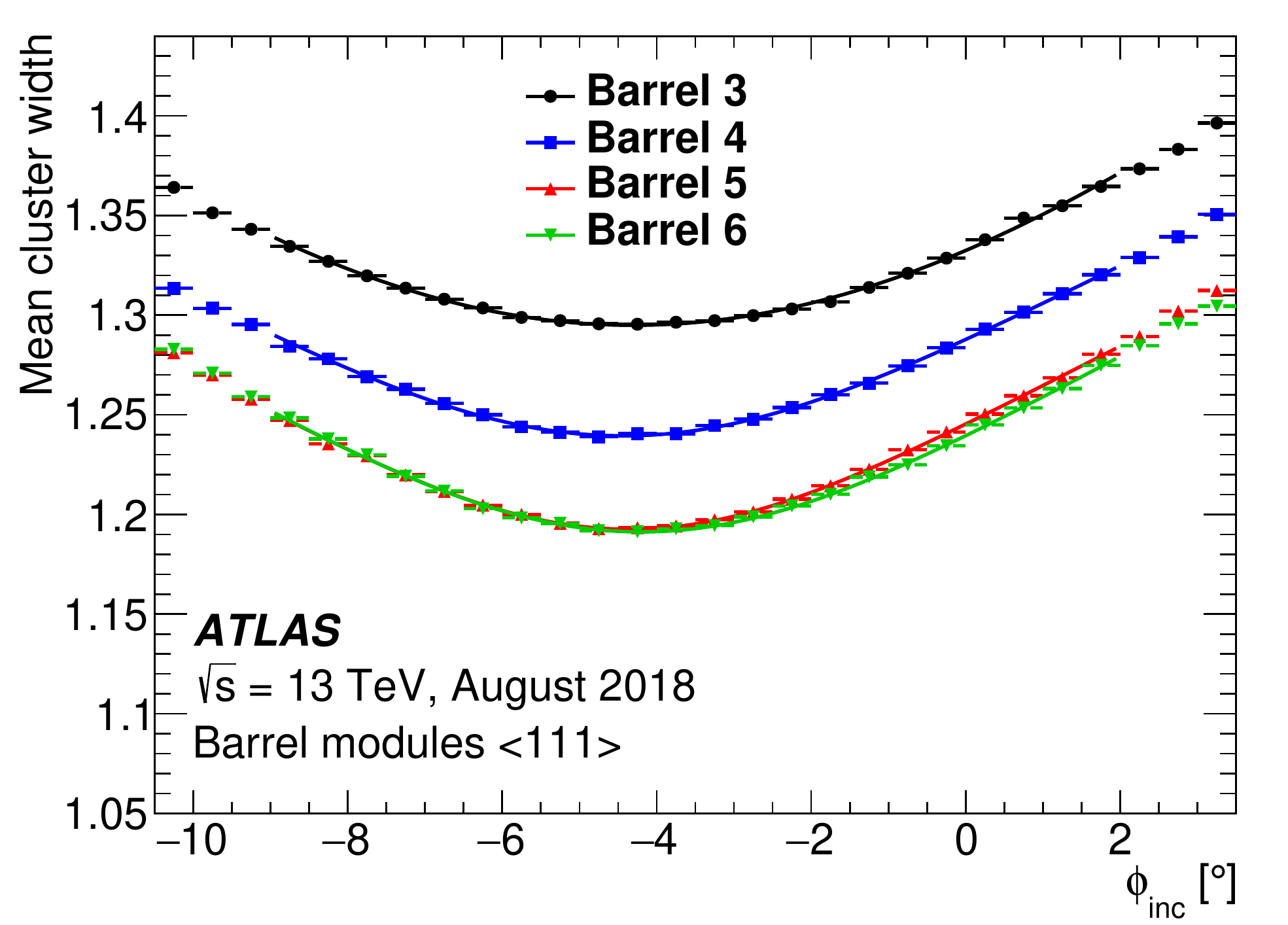}
\label{fig:clusize_vs_phiin_LA:a}
}
\subfloat[]{
\includegraphics[width=0.45\textwidth]{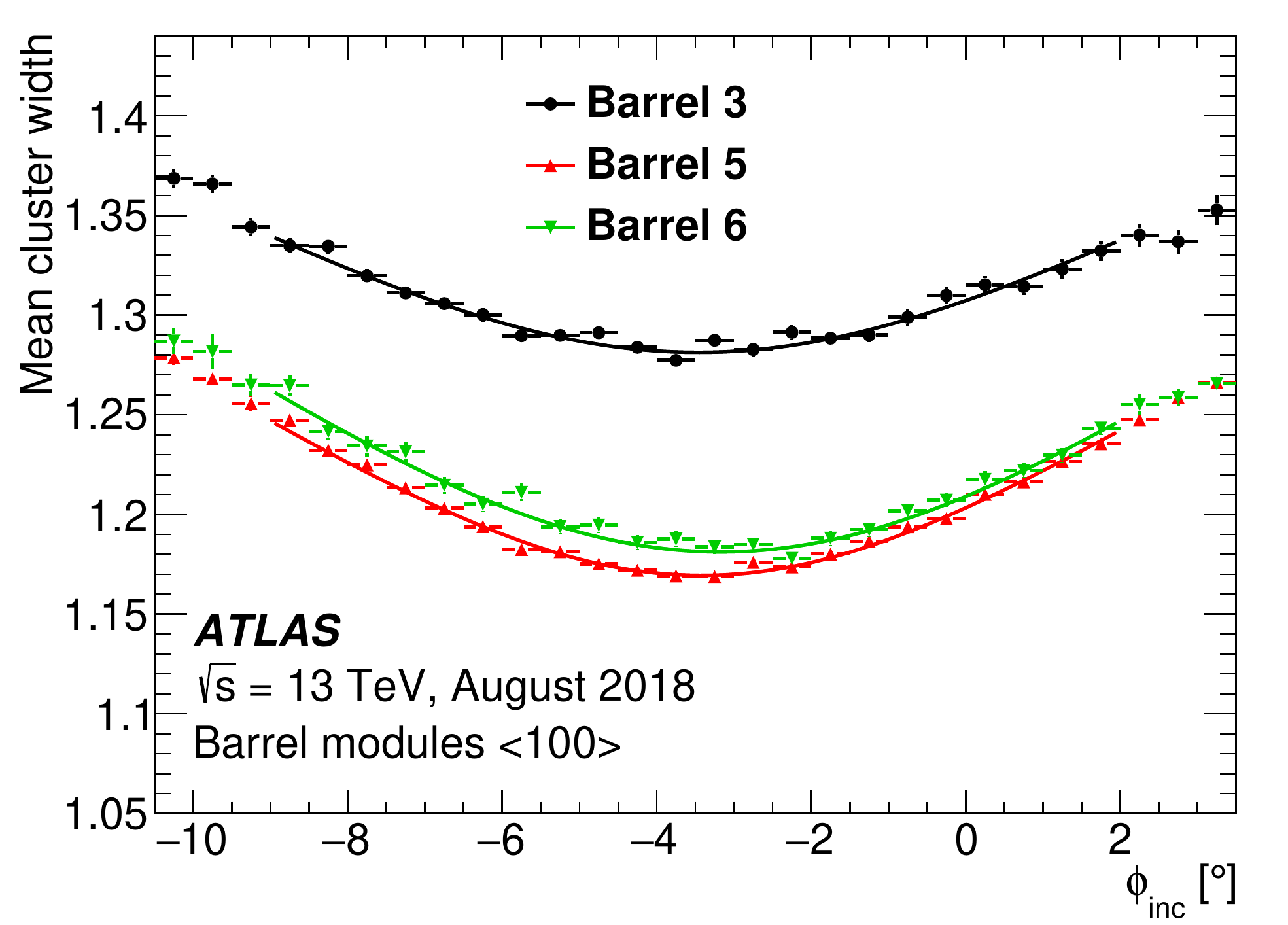}
\label{fig:clusize_vs_phiin_LA:b}
}
\caption{
Mean cluster width as a function of $\phiin$ for barrel layers~3 to~6 in a typical $pp$ collision run in August 2018.
Modules with crystal orientation \protect\subref{fig:clusize_vs_phiin_LA:a} <111> and \protect\subref{fig:clusize_vs_phiin_LA:b} <100> are shown separately.
For geometrical simplicity, only the sides with no stereo angle are included.
}
\label{fig:clusize_vs_phiin_LA}
\end{figure}
 
The data are fitted using a convolution of the function~\cite{IDET-2013-01}
\begin{eqnarray}
f(\phiin) &= a |\tan \phiin - \tan \phiMCW| + b
\label{equation:convfunc}
\end{eqnarray}
with a Gaussian distribution.
The fitted parameters are $\phiMCW$, $a$, $b$ and the width of the Gaussian distribution.
Typical fit results are shown in Figure~\ref{fig:clusize_vs_phiin_LA}; modules with crystal orientations of <111> and <100> are shown separately, because differences in the charge-carrier mobilities between these two types of sensor are expected to result in different values for the Lorentz angle.
The displacements of $\phiMCW$ from zero correspond to a shift of $\sim$10~$\mu$m in the position of the cluster on the sensor surface. The effect of the Lorentz angle must therefore be taken into account for precise tracking with the SCT.
The layer-dependence of the minimum cluster width was marginal in Run~1~\cite{IDET-2013-01}.
However, Figure~\ref{fig:clusize_vs_phiin_LA} clearly shows a large layer-dependence during Run~2, because the hit occupancy was significantly higher, resulting in a greater probability that hits induced by multiple charged particles are merged into a single large cluster.
Although the four barrel layers are almost equally spaced, the minimum cluster width in barrel layer~6 is almost equal to that in barrel layer~5.
This is due to the higher temperature of barrel layer 6 (see Section~\ref{sec:intro}), which induces faster diffusion of the charge carriers, and consequently wider dispersion over the sensor surface.
 
\begin{figure}[t!]
\centering
\includegraphics[width=0.55\textwidth]{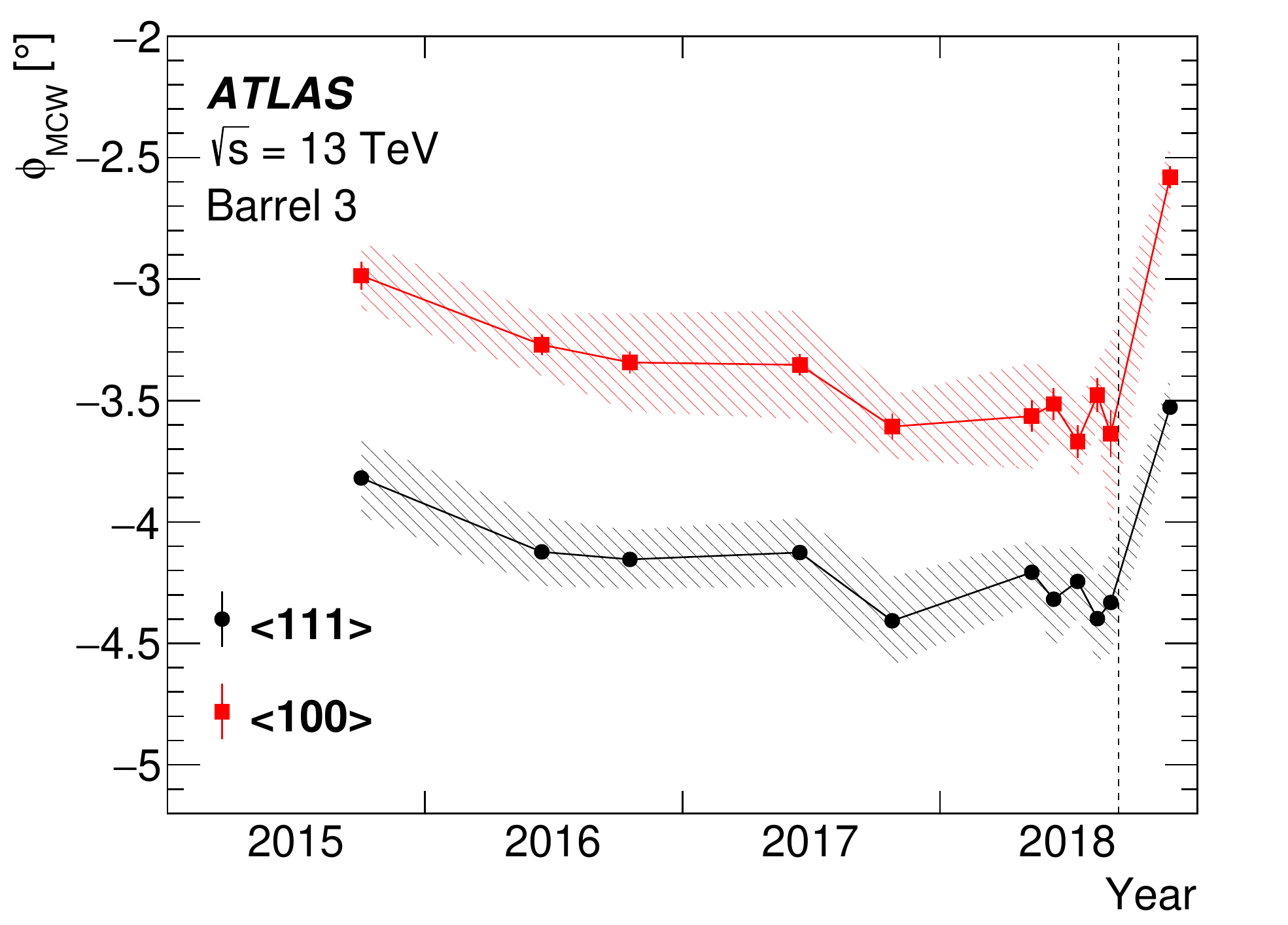}
\caption{
Time dependence of $\phiMCW$ between 2015 and 2018 in barrel layer~3.
Only the sides with no stereo angle are shown.
The error bars are statistical, while the shaded bands show the systematic uncertainties.
The time when the nominal operational HV was raised from 150~V to 250~V is indicated by the vertical dashed line.
}
\label{fig:phiMCW_time}
\end{figure}
 
Figure~\ref{fig:phiMCW_time} shows the variation of $\phiMCW$ measured in barrel layer~3 between 2015 and 2018, corresponding to the evolution of the Lorentz angle.
No corrections have been made to account for changes in parameters such as the HV setting or the cooling temperature; however, these were stable up to September 2018, when the operational HV was intentionally raised from 150~V to 250~V.
The systematic uncertainty in $\phiMCW$ is estimated by fitting an alternative asymmetric function, in which the parameter $a$ in Eq.~(\ref{equation:convfunc}) is allowed to have different values above and below the minimum.
The deviation from the nominal fit is taken to be the systematic uncertainty.
During Run~2, $|\phiMCW|$ increased continually, with a change of about 0.5$^\circ$ between 2015 and 2018.
This is presumably induced by radiation damage to the sensors, which modifies the space-charge distribution and changes the mobility.
For a sensor thickness of 285~$\mu$m, this change of 0.5$^\circ$ corresponds to a shift of $\sim$1~$\mu$m in the cluster position.
This is negligibly small because the intrinsic position resolution of the SCT is about 20~$\mu$m~\cite{IDET-2013-01}.
However, in Run~3, where the total fluence of radiation will increase by a factor of two, monitoring of changes in the Lorentz angle may become more important.
 
The final $\phiMCW$ measurement shown in Figure~\ref{fig:phiMCW_time} corresponds to the period after the operational HV of all modules in barrel layer~3 was raised to 250~V, which reduced $|\phiMCW|$ by about 0.7$^\circ$.
This behaviour is expected, because increased HV results in lower charge-carrier mobility~\cite{Solid-State-Elec-20-77}, and thus a smaller deflection in the magnetic field.
The change in the HV setting is taken into account when the SCT cluster position is used to determine a space point for tracking.
 
\subsection{Noise measurement}
\label{sec:Noise}
One of the crucial conditions for maintaining high tracking efficiency is to keep the readout thresholds as low as possible.
This is possible only when the input noise levels are sufficiently low.
It is therefore important to understand the noise levels of radiation-damaged sensors during detector operation.
Two methods are used to evaluate the sensor noise in dedicated SCT stand-alone calibration runs: (i) the response curve test, which measures the input equivalent noise charge (ENC, with the unit `number of electrons'), and (ii) the noise occupancy test.
The former method reflects the width of the Gaussian-like noise distribution at around one standard deviation ($1\sigma$), while the latter measures the tail of the noise distribution at around $4\sigma$ when a threshold is set at 1~fC, and it is therefore more sensitive to external noise sources.
 
\begin{figure}[t]
\centering
\subfloat[]{
\includegraphics[width=0.30\textwidth]{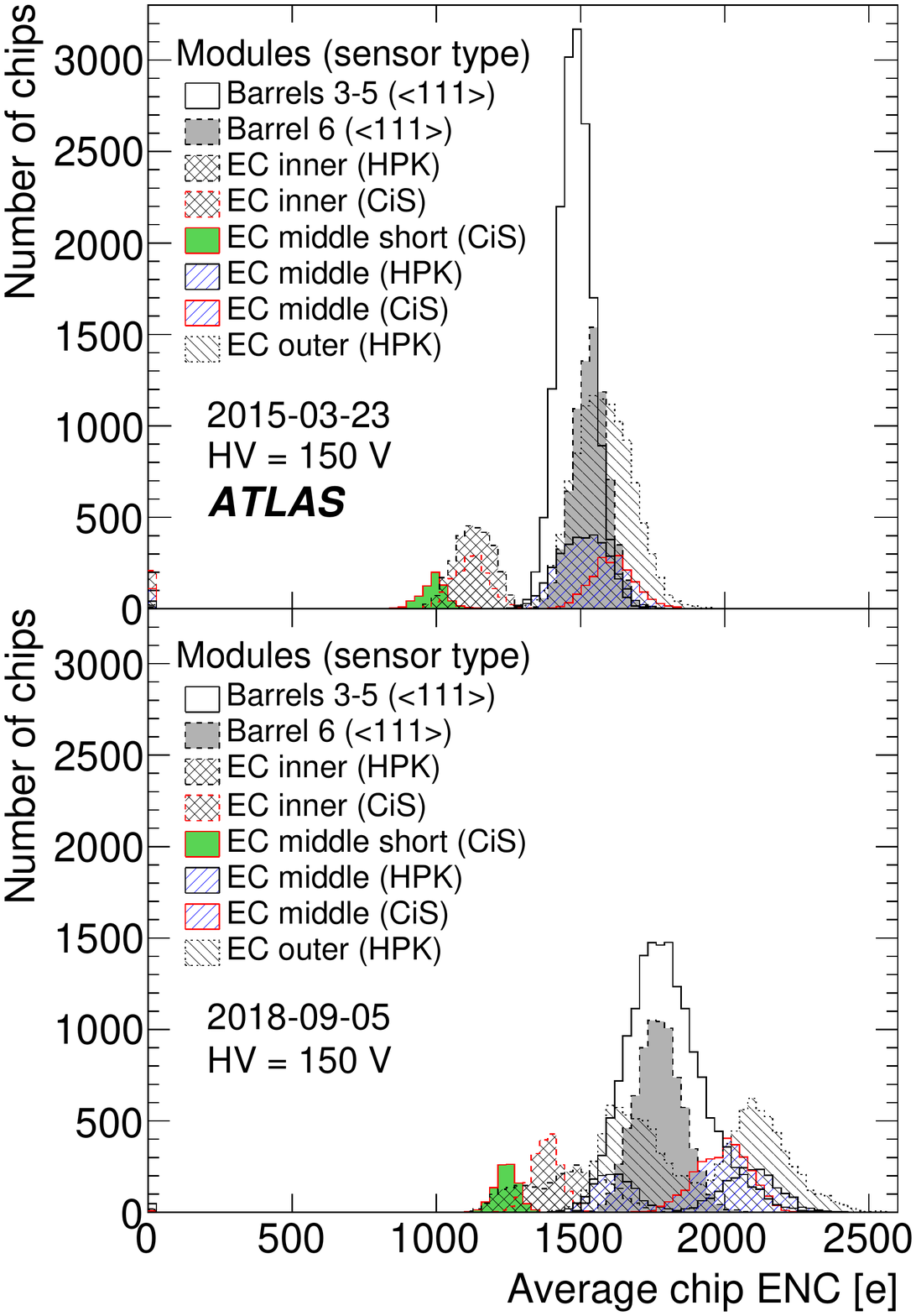}
\label{fig:chip_noise:a}
}
\subfloat[]{
\includegraphics[width=0.30\textwidth]{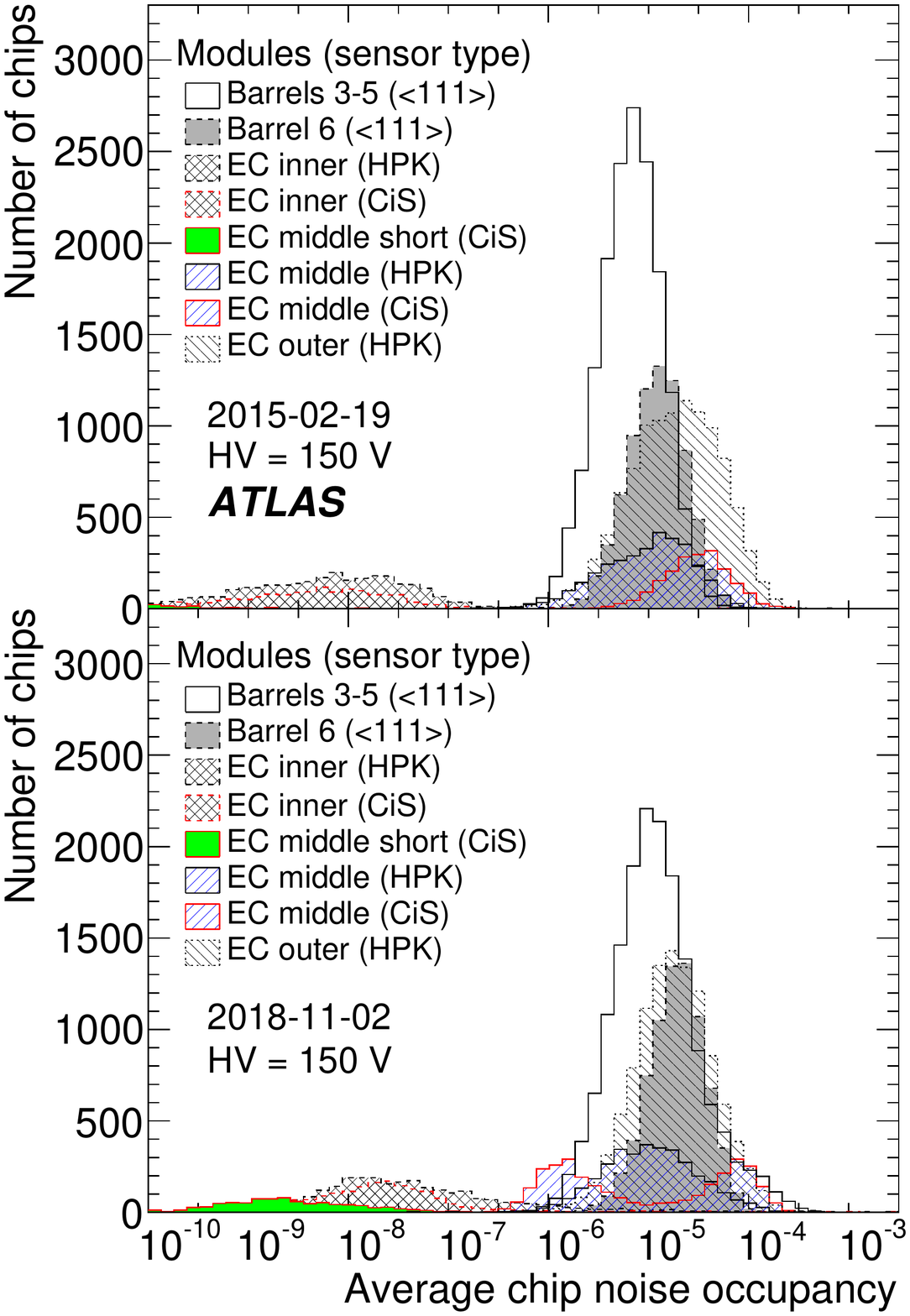}
\label{fig:chip_noise:b}
}
\subfloat[]{
\includegraphics[width=0.30\textwidth]{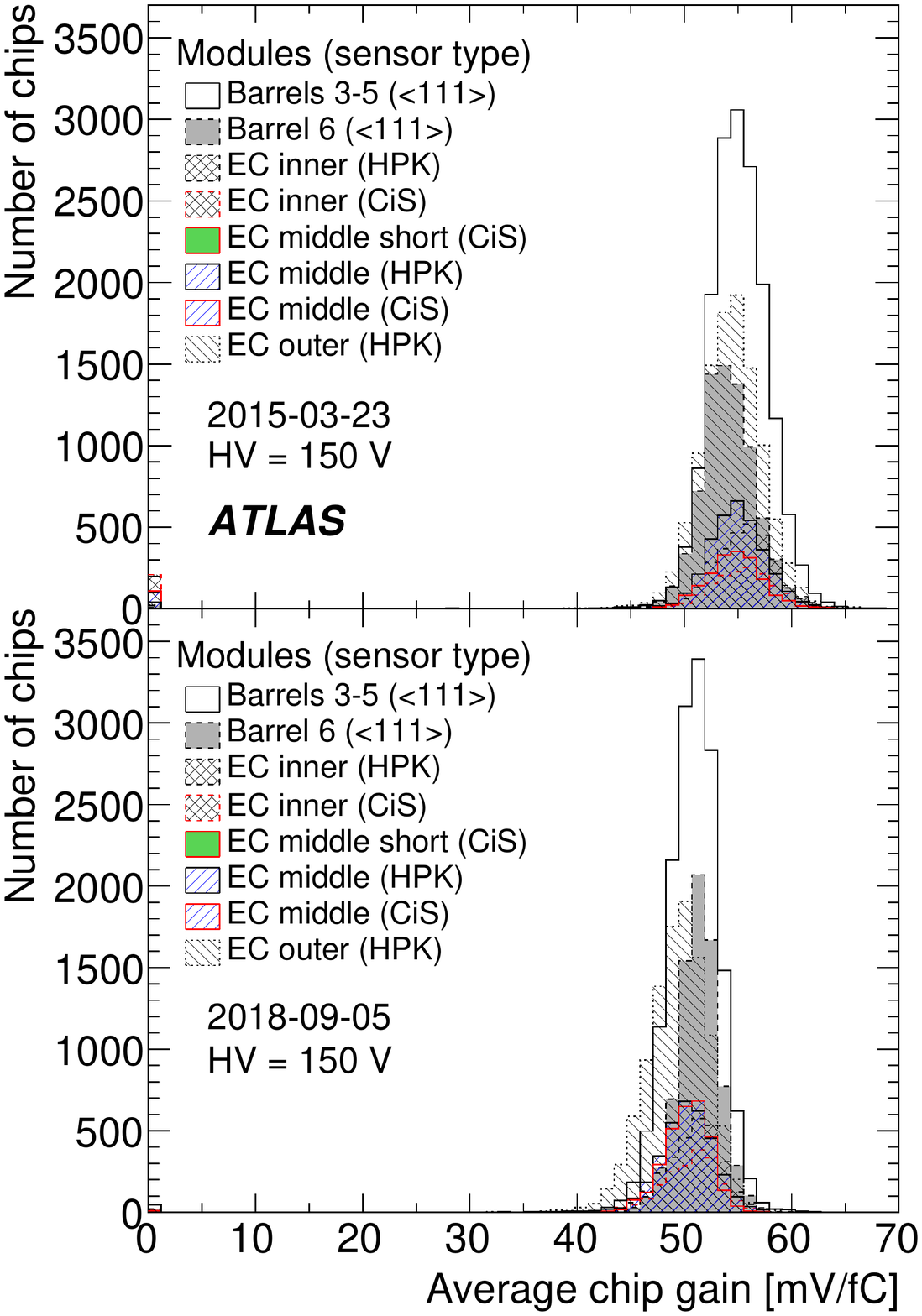}
\label{fig:chip_noise:c}
}
\caption{
Distributions of chip-averaged \protect\subref{fig:chip_noise:a} response-curve noise, \protect\subref{fig:chip_noise:b} noise occupancy and \protect\subref{fig:chip_noise:c} amplifier gain.
Data from February or March 2015 (top) and from September or November 2018 (bottom) are compared for different types of modules.
For the barrel, only modules with sensors of crystal orientation <111> are plotted; barrel layer~6, which has a higher temperature and thus higher noise, is shown separately.
Other barrel modules are combined regardless of the layer because there is no significant difference.
For the endcaps, chips reading out HPK and CiS sensors are shown separately.
}
\label{fig:chip_noise}
\end{figure}
 
Distributions of the chip-averaged response-curve noise and noise occupancy at a HV of 150~V are shown in Figure~\ref{fig:chip_noise}\subref{fig:chip_noise:a} and Figure~\ref{fig:chip_noise}\subref{fig:chip_noise:b}, respectively.
Results from February or March of 2015 and from September or November 2018 are compared, corresponding to the start and end of the Run-2 physics data-taking period.
Results from the response-curve test are also summarised numerically in Table~\ref{tab:response_curve}.
During Run~2, most of the chip-averaged noise levels increased marginally by 10\%--20\%.
The response curve test also provides measurements of the amplifier gain, as shown in Figure~\ref{fig:chip_noise}\subref{fig:chip_noise:c}.
The gain decreased gradually by around 10\% during Run~2.
Since calibration runs were carried out several times each year (see Section~\ref{sec:calibration}) to adjust the threshold to be 1~fC, the observed decrease in gain had only a marginal impact on the hit efficiency.
 
\begin{sidewaystable}
\centering
\caption{Numerical results for noise and amplifier gain from the response curve test.
Mean values from March 2015 and from September 2018 are compared for two FE-links on different types of modules, corresponding to the groups in Figure~\ref{fig:chip_noise}.
}
\label{tab:response_curve}
\begin{tabular}{lcc|cccc|cc|cccc|cc}
\hline
\hline
& & & \multicolumn{6}{c|}{ENC [$\times 10^3$ electrons]} & \multicolumn{6}{c}{Gain [mV/fC]}\\
\hline
\multicolumn{3}{c|}{\multirow{2}{*}{Module type}} & \multicolumn{2}{c}{March 2015} & \multicolumn{2}{c|}{September 2018} & \multicolumn{2}{c|}{Ratio} & \multicolumn{2}{c}{March 2015} & \multicolumn{2}{c|}{September 2018} & \multicolumn{2}{c}{Ratio}\\
& & &                    Link-0 & Link-1 & Link-0 & Link-1 & Link-0 & Link-1 & Link-0 & Link-1 & Link-0 & Link-1 & Link-0 & Link-1\\
\hline
\multirow{2}{*}{Barrels 3--5}    & <111> & HPK & 1.49 & 1.48 & 1.84 & 1.76 & 1.24 & 1.19 & 55.0 & 55.2 & 50.7 & 51.0 & 0.92 & 0.92\\
& <100> & HPK & 1.39 & 1.39 & 1.57 & 1.54 & 1.12 & 1.11 & 55.3 & 55.6 & 52.1 & 52.2 & 0.94 & 0.94\\
\multirow{2}{*}{Barrel 6}        & <111> & HPK & 1.53 & 1.54 & 1.78 & 1.77 & 1.16 & 1.15 & 54.1 & 54.2 & 51.3 & 51.3 & 0.95 & 0.95\\
& <100> & HPK & 1.44 & 1.43 & 1.59 & 1.59 & 1.11 & 1.11 & 54.5 & 55.0 & 51.9 & 52.3 & 0.95 & 0.95\\
\multirow{2}{*}{EC inner}        & <111> & HPK & 1.16 & 1.10 & 1.53 & 1.35 & 1.31 & 1.23 & 55.6 & 55.7 & 50.5 & 51.1 & 0.91 & 0.92\\
& <111> & CiS & 1.09 & 1.15 & 1.39 & 1.37 & 1.27 & 1.19 & 55.7 & 55.5 & 50.7 & 50.5 & 0.91 & 0.91\\
\multirow{2}{*}{EC middle}       & <111> & HPK & 1.57 & 1.46 & 2.10 & 1.69 & 1.34 & 1.16 & 54.5 & 55.2 & 49.2 & 51.2 & 0.90 & 0.93\\
& <111> & CiS & 1.59 & 1.63 & 2.04 & 1.94 & 1.28 & 1.19 & 54.7 & 54.6 & 50.0 & 50.4 & 0.91 & 0.92\\
EC middle short                  & <111> & CiS & 0.99 & 1.00 & 1.25 & 1.23 & 1.27 & 1.23 & 56.6 & 56.5 & 52.2 & 52.0 & 0.92 & 0.92\\
EC outer                         & <111> & HPK & 1.62 & 1.54 & 2.05 & 1.76 & 1.27 & 1.14 & 54.2 & 54.5 & 48.8 & 49.9 & 0.90 & 0.92\\
\hline
\hline
\end{tabular}
\end{sidewaystable}
 
Figure~\ref{fig:noise_evolution}\subref{fig:noise_evolution:a} shows the evolution of ENC for barrel layer~3 and for endcap~A disk~4, where measurements made through each of the two FE-links on either side of a module (referred to as `link-0' and `link-1') are shown separately.
The endcap modules show significantly higher noise in link-0 than in link-1; the difference creates a double peak in the noise distributions shown in Figure~\ref{fig:chip_noise}.
Figure~\ref{fig:noise_evolution}\subref{fig:noise_evolution:b} shows the ENC for all modules in endcap~A.
Almost all of the HPK modules, except those in disks 1 and~9, have more noise in link-0, while the CiS modules show only a marginal difference.
In the endcaps, all the modules are oriented so that the side with link-1 is attached to the support frame, while the other side with link-0 is generally exposed to the air.\footnote{How the endcap modules are attached to the disks is shown in Figure~\ref{fig:SCT_quadrant}. Modules staggered to the left are attached to the disk and face the collision point, while modules staggered to the right are attached to the disk and face away from the collision point. The side of each module on which link-1 is located is the side attached to the disk.}
In the case of disks~1 and~9, however, the side with link-0 also faces the support structure or the plate of the thermal enclosure.
It is therefore evident that the exposed HPK sensors show systematically higher levels of noise; however, its source is not yet clear.
Nevertheless, the ENC level is at most 2000 electrons (0.32~fC), which is sufficiently low compared to the threshold of 1~fC.
 
In contrast to the endcap modules, barrel modules do not show a noise-level difference between link-0 and link-1.
The noise from sensors with the <100> crystal orientation is consistently lower than that from sensors with the <111> orientation, reproducing the results from earlier studies presented in Ref.~\cite{NIMA-466-354-2001}.
 
\begin{figure}[t!]
\centering
\subfloat[]{
\includegraphics[width=0.42\textwidth]{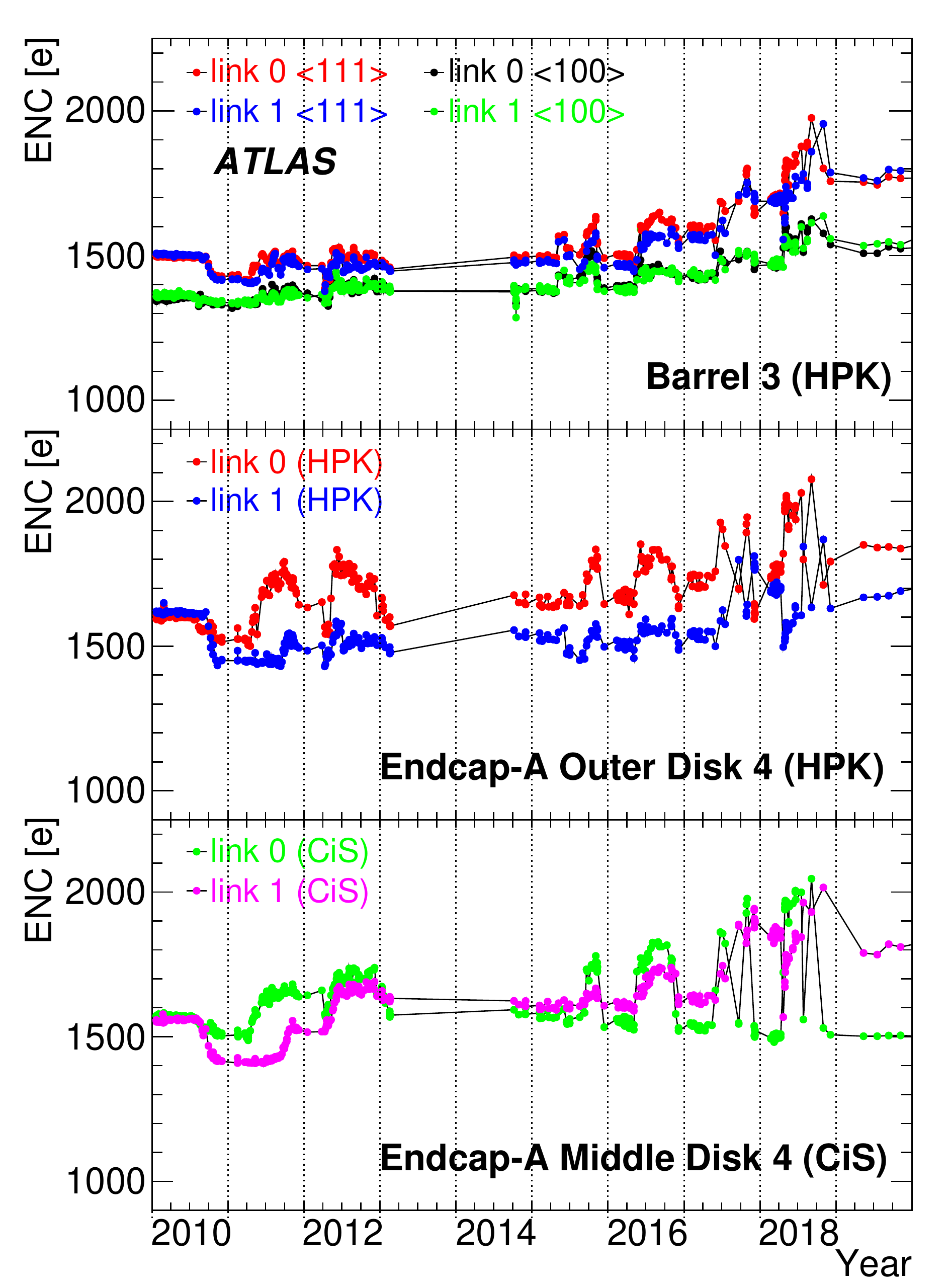}
\label{fig:noise_evolution:a}
}
\subfloat[]{
\includegraphics[width=0.42\textwidth]{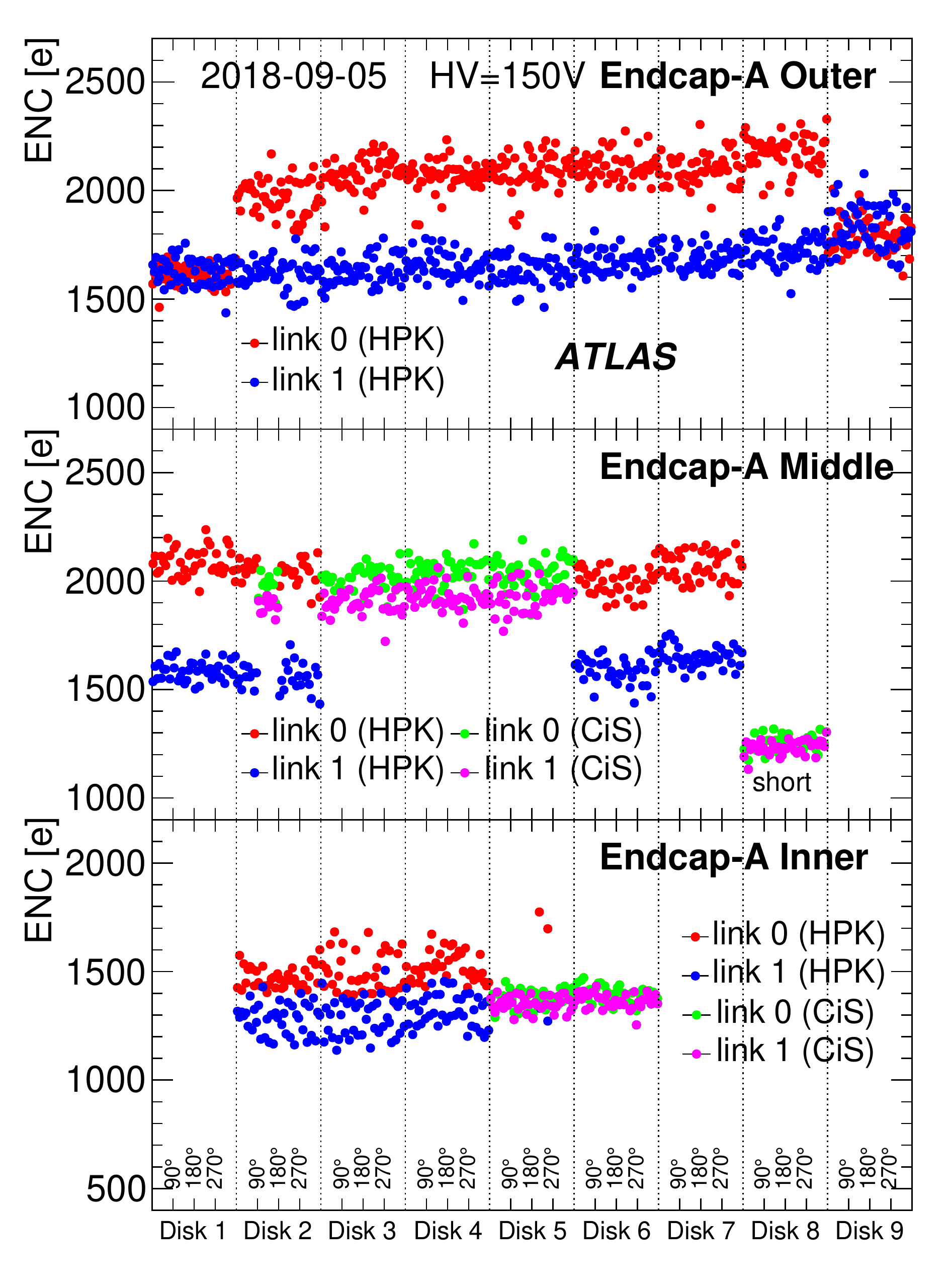}
\label{fig:noise_evolution:b}
}
\caption{
Chip-averaged noise. Only modules with HV between 145~V and 155~V are shown.
\protect\subref{fig:noise_evolution:a} Evolution of the response-curve noise for barrel layer~3 (top), the outer ring of endcap~A disk~4 (middle) and the middle ring of endcap~A disk~4 (bottom).
Measurements from link-0 and link-1, from modules with different crystal orientations and also from modules manufactured by HPK and CiS, are shown separately.
\protect\subref{fig:noise_evolution:b} Chip-averaged response-curve noise for all modules in endcap~A.
From top to bottom, outer, middle and inner rings are shown.
Measurements from link-0 and link-1, and also from HPK modules and CiS modules, are shown separately.
The nine segments separated by vertical dashed lines correspond to disks~1 to~9.
}
\label{fig:noise_evolution}
\end{figure}
 
\subsubsection{HV dependence of noise}
 
\begin{figure}[t!]
\centering
\subfloat[]{
\includegraphics[width=0.72\textwidth]{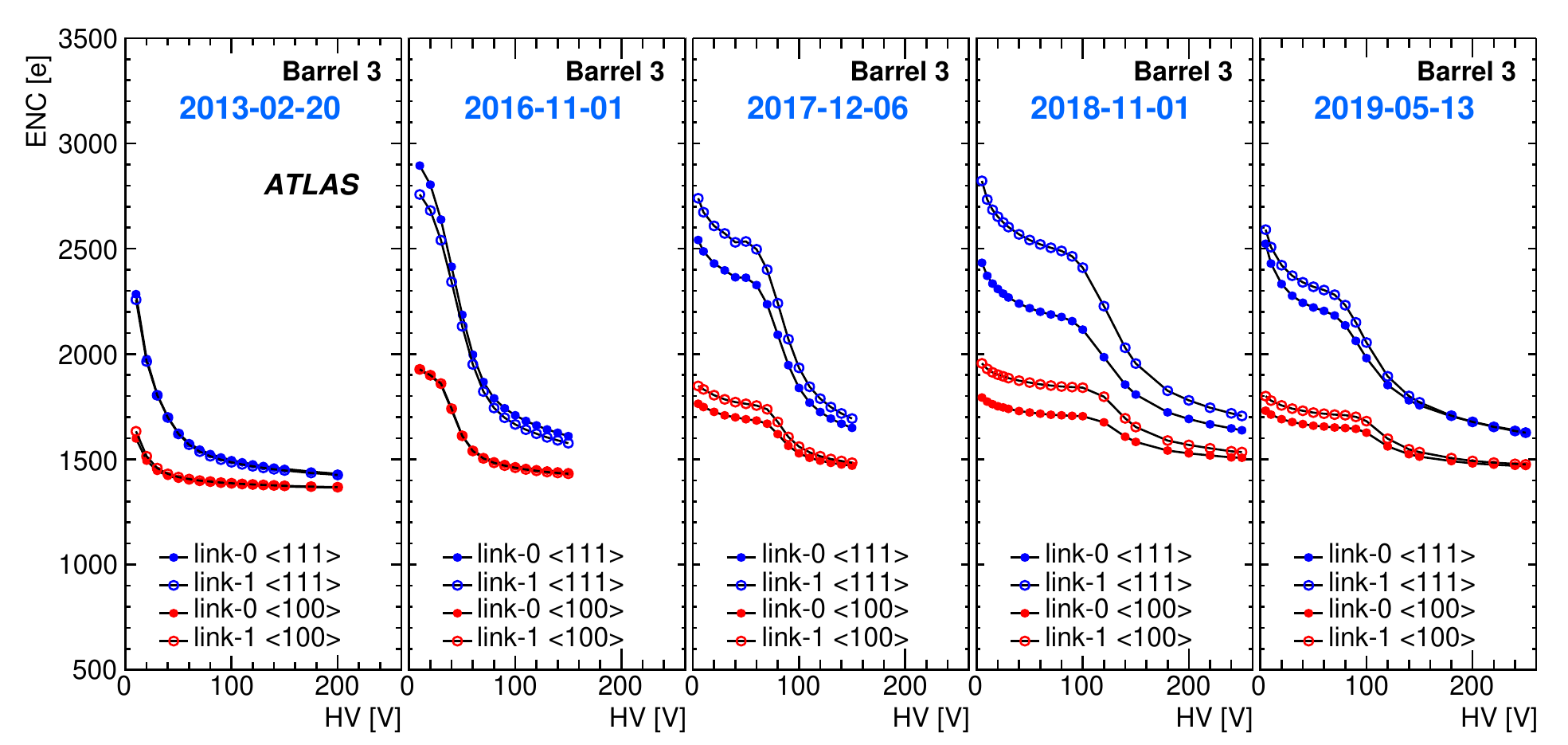}
\label{fig:IVsum_noise:a}
}
\subfloat[]{
\includegraphics[width=0.18\textwidth]{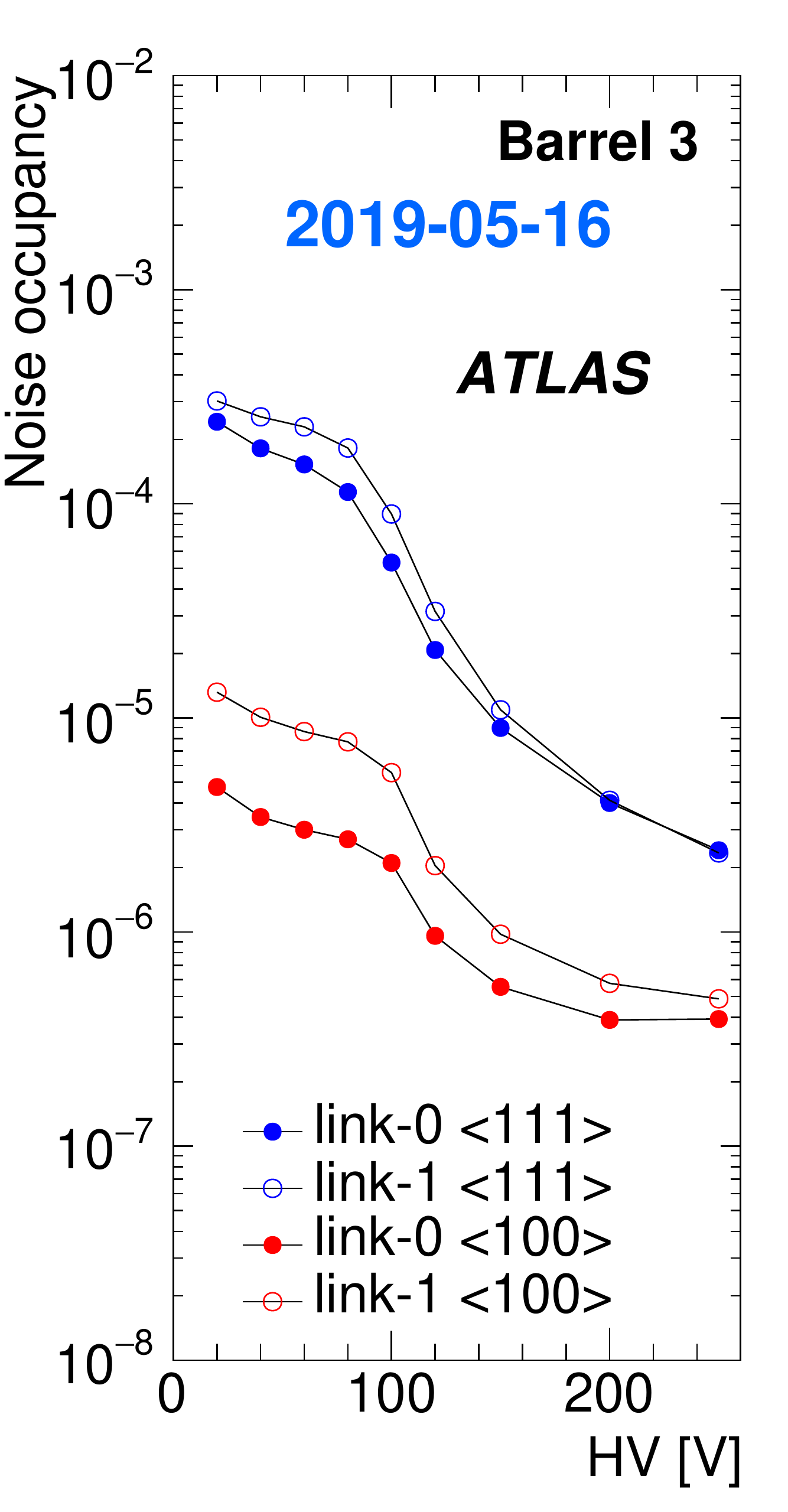}
\label{fig:IVsum_noise:b}
}
\caption{
HV dependence of the noise for modules in barrel layer~3.
\protect\subref{fig:IVsum_noise:a} Response-curve noise measured between February 2013 and May 2019.
\protect\subref{fig:IVsum_noise:b} Noise occupancy measured in May 2019.
}
\label{fig:IVsum_noise}
\end{figure}
 
The HV dependence of the noise levels was measured periodically throughout Run~2 using response curve tests.
Figure~\ref{fig:IVsum_noise}\subref{fig:IVsum_noise:a} shows results from tests performed between February 2013 and May 2019.
In November 2016 a knee-like structure started to appear in these curves, and the HV value corresponding to this knee increased with time until November 2018.
The same trend is observed for both crystal orientations <111> and <100>, although the absolute level of the noise is significantly different.
This knee-like structure also appears in a noise occupancy scan performed in May 2019, which is shown in Figure~\ref{fig:IVsum_noise}\subref{fig:IVsum_noise:b}.
 
The noise level is largely determined by the interstrip capacitance, which is the dominant component of the overall sensor capacitance~\cite{NIMA-578-98-2007}, and it is sensitive to the properties of the silicon at the surface of the sensor. The development of the depletion region near the surface shows a complex evolution during type inversion, because of the non-uniform interstrip space-charge distribution. The appearance of the knee-like structure is predicted to be a qualitative indicator of type inversion; such behaviour was reported, for example, in Ref.~\cite{NIMA-466-354-2001}. Its evolution results from changes in the full depletion voltage.
As shown in Section~\ref{sec:FullDepletionVoltage}, type inversion in late 2016 is consistent with a more quantitative analysis using the $I$--$V$ relation.

 
\section{Radiation damage}
\label{sec:radiation}
 
During LHC operations, the SCT silicon sensors are irradiated by various species of particles with a wide range of energies, from the \TeV-scale to thermal neutrons, resulting from $pp$ collisions at the interaction point, or from back-scattered secondary particles from the calorimeters or the accelerator collimators.
After receiving radiation, the SCT performance is changed due to the creation of additional states within the semiconductor band gap.
The SCT was designed to endure up to about 700~fb$^{-1}$ at a $pp$ collision energy of 14~\TeV~\cite{ATLAS-TDR-05}.
Up to the end of 2018 the accumulated integrated luminosity was 156~fb$^{-1}$ at 13~\TeV\ during Run~2 and 29~fb$^{-1}$ at 7--8~\TeV\ during Run~1, so a safe margin for further operation remains.
 
Nevertheless, the intense radiation flux in the environment of the ID has modified various SCT parameters such as the leakage current, the full depletion voltage and the probability of charge trapping.
The expected fluence around the SCT is shown in Figure~\ref{fig:fluences}, with the numerical values calculated at the positions of each barrel layer and endcap disk listed in Table~\ref{tab:fluences}.
These are based on simulation using the \textsc{Fluka}~\cite{FLONT-ONCOL-6-116} software package and a detailed description of the ATLAS detector geometry~\cite{SOFT-2010-01}.
Inelastic $pp$ collisions were simulated using the \PYTHIA[8] event generator~\cite{Sjostrand:2006za,Sjostrand:2014zea} with the \MSTW[lo] parton distribution functions~\cite{Martin:2009iq} and the A3 set of tuned parameters~\cite{ATL-PHYS-PUB-2016-017}.
The integrated luminosities delivered by the LHC at $\sqrt{s}$ = 7, 8 and 13~\TeV\ correspond to 5.6, 23.2 and 159.4~fb$^{-1}$, respectively.
These slightly larger luminosity values than in Ref.~\cite{ATLAS-DAPR-2018-01-001} result from the inclusion of beam collision periods outside the stable beam time when data collection occurred.
 
\begin{figure}[t]
\centering
\includegraphics[width=0.7\textwidth]{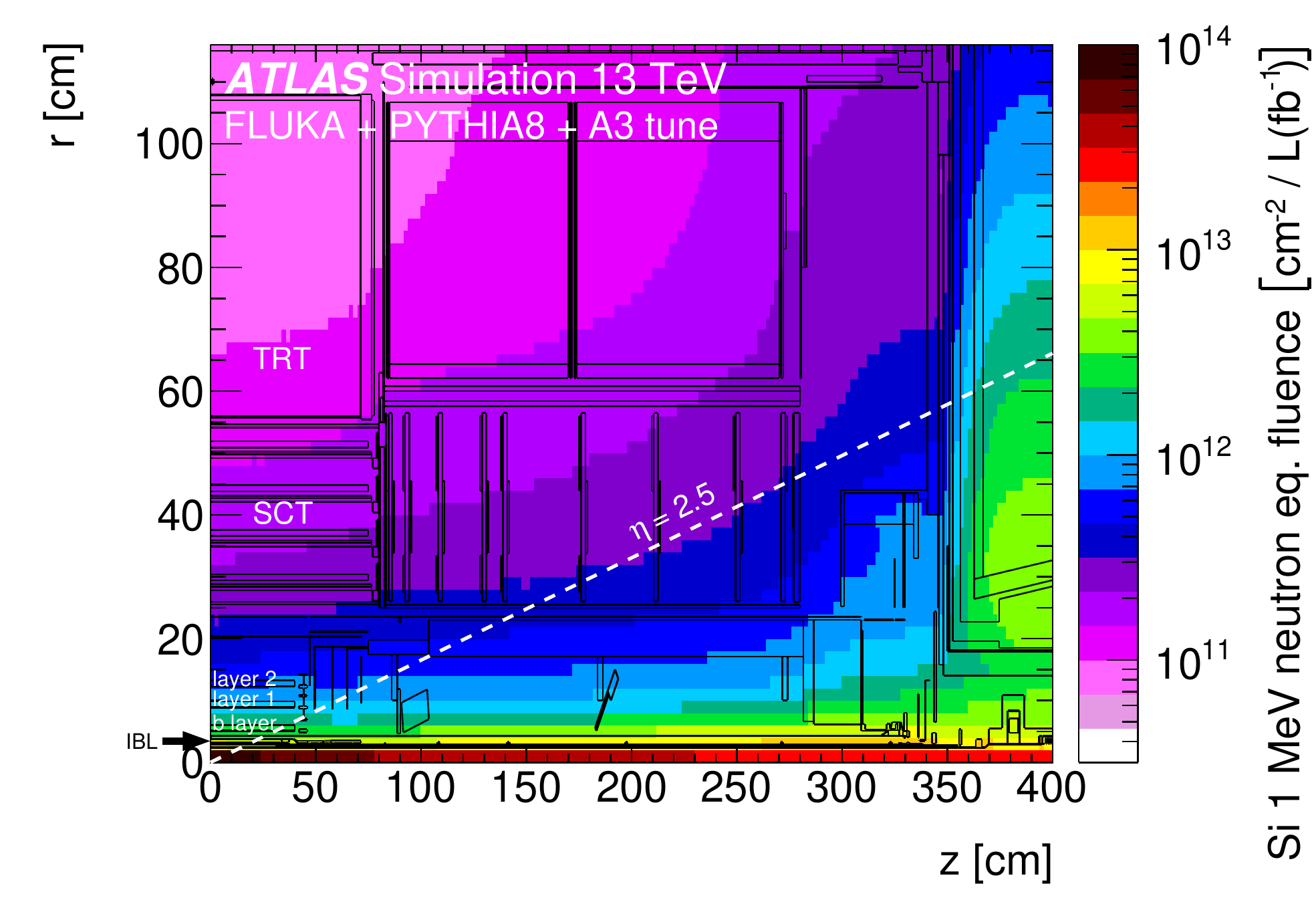}
\caption{
Simulated map of the 1-\MeV-neutron equivalent fluence per fb$^{-1}$ of integrated luminosity at 13~\TeV\ centre-of-mass energy around the ID region with the ATLAS Run-2 detector geometry.
}
\label{fig:fluences}
\end{figure}
 
\begin{table}[t!]
\caption{Simulated 1-\MeV-neutron equivalent fluence in units of $10^{13}$~cm$^{-2}$ up to the end of Run~2.
Radiation backgrounds originating from $pp$ collisions are assumed to be symmetric about the origin along the $z$-axis.
The $r$ and $z$ positions listed correspond to the centres of the modules.
}
\label{tab:fluences}
\centering
\begin{tabular}{crccccc}
\hline
\hline
$|\etaindex| $ & $|z|$ [mm] & Barrel 3 & Barrel 4 & Barrel 5 & Barrel 6 \\
&          & ($r$ = 299~mm) & ($r$ = 371~mm) & ($r$ = 443~mm) & ($r$ = 514~mm)\\
\hline
1 & 61.9  & 4.30 & 3.41 & 2.85 & 2.45 \\
2 & 184.0 & 4.37 & 3.46 & 2.87 & 2.47 \\
3 & 308.7 & 4.47 & 3.51 & 2.92 & 2.52 \\
4 & 431.2 & 4.53 & 3.58 & 3.00 & 2.57 \\
5 & 555.3 & 4.65 & 3.71 & 3.09 & 2.66 \\
6 & 679.1 & 4.80 & 3.84 & 3.20 & 2.76 \\
\hline
\hline
Disk & $|z|$ [mm] & Inner & Middle & Outer & \\
&            & ($r$ = 304.6~mm) & ($r$ = 396.5~mm) & ($r$ = 499.4~mm) & \\
\hline
1 & 853.8  & $-$  & 3.86 & 3.07 & \\
2 & 934.0  & 5.12 & 3.89 & 3.08 & \\
3 & 1091.5 & 5.20 & 3.98 & 3.19 & \\
4 & 1299.9 & 5.27 & 4.14 & 3.31 & \\
5 & 1399.7 & 5.37 & 4.21 & 3.40 & \\
6 & 1771.4 & 5.69 & 4.47 & 3.67 & \\
7 & 2115.2 & $-$  & 4.94 & 4.01 & \\
8 & 2505.0 & $-$  & 5.42 & 4.66 & \\
9 & 2720.2 & $-$  & $-$  & 5.33 & \\
\hline
\hline
\end{tabular}
\end{table}
 
Both the total ionising dose (TID) and the non-ionising energy loss (NIEL) affect the performance of the SCT sensors.
The TID as well as neutron fluences around the SCT were monitored by a dedicated online radiation monitoring system~\cite{IEEE-54-1143-2007}, in which the TID was measured using radiation-sensitive $p$-MOSFET transistors, while the neutron fluences were measured using forward- and reverse-biased $p$-$i$-$n$ diodes.
There is good agreement between these measurements and the simulated radiation background in Run~2, as presented in Ref.~\cite{ANIMMA-2019}.
 
The instantaneous leakage current increases are approximately proportional to the NIEL.
Too large a leakage current might damage the SCT sensors due to thermal runaway.
Radiation damage also modifies the doping concentration, which is proportional to $\vfd$.
As damage accumulates, the number of donors in the $n$-bulk material decreases, while the number of acceptors effectively increases.
This leads to type inversion of the $n$-bulk into $p$-bulk material, with a further increase of $\vfd$.
After type inversion, the HV applied needs to be kept above $\vfd$ to achieve full hit efficiency.
If $\vfd$ becomes too high, this might therefore lead to degradation of the tracking performance.
Although these changes in the bulk properties were taken into account in the design of the SCT, improved understanding of these characteristics is important for safe operation of the SCT.
 
In this section, the evolution of the leakage current and the full depletion voltage are discussed; leakage currents from both the pixel detector and the SCT are reported in Ref.~\cite{IDET-2020-01}.
Other effects of radiation damage, such as SEUs and changes in hit efficiency, cluster width and noise, are discussed in Sections~\ref{sec:operation} and \ref{sec:performance}, respectively.
Variation of the charge-trapping probability is also important, but its impact on the SCT at the end of Run~2 is expected to be marginal~\cite{Krasel:2004mi,Moll:2018fol} and it is therefore not discussed in this paper.
 
\subsection{Leakage current}
 
The main source of leakage current is thermally generated electron--hole pairs.
The leakage current is therefore highly dependent on the sensor temperature, $\Tsensor$, as well as on the accumulated radiation damage.
In order to evaluate the evolution of the leakage current under the same conditions throughout all periods, the measured current on the HV supply line (HV current), $\Imeas$, is normalised to a reference temperature, $\Tnorm = 0\degC$, using the relationship
\begin{eqnarray}
\left( \frac{\Ileak}{\Imeas} \right) = \left( \frac{\Tnorm}{\Tsensor} \right)^2 \exp\left[ - \frac{\Egen}{2 \kB} \left( \frac{1}{\Tnorm} - \frac{1}{\Tsensor} \right) \right],\label{eq:IT}
\end{eqnarray}
where $\Ileak$ is the normalised leakage current, $\Egen$ is the effective electron--hole pair generation energy of $1.21 \pm 0.01$~\eV~\cite{Chilingarov_2013}, and $\kB$ is Boltzmann's constant.
 
The value of $\Tsensor$ for each module is determined from a hybrid-board temperature, $\Thybrid$, measured using negative temperature coefficient thermistors mounted on the hybrid boards.
However, there is an offset between these temperatures, $\delta T = \Thybrid - \Tsensor$, due to the thermal impedance between the sensor and the thermistor, which depends strongly on the mechanical and thermal structure of each board.
It differs from module to module, due to small variations in the thermal impedances among the hybrid circuits, sensors and cooling pipes.
In Run 1~\cite{IDET-2013-01}, $\delta T$ was estimated using thermal simulations based on the finite element method (FEM).
Compared with the barrel modules, the endcap modules had larger uncertainties on these estimates.
Moreover, a systematic difference in the raw leakage current between the two endcaps existed, which could not be explained by the FEM estimation.
It was therefore difficult in Run~1 to estimate the correct endcap sensor temperatures using the thermistors.
In order to improve the estimate of $\Ileak$ for the endcap modules, a method to determine $\delta T$ using cooling temperature scans was developed during Run~2.
 
\subsubsection{Development of temperature correction method}
 
A value of $\delta T$ for each module was estimated using dedicated measurements, in which the cooling temperature was varied.
With the LV supplies to the hybrid circuits turned off, $\Tsensor$ was considered to be equal to $\Thybrid$, due to the absence of any hybrid heating sources.
After turning on only the HV, the cooling temperature was then increased from approximately $-20\degC$ to $0\degC$.
In this scan, the relation between $\Thybrid \simeq \Tsensor$ and $\Imeas$ at various temperatures was measured for each module.
Two representative examples are shown in Figure~\ref{fig:Tscan}\subref{fig:Tscan:a}, together with the results of fits using Eq.~(\ref{eq:IT}).
Once the cooling temperature had returned to its nominal value, the LV was then turned on.
Since heat was generated by the hybrid circuit, $\Thybrid$ increased to become significantly higher than $\Tsensor$, while $\Tsensor$ remained sensitive to $\Imeas$ and so could be deduced for each module using the fitted curve.
Hence values of $\delta T$ were obtained.
 
A systematic uncertainty of $1\degC$ in $\delta T$ is assigned for all modules.
This was estimated by performing an alternative fit using only the lowest pair of $\Thybrid$ data points and comparing the result with that from the full fitted curve.
Smaller contributions arising from the stability of the leakage current measurements, the digitisation of the temperature measurements and the uncertainty in $\Egen$ are also included within this $1\degC$ uncertainty.
 
\begin{figure}[t!]
\centering
\subfloat[]{
\includegraphics[width=0.49\textwidth]{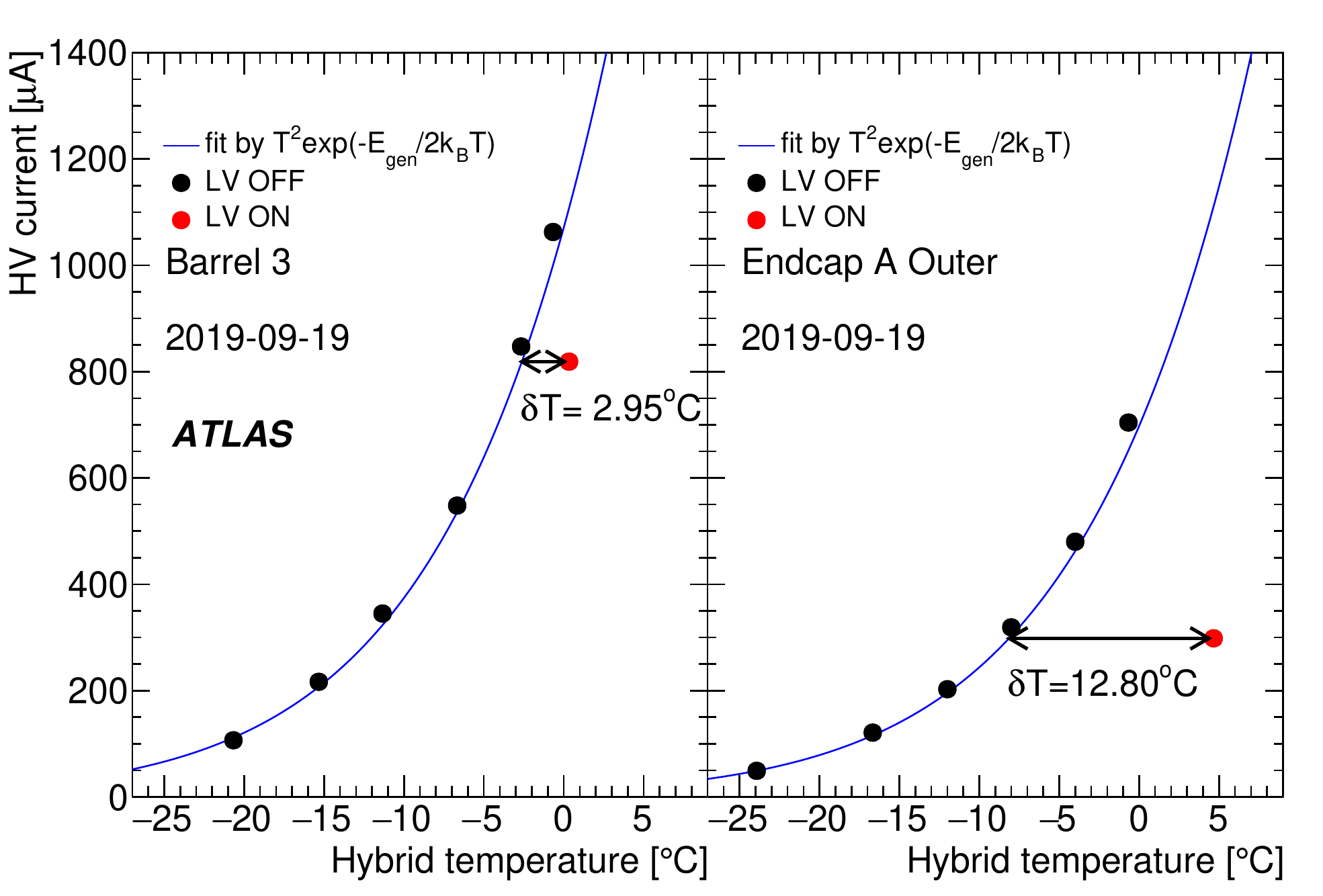}
\label{fig:Tscan:a}
}
\subfloat[]{
\includegraphics[width=0.49\textwidth]{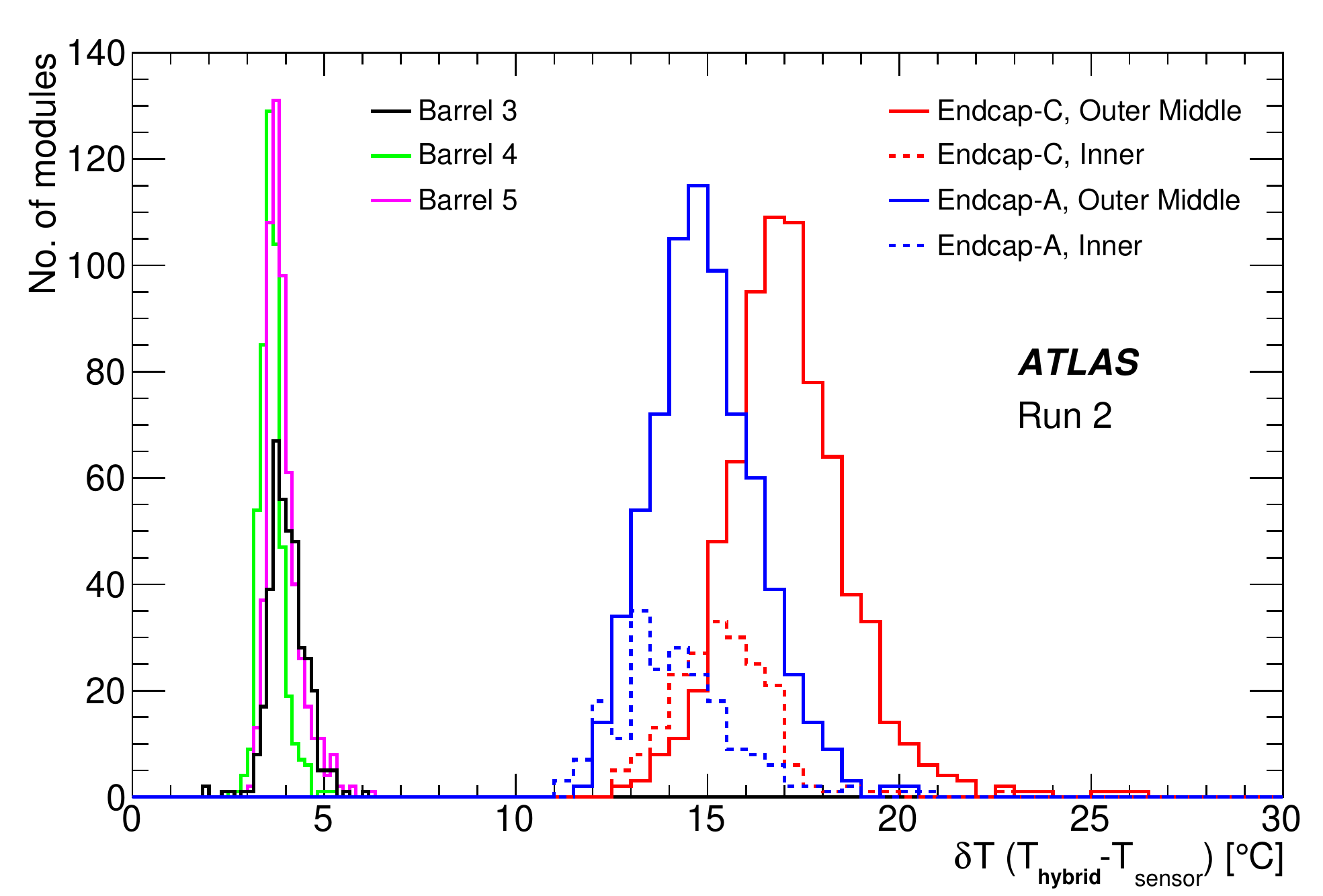}
\label{fig:Tscan:b}
}
\caption{
\protect\subref{fig:Tscan:a} Measured currents as a function of $\Thybrid$ for typical barrel and endcap modules with the LV off.
The blue curves are obtained from fits to the measured points using Eq.~(\ref{eq:IT}).
The red point displaced from each curve shows the hybrid temperature measured with the LV on; the temperature differences $\delta T$ are indicated by the arrows.
\protect\subref{fig:Tscan:b} Distributions of $\delta T$ for different categories of SCT modules.
}
\label{fig:Tscan}
\end{figure}
 
Distributions of $\delta T$ for modules in barrel layers~3, 4 and ~5, and in each of the endcap disks are shown in Figure~\ref{fig:Tscan}\subref{fig:Tscan:b}.
Barrel layer~6 cannot be cooled to a temperature lower than 6$\degr$C, as explained in Section~\ref{sec:intro}, so modules in layer~6 were excluded from the $\delta T$ measurement.
The average $\delta T$ for barrel modules is about $4\degC$.
As expected, $\delta T$ is larger for the endcap modules, with broader distributions and averages in the range $\sim$15$\degC$--17$\degC$.
There is a difference of about $2\degC$ between the average values of $\delta T$ measured in endcap~C and in endcap~A.
There has been a difference in the cooling temperatures set for endcap~A and endcap~C since the initial commissioning of the SCT, in order to compensate for a measured difference in $\Thybrid$ between the two endcaps.
The cause of this difference in $\Thybrid$ is unknown, but it may also result in the systematic difference in $\delta T$ between the two endcaps.
 
\subsubsection{Leakage currents at the end of Run~2}
 
\begin{figure}[t!]
\centering
\subfloat[]{
\includegraphics[width=0.48\columnwidth]{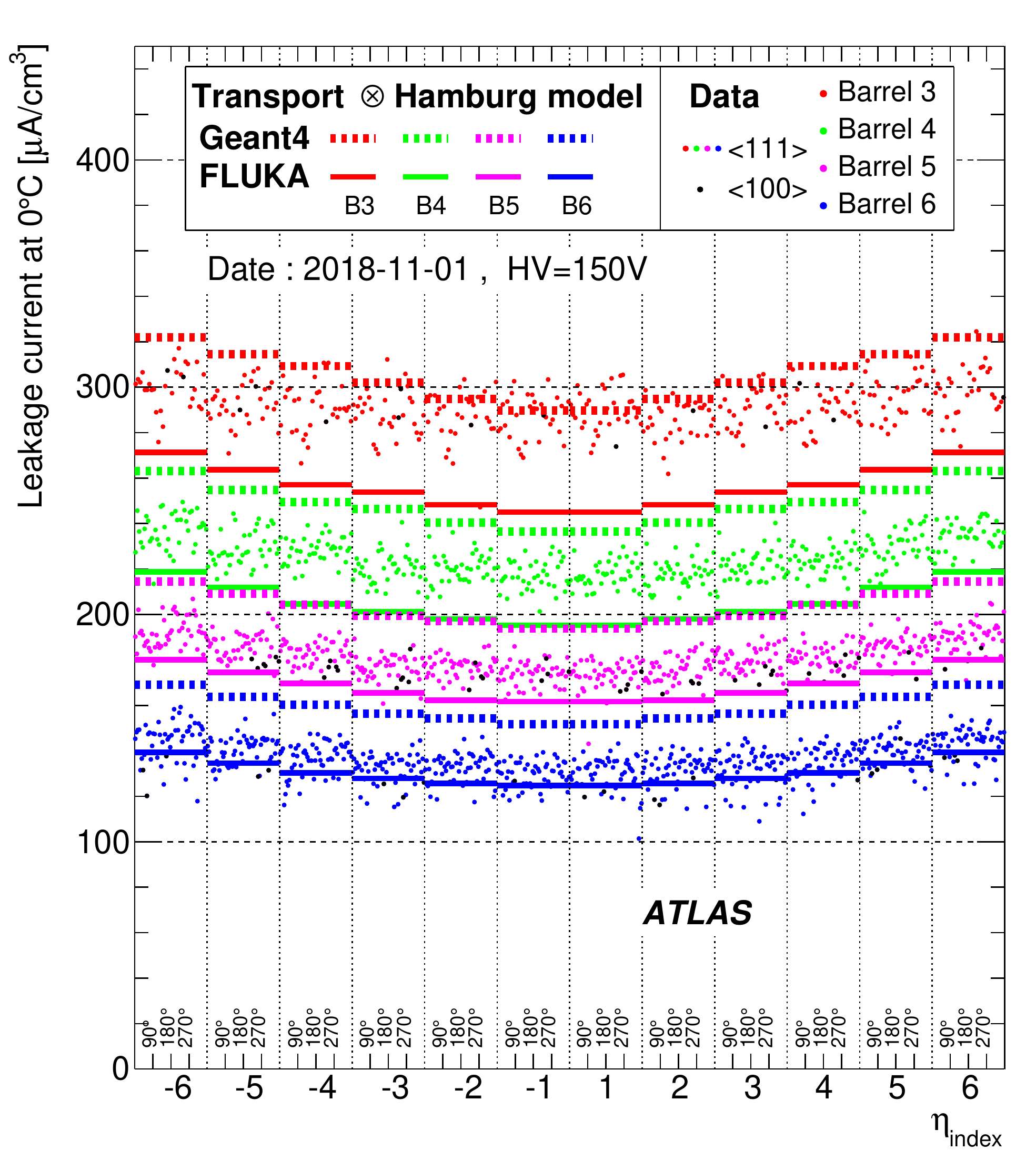}
\label{fig:Lateral_distributions:a}
}
\subfloat[]{
\includegraphics[width=0.48\columnwidth]{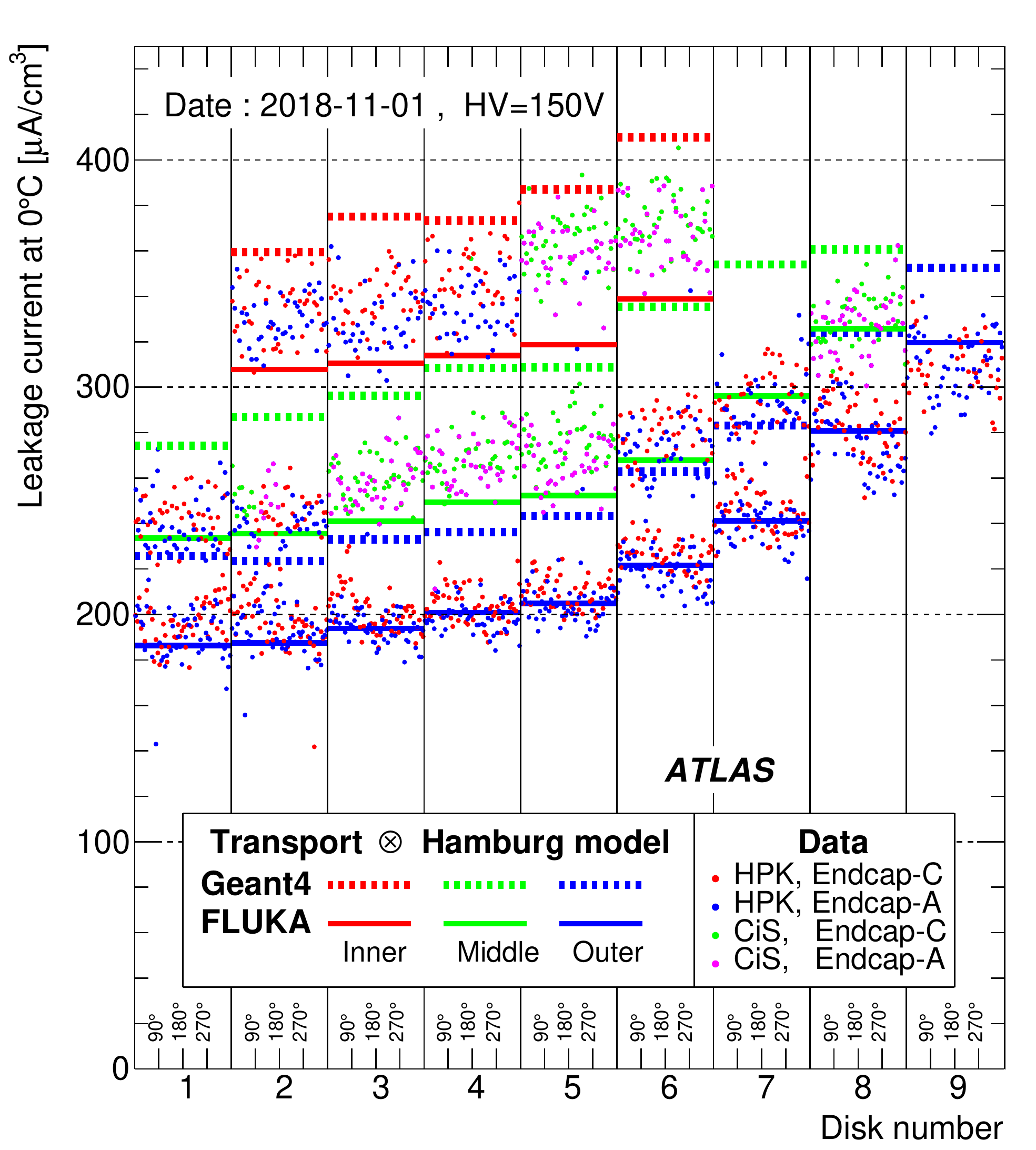}
\label{fig:Lateral_distributions:b}
}
\caption{
Measured leakage currents per unit volume, normalised to 0$\degC$, with HV at 150~V, for \protect\subref{fig:Lateral_distributions:a} barrel and \protect\subref{fig:Lateral_distributions:b} endcap modules in November 2018.
Modules with the same $r$- and $z$-coordinates, but different azimuthal locations, are grouped together within the same vertical bin, arranged from $\phi=0^\circ$ on the left to $\phi=360^\circ$ on the right.
Modules in endcaps~A and~C, as well as from different sensor manufacturers, are plotted in different colours.
Horizontal solid and dashed bars indicate model predictions from the Hamburg model, using conversion factors estimated from \textsc{Fluka} and \GEANT transport simulations.
The main difference arises from the simplified detector geometry used by \textsc{Fluka}, in comparison with the \GEANT simulation.
The uncertainties in the predictions are about 15\%, and are not shown in the plot to maintain visibility.
}
\label{fig:Lateral_distributions}
\end{figure}
 
Figure~\ref{fig:Lateral_distributions} shows normalised leakage currents per unit volume for all barrel and endcap modules in November 2018, at the end of Run~2, with the HV set to 150~V.
Four modules with abnormally high currents and 42 permanently disabled modules have been excluded.
These plots display a lateral view of the SCT, with the horizontal axis corresponding to the $z$-coordinate, and modules with the same $r$- and $z$-coordinates, but different azimuthal locations, grouped together within the same vertical bin.
 
Almost all modules in the same group have consistent leakage currents to within about 3\%.
Despite the difference of about 2$\degC$ between the average values of $\delta T$ found in the two endcaps, as shown in Figure~\ref{fig:Tscan}, their leakage currents agree once they are normalised to 0$^\circ$C, giving confidence in the validity of this method for correcting the sensor temperatures.
No appreciable differences in leakage currents are observed among sensors from different manufacturers, with different crystal orientations (\textless111\textgreater \ vs \textless100\textgreater ), or between the standard or oxygenated silicon wafers.
 
There is a trend toward larger leakage currents in regions at higher $|z|$ for all radii.
In the barrel layers, the normalised leakage currents of modules at $|\etaindex| = 1$ (closest to $z = 0$~mm) are about 3\% smaller than those of modules at $|\etaindex| = 6$ (near $|z|$ = 680~mm).
This is in contrast to the end of Run~1, when a small excess in leakage current was measured in the central region of barrel layer~3 and uniform distributions were observed in the other barrel layers~\cite{IDET-2013-01}.
The trend seen in Figure~\ref{fig:Lateral_distributions} is consistent with the radiation fluences listed in Table~\ref{tab:fluences}.
These results are compared with predictions from the Hamburg model\footnote{The current-related Hamburg model has been commonly used to describe the annealing effects of the radiation-induced leakage current. It has an exponential and logarithmic term with parameters obtained experimentally in a series of accelerated annealing tests. Details of the parameterisation are discussed in Ref.~\cite{Krasel:2004mi}.}~\cite{Moll:425274, Krasel:2630423} with conversion factors estimated from \textsc{Fluka} and \GEANT~\cite{Agostinelli:2002hh} transport simulations, where about 15\% uncertainty in each simulation is expected.
Within this uncertainty, the predictions agree with the results.
 
\subsubsection{Evolution of leakage currents}
 
Hybrid temperatures, and the voltages and currents of all HV power supplies were monitored continuously since the start of data collection and their values were stored in a database.
The normalised leakage currents can therefore be calculated precisely throughout Run~2, with the $\delta T$ correction applied, and compared with variations in operating conditions.
In LS1 and other technical shutdowns, as well as during unexpected power outages, when $\Thybrid$ data were not available, the temperature of the closest cooling pipe is used instead.
This is a reasonable approximation, because there was no heat generation in the SCT volume during these periods.
 
Typically, 150 $\Imeas$ data points are recorded for each module every hour.
During most physics runs, $\Imeas$ drops by 0.2\%--2\% as the instantaneous luminosity decreases, consistent with expectations from self-heating of the sensor and heat from the front-end chips.
A simple time-weighted average is used to determine a single value of the normalised leakage current for each module in each physics run.
During LS1, technical stops and periods with no physics runs, $\Imeas$ values recorded during calibration runs are used instead. These two sources of data show consistency to within~1\%.
 
The HV current measured for a module depends on the HV applied to the sensor.
Above the full depletion voltage, the HV current increases by several percent for each additional 100~V.
Also, due to the filter resistance of about 12~k$\Omega$ on the HV supply line, the true voltage on a sensor is lower than the voltage applied by the power supply; the difference is about 20~V for modules in barrel layer~3, for which the operational HV was increased to 250~V in late 2018.
This introduces an additional uncertainty of up to about 1\% in these $\Imeas$ values.
 
Figure~\ref{fig:Evolution_leakcurrent} shows the evolution of normalised leakage currents during Run~2 for four representative groups of modules: modules at $\etaindex = 1$ in barrel layer~3, modules at $\etaindex = 6$ in barrel layer~6, outer ring of endcap~C disk~9 and inner ring of endcap~A disk~5.
Values of $\Tsensor$ were around $-1\degC$, $+5\degC$ and $-7\degC$ for modules in barrel layer~3, barrel layer~6 and the two endcaps, respectively, during data-taking periods, while they were ${>}15\degC$ during LS1 and winter technical shutdowns.
As described in Section~\ref{sec:operation_overview}, in 2015 all modules except those in barrel layer~6 were set about $5\degC$ warmer to avoid condensation in the SCT volume.
Leakage currents increased by more than one order of magnitude over the course of Run~2.
The leakage current decreased by 20\%--30\% during each winter shutdown due to annealing.
 
\begin{figure}[t!]
\begin{center}
\includegraphics[width=1.00\textwidth]{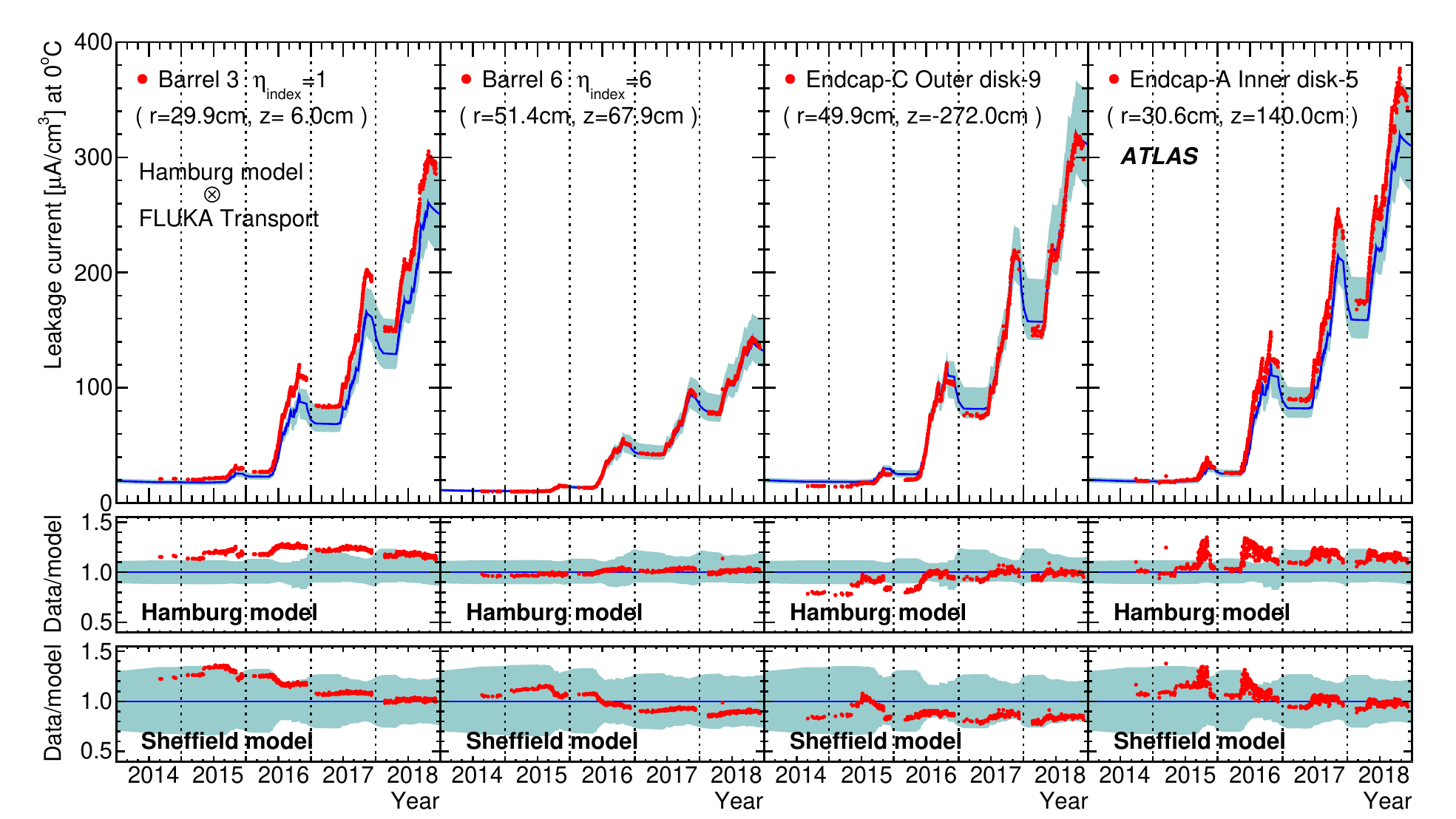}
\caption{
Evolution of normalised leakage currents for four groups of modules: from left to right, barrel layer~3 (modules at $\etaindex = 1$), barrel layer~6 (modules at $\etaindex = 6$), endcap~C (outer ring of disk~9) and endcap~A (inner ring of disk~5).
The main plots show leakage current data (red points) compared with predictions from the Hamburg model (blue line with uncertainty bands).
Data points were measured during physics runs or in SCT stand-alone calibration runs.
The lower two sets of plots show ratios of data to model predictions from the Hamburg and Sheffield models.
Both models use the same conversion factors estimated from \textsc{Fluka} and \GEANT transport simulations.
Coloured bands correspond to $\pm1\sigma$ uncertainties in the model prediction, but uncertainties from the \textsc{Fluka} simulation are not included.
}
\label{fig:Evolution_leakcurrent}
\end{center}
\end{figure}
 
Leakage current measurements are compared with two model predictions in Figure~\ref{fig:Evolution_leakcurrent}: the Hamburg model~\cite{Moll:425274, Krasel:2630423} and the Sheffield model~\cite{HARPER2002548,Harper}.\footnote{This is an alternative model to the current-related Hamburg model. Its current-related damage constant was obtained using actual SCT sensors irradiated at $-10\degC$. It parameterises the annealing behaviour using a sum of five exponential functions with time constants given by the proton and neutron irradiation tests. Exact formulae for both the Hamburg and Sheffield models are briefly summarised in Appendix A of Ref.~\cite{IDET-2013-01}.} The Sheffield model predicts about 15\% more leakage current than the Hamburg model.
Systematic uncertainties in these model predictions are estimated by varying each parameter of the respective model by $\pm1$ standard deviation and summing the components in quadrature.
Additional uncertainties are also included in the temperature measurements (1$\degC$) and the delivered luminosities (1.7\%).
Uncertainties from the \textsc{Fluka} simulation are not included; however, these comparisons provide an estimate of the quality of the simulation as well as the uncertainty in the current-related damage factors.
The model predictions are consistent with the data at a level of around 30\%.
The partial annealing which occurred during the winter shutdown periods is also modelled with similar precision.
Discontinuities in the data-to-model ratios between years may be due to errors in the estimation of $\Tsensor$, for which the $\delta T$ correction was performed only once a year during shutdowns, as well as from other imperfections in the model predictions.
 
\subsection{Full depletion voltage}
\label{sec:FullDepletionVoltage}
\begin{figure}[t!]
\centering
\subfloat[]{
\includegraphics[width=0.49\textwidth]{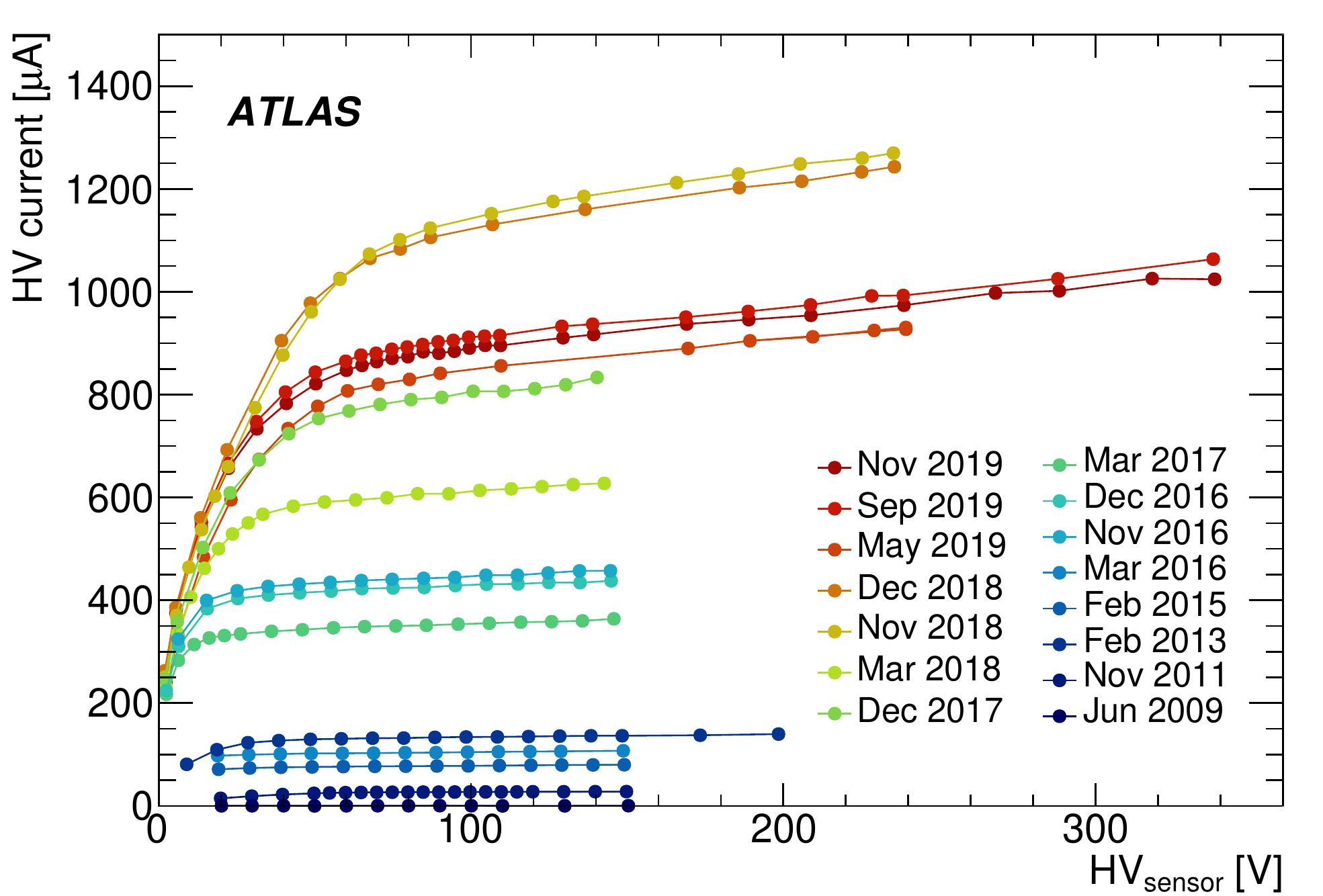}
\label{fig:Leakagecurrent_fit:a}
}
\subfloat[]{
\includegraphics[width=0.49\textwidth]{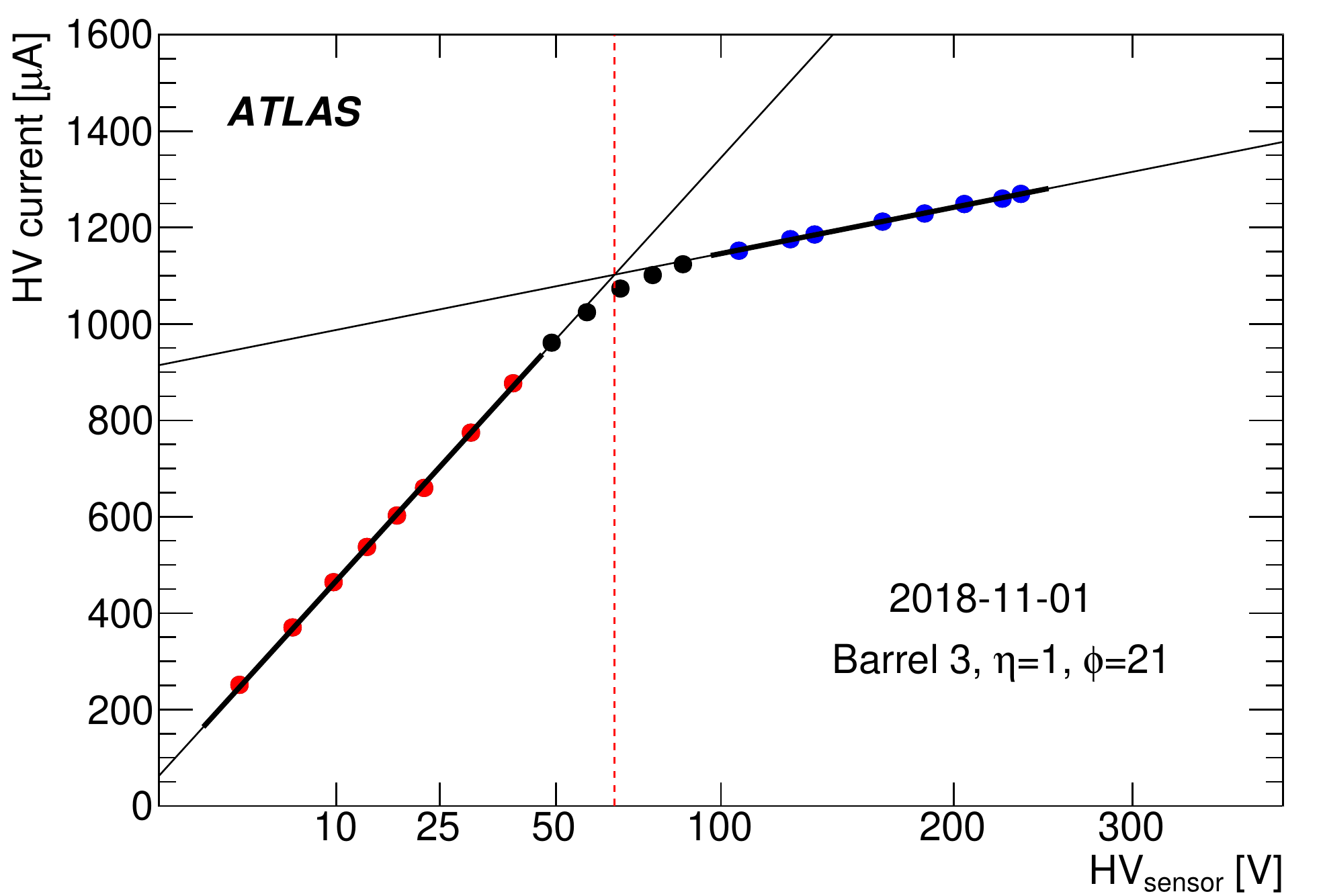}
\label{fig:Leakagecurrent_fit:b}
}
\caption{
Measurement of $\vfd$.
\protect\subref{fig:Leakagecurrent_fit:a} Evolution of the relation between the HV current and the HV ($I$--$V$ curve) for one module in barrel layer~3, from June 2009 to November 2019.
\protect\subref{fig:Leakagecurrent_fit:b} Example of linear fits made to an $I$--$V$ curve.
The scale of the horizontal axis is linear for the square-root of the HV.
The red and blue dots are the data points used for the two linear fits.
}
\label{fig:Leakcurrent_fit}
\end{figure}
 
\begin{figure}[t!]
\centering
\includegraphics[width=0.45\textwidth]{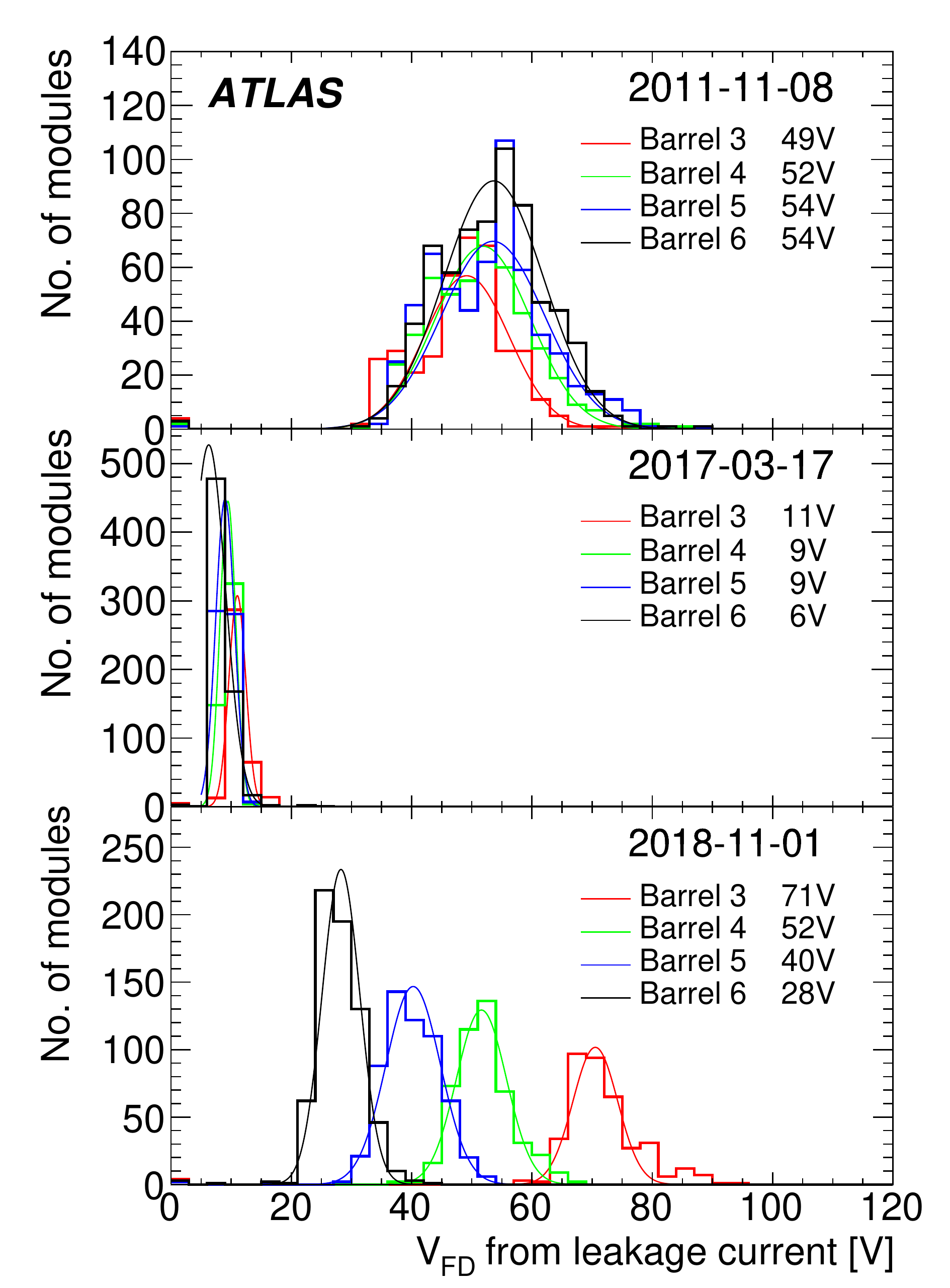}
\caption{
Distributions of $\vfd$ for modules in the four barrel layers.
From top to bottom, results from November 2011, March 2017 and November 2018 are shown.
}
\label{fig:IV_VFD}
\end{figure}

\begin{figure}[t!]
\centering
\includegraphics[width=0.8\textwidth]{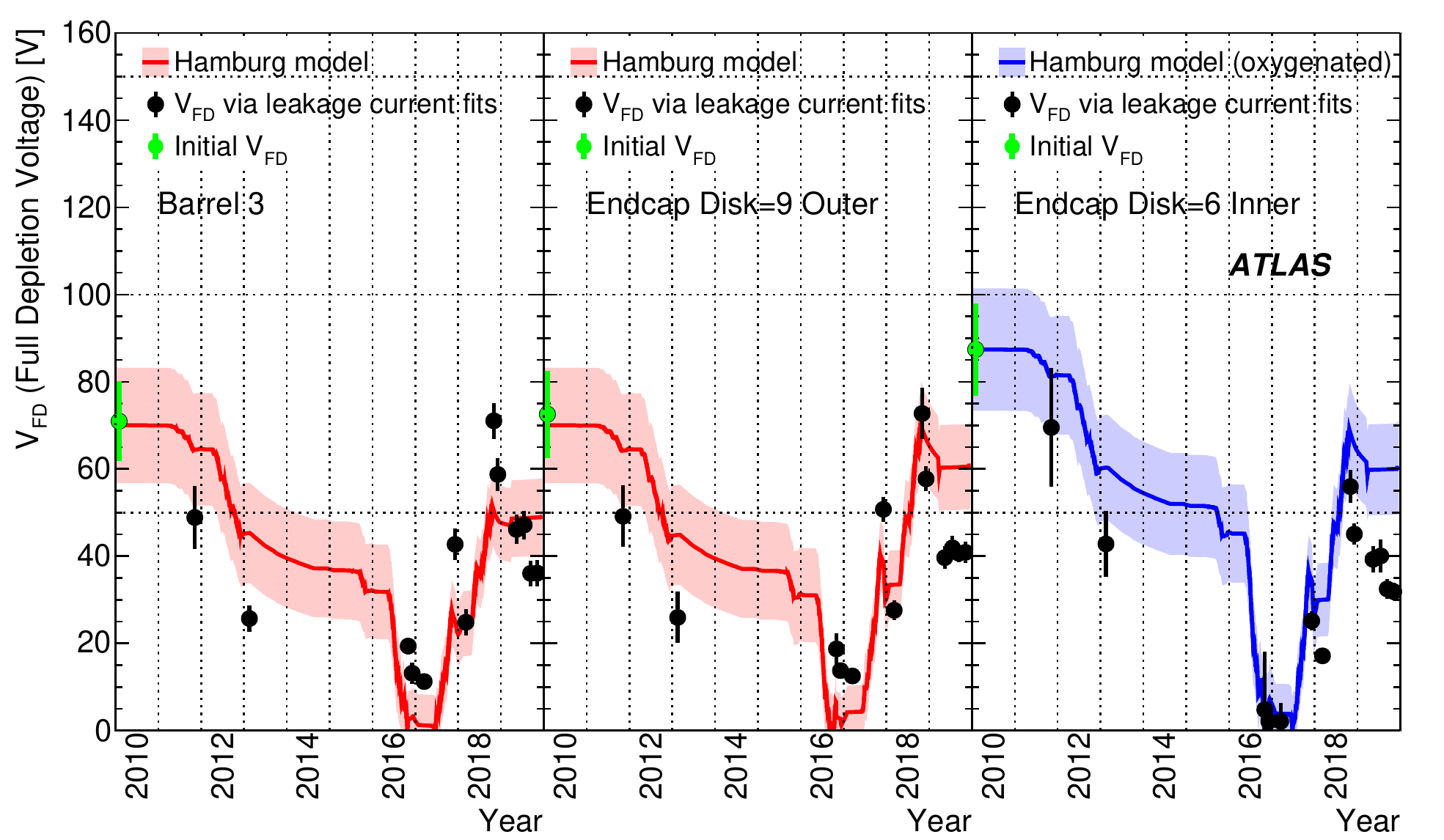}
\caption{
Evolution of $\vfd$ estimated from the $I$--$V$ curves.
From left to right, results from barrel layer~3, outer ring of endcap disk~9 and inner ring of endcap disk~6 are shown.
Green points show initial values of $\vfd$ determined from $C$--$V$ measurements.
Predictions from the Hamburg model are shown by lines with model
uncertainty bands for standard (red) and
oxygenated (blue) silicon sensors.
}
\label{fig:Hamburg_VFD_3}
\end{figure}
 
Since all SCT sensors are expected to have been type-inverted, it is essential to understand $\vfd$ and to predict its future evolution.
If space charge is uniformly distributed within the bulk silicon, then the leakage current should be proportional to the depleted volume, which develops as a function of the square-root of the HV.
Once full depletion has been reached, the leakage current remains nearly constant.
Therefore, a point of abrupt change in the variation of $\Imeas$ as a function of the HV ($I$--$V$ curve) provides an estimate of the value of~$\vfd$.
 
Figure~\ref{fig:Leakcurrent_fit}\subref{fig:Leakagecurrent_fit:a} shows the evolution of the $I$--$V$ curve from one of the modules in barrel layer~3, with 15 different measurements made between 2009 and 2019 superimposed.
As expected, the $I$--$V$ curve consists of a lower region, in which $\Imeas$ rises sharply, and a higher region where the rate of increase is less.
As shown in Figure~\ref{fig:Leakcurrent_fit}\subref{fig:Leakagecurrent_fit:b}, separate linear fits are made to the data in each region, and the crossing point of these two lines is considered to be~$\vfd$.
Values of $\vfd$ obtained from all modules in the barrel region on three different dates are shown in Figure~\ref{fig:IV_VFD}.
In November 2011, all barrel layers had similar values of $\vfd$ around 50~V.
In 2016 and early 2017, almost all modules had small values of $\vfd$, indicating a significant decrease in the donor concentration in the silicon bulk.
By late 2018, values of $\vfd$ had increased again, clearly indicating type inversion.
Barrel layer~3 (innermost layer) has the largest values of $\vfd$, around 70~V, while barrel layer~6 (outermost layer) has the lowest, around 30~V.
 
A similar analysis was performed for various periods during Run~2. The results are shown in Figure~\ref{fig:Hamburg_VFD_3}, separately for three representative groups of modules: barrel layer~3, the outer ring of endcap disk~9 and the inner ring of endcap disk~6 (oxygenated sensors).
The evolution of $\vfd$ is compared with predictions from the Hamburg model, where radiation-induced changes of the space-charge concentration are expressed by four main components: removal of initial donors, creation of stable acceptors, creation of short-term acceptors and a reverse annealing term, where a constant space-charge distribution over the bulk volume is assumed.
Overall, the model reproduces qualitatively the $\vfd$ changes observed in the data, including the type inversion in 2016--17.
Short-term annealing effects are more prominent in the data, however.
The thermal history of the detector is included in the model.
For example, the SCT was kept at room temperature for two weeks in March 2019, because of maintenance work on the evaporative cooling system.
Annealing occurred at a faster rate than predicted during this period, as can be seen in Figure~\ref{fig:Hamburg_VFD_3}.
Although this behaviour is not fully understood, the larger than expected decrease of $\vfd$ helps to maintain safe detector operating conditions.
The expected increase of $\vfd$ is at most 150~V, so this result allows continued SCT operation during LHC Run~3.

 
\section{Conclusion}

Operating conditions for the SCT detector during LHC Run~2 were more challenging than they had been in Run~1, because of the increased instantaneous luminosity and rates of pile-up, resulting in higher data transmission rates and more radiation damage.
Nevertheless, the SCT was available for 99.9\% of the integrated luminosity and achieved a data-quality efficiency of 99.85\%. This was due to various improvements made to the SCT DAQ system: introduction of additional RODs and BOCs, more aggressive data compression, and automatic recovery mechanisms for SCT modules and RODs.
The fraction of noisy strips often increased up to 0.3\%, but these increases were temporary as recovery could be achieved by regular detector calibrations.
The SCT calibration time was significantly reduced to fit into the shorter intervals available between LHC fills in the later years of Run~2.
 
Key parameters that determine SCT performance are high hit efficiency and low noise occupancy.
During Run~2 the hit efficiency of the SCT remained around 99\% or higher.
Noise occupancy was kept below $5 \times 10^{-4}$, although noise levels increased by 10\%--20\% during Run~2.
Noise levels and their evolution were observed to differ between the different types of sensors.
 
It is important to understand the effects of radiation damage to the SCT sensors.
This was studied in terms of the leakage current and the full depletion voltage, which were monitored throughout Run~2.
A method was developed to correct for the offset between the sensor temperatures and the temperatures measured using thermistors on the hybrid boards.
This enabled leakage currents in the endcap modules to be determined more accurately.
Leakage currents in the SCT modules increased by more than a factor of ten during the four years of Run-2 operation.
Measured values agree with predictions made by the Hamburg model to within $\sim$30\%.
The full depletion voltage was measured from the HV dependence of the leakage current.
Overall trends, such as type inversion, are consistent with predictions from the Hamburg model, while short-term annealing effects were more prominent in the data.
 
In summary, a number of new operational developments and improvements were made, which allowed stable operation of the SCT with high efficiency during Run~2.
Performance parameters such as hit efficiency, cluster width and noise levels, as well as radiation damage, have been measured and understood.
In LHC Run~3 an integrated luminosity of about 200~fb$^{-1}$ is expected with a level of pile-up similar to that in Run~2.
The experience and improved understanding of the detector acquired during Run~2 will be important to allow the safe and stable operation of the SCT until the end of Run~3.
 
\section*{Acknowledgements}
 

We thank CERN for the very successful operation of the LHC, as well as the
support staff from our institutions without whom ATLAS could not be
operated efficiently.
 
We acknowledge the support of
ANPCyT, Argentina;
YerPhI, Armenia;
ARC, Australia;
BMWFW and FWF, Austria;
ANAS, Azerbaijan;
SSTC, Belarus;
CNPq and FAPESP, Brazil;
NSERC, NRC and CFI, Canada;
CERN;
ANID, Chile;
CAS, MOST and NSFC, China;
Minciencias, Colombia;
MSMT CR, MPO CR and VSC CR, Czech Republic;
DNRF and DNSRC, Denmark;
IN2P3-CNRS and CEA-DRF/IRFU, France;
SRNSFG, Georgia;
BMBF, HGF and MPG, Germany;
GSRI, Greece;
RGC and Hong Kong SAR, China;
ISF and Benoziyo Center, Israel;
INFN, Italy;
MEXT and JSPS, Japan;
CNRST, Morocco;
NWO, Netherlands;
RCN, Norway;
MEiN, Poland;
FCT, Portugal;
MNE/IFA, Romania;
JINR;
MES of Russia and NRC KI, Russian Federation;
MESTD, Serbia;
MSSR, Slovakia;
ARRS and MIZ\v{S}, Slovenia;
DSI/NRF, South Africa;
MICINN, Spain;
SRC and Wallenberg Foundation, Sweden;
SERI, SNSF and Cantons of Bern and Geneva, Switzerland;
MOST, Taiwan;
TAEK, Turkey;
STFC, United Kingdom;
DOE and NSF, United States of America.
In addition, individual groups and members have received support from
BCKDF, CANARIE, Compute Canada and CRC, Canada;
COST, ERC, ERDF, Horizon 2020 and Marie Sk{\l}odowska-Curie Actions, European Union;
Investissements d'Avenir Labex, Investissements d'Avenir Idex and ANR, France;
DFG and AvH Foundation, Germany;
Herakleitos, Thales and Aristeia programmes co-financed by EU-ESF and the Greek NSRF, Greece;
BSF-NSF and GIF, Israel;
Norwegian Financial Mechanism 2014-2021, Norway;
NCN and NAWA, Poland;
La Caixa Banking Foundation, CERCA Programme Generalitat de Catalunya and PROMETEO and GenT Programmes Generalitat Valenciana, Spain;
G\"{o}ran Gustafssons Stiftelse, Sweden;
The Royal Society and Leverhulme Trust, United Kingdom.
 
The crucial computing support from all WLCG partners is acknowledged gratefully, in particular from CERN, the ATLAS Tier-1 facilities at TRIUMF (Canada), NDGF (Denmark, Norway, Sweden), CC-IN2P3 (France), KIT/GridKA (Germany), INFN-CNAF (Italy), NL-T1 (Netherlands), PIC (Spain), ASGC (Taiwan), RAL (UK) and BNL (USA), the Tier-2 facilities worldwide and large non-WLCG resource providers. Major contributors of computing resources are listed in Ref.~\cite{ATL-SOFT-PUB-2021-003}.
 
 
\printbibliography

\clearpage \input{atlas_authlist}

\end{document}

%% file: atlas_authlist.tex
 
\begin{flushleft}
\hypersetup{urlcolor=black}
{\Large The ATLAS Collaboration}

\bigskip

\AtlasOrcid[0000-0002-6665-4934]{G.~Aad}$^\textrm{\scriptsize 98}$,    
\AtlasOrcid[0000-0002-5888-2734]{B.~Abbott}$^\textrm{\scriptsize 124}$,    
\AtlasOrcid[0000-0002-7248-3203]{D.C.~Abbott}$^\textrm{\scriptsize 99}$,    
\AtlasOrcid[0000-0002-2788-3822]{A.~Abed~Abud}$^\textrm{\scriptsize 34}$,    
\AtlasOrcid[0000-0002-1002-1652]{K.~Abeling}$^\textrm{\scriptsize 51}$,    
\AtlasOrcid[0000-0002-2987-4006]{D.K.~Abhayasinghe}$^\textrm{\scriptsize 91}$,    
\AtlasOrcid[0000-0002-8496-9294]{S.H.~Abidi}$^\textrm{\scriptsize 27}$,    
\AtlasOrcid[0000-0002-9987-2292]{A.~Aboulhorma}$^\textrm{\scriptsize 33e}$,    
\AtlasOrcid[0000-0001-5329-6640]{H.~Abramowicz}$^\textrm{\scriptsize 157}$,    
\AtlasOrcid[0000-0002-1599-2896]{H.~Abreu}$^\textrm{\scriptsize 156}$
% ...
\bigskip
\\

% ...
$^{27}$Physics Department, Brookhaven National Laboratory, Upton NY; United States of America.\\
% ...
$^{33}$$^{(a)}$Facult\'e des Sciences Ain Chock, R\'eseau Universitaire de Physique des Hautes Energies - Universit\'e Hassan II, Casablanca;$^{(b)}$Facult\'{e} des Sciences, Universit\'{e} Ibn-Tofail, K\'{e}nitra;$^{(c)}$Facult\'e des Sciences Semlalia, Universit\'e Cadi Ayyad, LPHEA-Marrakech;$^{(d)}$LPMR, Facult\'e des Sciences, Universit\'e Mohamed Premier, Oujda;$^{(e)}$Facult\'e des sciences, Universit\'e Mohammed V, Rabat;$^{(f)}$Mohammed VI Polytechnic University, Ben Guerir; Morocco.\\
$^{34}$CERN, Geneva; Switzerland.\\
% ...
$^{51}$II. Physikalisches Institut, Georg-August-Universit\"{a}t G\"ottingen, G\"ottingen; Germany.\\
% ...
$^{91}$Department of Physics, Royal Holloway University of London, Egham; United Kingdom.\\
% ...
$^{98}$CPPM, Aix-Marseille Universit\'e, CNRS/IN2P3, Marseille; France.\\
$^{99}$Department of Physics, University of Massachusetts, Amherst MA; United States of America.\\
% ...
$^{124}$Homer L. Dodge Department of Physics and Astronomy, University of Oklahoma, Norman OK; United States of America.\\
% ...
$^{156}$Department of Physics, Technion, Israel Institute of Technology, Haifa; Israel.\\
$^{157}$Raymond and Beverly Sackler School of Physics and Astronomy, Tel Aviv University, Tel Aviv; Israel.\\
% ...
$^{a}$ Also at Borough of Manhattan Community College, City University of New York, New York NY; United States of America.\\
$^{b}$ Also at Bruno Kessler Foundation, Trento; Italy.\\
$^{c}$ Also at Center for High Energy Physics, Peking University; China.\\
$^{d}$ Also at Centro Studi e Ricerche Enrico Fermi; Italy.\\
$^{e}$ Also at CERN, Geneva; Switzerland.\\
$^{f}$ Also at D\'epartement de Physique Nucl\'eaire et Corpusculaire, Universit\'e de Gen\`eve, Gen\`eve; Switzerland.\\
% ...

\end{flushleft}


%% file: ANA-SCTD-2019-01-PAPER.bib
@Article{NIMA-568-642-2006,
    author         = "Abdesselam, A. and others",
    title          = "{The barrel modules of the ATLAS semiconductor tracker}",
    journal        = "Nucl. Instrum. Meth. A",
    volume         = "568",
    year           = "2006",
    pages          = "642-671",
    doi            = "10.1016/j.nima.2006.08.036",
%    url            = "https://www.sciencedirect.com/science/article/pii/S016890020601388X",
}

@Article{NIMA-575-353-2007,
    author         = "Abdesselam, A. and others",
    title          = "{The ATLAS semiconductor tracker end-cap module}",
    journal        = "Nucl. Instrum. Meth. A",
    volume         = "575",
    year           = "2007",
    pages          = "353-389",
    doi            = "10.1016/j.nima.2007.02.019",
%    url            = "https://www.sciencedirect.com/science/article/pii/S0168900207003270",
}

@Article{NIMA-578-98-2007,
    author         = "Ahmad, A. and others",
    title          = "{The silicon microstrip sensors of the ATLAS semiconductor tracker}",
    journal        = "Nucl. Instrum. Meth. A",
    volume         = "578",
    year           = "2007",
    pages          = "98-118",
    doi            = "10.1016/j.nima.2007.04.157",
%    url            = "https://www.sciencedirect.com/science/article/pii/S0168900207007644",
}

@Article{NIMA-552-292-2005,
    author         = "Campabadal, F. and others",
    title          = "{Design and performance of the ABCD3TA ASIC for readout of silicon strip detectors in the ATLAS semiconductor tracker}",
    journal        = "Nucl. Instrum. Meth. A",
    volume         = "552",
    year           = "2005",
    pages          = "292-328",
    doi            = "10.1016/j.nima.2005.07.002",
%    url            = "https://www.sciencedirect.com/science/article/pii/S0168900205013926",
}

@Article{JINST-3-P07003-2008,
	author = "Attree, D. and others",
	title = "The evaporative cooling system for the {ATLAS} inner detector",
	journal = "JINST",
	volume = "3",
	year = "2008",
	pages = "P07003",
	doi = "10.1088/1748-0221/3/07/p07003",
%	url = "https://iopscience.iop.org/article/10.1088/1748-0221/3/07/P07003",
}

@Article{IJCRE-13-511-2015,
	author = "Battistin, M. and others",
	title = "{The Thermosiphon Cooling System of the ATLAS Experiment at the CERN Large Hadron Collider}",
	journal = "Int. J. Chem. React. Eng.",
	volume = "13",
	year = "2015",
	pages = "511-521",
	doi = "10.1515/ijcre-2015-0022",
%	url = "https://www.degruyter.com/view/j/ijcre.2015.13.issue-4/ijcre-2015-0022/ijcre-2015-0022.xml",
}

@Report{ATL-INDET-PROC-2010-037,
    author         = "Gibson, S M and Dehchar, M and Horton, K and Lewis, A and Liang, Z and Livermore, S and Mattravers, C and Nickerson, R B",
    title          = "A novel method for ATLAS FSI alignment based on rapid, direct phase monitoring",
    type           = "ATL-INDET-PROC-2010-037",
    year           = "2010",
    url            = "https://cds.cern.ch/record/1305878",
}

@Report{Baranov:2005ewa,
      author         = "Baranov, S. and Bosman, M. and Dawson, I. and Hedberg, V. and Nisati, A. and Shupe, M.",
      title          = "{Estimation of Radiation Background, Impact on Detectors, Activation and Shielding Optimization in ATLAS}",
      year           = "2005",
      reportNumber   = "ATL-GEN-2005-001, ATL-COM-GEN-2005-001, CERN-ATL-GEN-2005-001",
      type           = "ATL-GEN-2005-001; ATL-COM-GEN-2005-001; CERN-ATL-GEN-2005-001",
      url            = "https://cds.cern.ch/record/814823"
%      SLACcitation   = "%%CITATION = ATL-GEN-2005-001;%%"
}

@Article{JINST-3-P02004-2008,
    author         = "Cindro, V. and others",
    title          = "{The ATLAS Beam Conditions Monitor}",
    journal        = "JINST",
    volume         = "3",
    year           = "2008",
    pages          = "P02004",
    doi            = "10.1088/1748-0221/3/02/P02004",
%    url            = "https://iopscience.iop.org/article/10.1088/1748-0221/3/02/P02004",
    primaryClass   = "hep-ex",
}

@Article{JINST-3-P01003-2008,
	author = "Abdesselam, A and others",
	title = "The data acquisition and calibration system for the ATLAS Semiconductor Tracker",
	journal = "JINST",
	volume = "3",
	year = "2008",
	pages = "P01003",
	doi = "10.1088/1748-0221/3/01/P01003",
%	url = "https://iopscience.iop.org/article/10.1088/1748-0221/3/01/P01003",
}

@Article{JINST-2-P09003-2007,
	author = "Abdesselam, A and others",
	title = "The optical links of the ATLAS SemiConductor Tracker",
	journal = "JINST",
	volume = "2",
	year = "2007",
	pages = "P09003",
	doi = "10.1088/1748-0221/2/09/P09003",
%	url = "https://iopscience.iop.org/article/10.1088/1748-0221/2/09/P09003",
}

@Article{NIMA-515-415-2003,
    author         = "Eklund, L and others",
    title          = "{SEU rate estimates for the ATLAS/SCT front-end ASIC}",
    journal        = "Nucl. Instrum. Meth. A",
    volume         = "515",
    year           = "2003",
    pages          = "415-421",
    doi            = "10.1016/j.nima.2003.07.003",
%    url            = "https://www.sciencedirect.com/science/article/pii/S0168900203023003",
}

@Article{ATLAS-DAPR-2018-01-001,
    author         = "{ATLAS Collaboration}",
    title          = "{ATLAS data quality operations and performance for 2015--2018 data-taking}",
    journal        = "JINST",
    volume         = "15",
    year           = "2019",
    pages          = "P04003",
    doi            = "10.1088/1748-0221/15/04/P04003",
    reportNumber   = "CERN-EP-2019-207",
    eprint         = "1911.04632",
    archivePrefix  = "arXiv",
    primaryClass   = "physics.ins-det",
}

@Article{JPhysConf-219-032037-2010,
    author         = "Dotti, A. and others",
    title          = "{The Online Histogram Presenter for the ATLAS experiment: A modular system for histogram visualization}",
    journal        = "J. Phys. Conf. Ser.",
    volume         = "219",
    year           = "2010",
    pages          = "032037",
    doi            = "10.1088/1742-6596/219/3/032037",
%    url            = "https://iopscience.iop.org/article/10.1088/1742-6596/219/3/032037",
    primaryClass   = "hep-ex",
}

@Article{JPhysConf-219-022035-2010,
    author         = "Ilchenko, Y. and others",
    title          = "{Data Quality Monitoring Display for ATLAS experiment at the LHC}",
    journal        = "J. Phys. Conf. Ser.",
    volume         = "219",
    year           = "2010",
    pages          = "022035",
    doi            = "10.1088/1742-6596/219/2/022035",
%    url            = "https://iopscience.iop.org/article/10.1088/1742-6596/219/2/022035",
    primaryClass   = "hep-ex",
}

@Article{EPJC-72-1960-2012,
    author         = "Golling, T. and others",
    title          = "{The ATLAS data quality defect database system}",
    journal        = "Eur. Phys. J. C",
    volume         = "72",
    year           = "2012",
    pages          = "1960",
    doi            = "10.1140/epjc/s10052-012-1960-y",
%    url            = "https://link.springer.com/article/10.1140/epjc/s10052-012-1960-y",
    primaryClass   = "hep-ex",
}

@Report{SOFT-PUB-2007-007,
    author         = "Cornelissen, T and Elsing, M and Fleischmann, S and Liebig, W and Moyse, E and Salzburger, A",
    title          = "Concepts, Design and Implementation of the ATLAS New Tracking (NEWT)",
    type           = "ATL-SOFT-PUB-2007-007, ATL-COM-SOFT-2007-002",
    year           = "2007",
    url            = "https://cds.cern.ch/record/1020106",
}

@Article{NIMA-538-384-2005,
    author         = "Campabadal, F. and others",
    title          = "{Beam tests of ATLAS SCT silicon strip detector modules}",
    journal        = "Nucl. Instrum. Meth. A",
    volume         = "538",
    year           = "2005",
    pages          = "384-407",
    doi            = "10.1016/j.nima.2004.08.133",
%    url            = "https://www.sciencedirect.com/science/article/pii/S0168900204021059",
}

@Article{Solid-State-Elec-20-77,
    author         = "Jacobini, C. and Canali, C. and Ottaviani, G. and Alberigi~Quaranta, A.",
    title          = "{A review of some charge transport properties of silicon}",
    journal        = "Solid-State Elec.",
    volume         = "20",
    year           = "1977",
    pages          = "77",
    doi            = "10.1016/0038-1101(77)90054-5",
%    url            = "https://www.sciencedirect.com/science/article/pii/0038110177900545",
}

@Article{NIMA-466-354-2001,
    author         = "Akimono, T. and others",
    title          = "{Characteristics of irradiated silicon microstrip detectors with <100> and <111> substrates}",
    journal        = "Nucl. Instrum. Meth. A",
    volume         = "446",
    year           = "2001",
    pages          = "354-358",
    doi            = "10.1016/S0168-9002(01)00588-5",
%    url            = "https://www.sciencedirect.com/science/article/pii/S0168900201005885",
}

@Article{FLONT-ONCOL-6-116,
      author         = "Battistoni, Giuseppe and others",
      title          = "{The FLUKA Code: An Accurate Simulation Tool for Particle Therapy}",
      journal        = "Front. Oncol.",
      volume         = "6",
      year           = "2016",
      pages          = "116",
      doi            = "10.3389/fonc.2016.00116",
%      url            = "https://www.frontiersin.org/articles/10.3389/fonc.2016.00116/full",
}

@Article{IEEE-54-1143-2007,
    author         = "Mandi\'c, I. and Cindro, V. and Gori\v{s}ek, A. and Kramberger, G. and Miku\v{z}, M.",
    journal        = "IEEE Trans. Nucl. Sci.", 
    title          = "Online Integrating Radiation Monitoring System for the ATLAS Detector at the Large Hadron Collider",
    year           = "2007",
    volume         = "54",
    number         = "4",
    pages          = "1143-1150",
    doi            = "10.1109/TNS.2007.895120",
}

@misc{ANIMMA-2019,
    author         = "Mandi\'c, I. and others",
    title          = "{Performance of the Online Radiation Dose Monitoring System in ATLAS experiment}",
%    year           = "2019",
    note           = "presented at ANNIMA 2019",
    reportNumber   = "ATL-INDET-SLIDE-2019-251",
    type           = "ATL-INDET-SLIDE-2019-251",
    url            = "https://cds.cern.ch/record/2677398/",
}

@Article{Krasel:2004mi,
    author = "Krasel, O. and Gossling, C. and Klingenberg, R. and Rajek, S. and Wunstorf, R.",
    title = "{Measurement of trapping time constants in proton-irradiated silicon pad detectors}",
    doi = "10.1109/TNS.2004.839096",
    journal = "IEEE Trans. Nucl. Sci.",
    volume = "51",
    number = "6",
    pages = "3055--3062",
    year = "2004"
}

@misc{Moll:425274,
    author        = "Moll, Michael",
    title         = "{Radiation damage in silicon particle detectors: microscopic defects and macroscopic properties}",
    url           = "http://www-library.desy.de/cgi-bin/showprep.pl?desy-thesis99-040",
    note          = "PhD thesis: Hamburg University, Germany",
    year          = "1999"
}

@misc{Krasel:2630423,
    author        = "Krasel, Olaf",
    title         = "{Charge collection in irradiated silicon detectors}",
    year          = "2004",
    url           = "https://cds.cern.ch/record/2630423",
    note          = "PhD thesis: Universitit{\"a}t Dortmund, Germany",
    doi           = "10.17877/DE290R-14839",
}

@Article{Moll:2018fol,
    author = "Moll, Michael",
    title = "{Displacement damage in silicon detectors for high energy physics}",
    doi = "10.1109/TNS.2018.2819506",
    journal = "IEEE Trans. Nucl. Sci.",
    volume = "65",
    number = "8",
    pages = "1561--1582",
    year = "2018"
}

@Article{Chilingarov_2013,
    author         = "Chilingarov, A.",
    title          = "{Temperature dependence of the current generated in Si bulk}",
    journal        = "JINST",
    volume         = "8",
    year           = "2013",
    pages          = "P10003",
    doi            = "10.1088/1748-0221/8/10/P10003",
%    url            = "https://iopscience.iop.org/article/10.1088/1748-0221/8/10/P10003",
}

@Article{HARPER2002548,
    author         = "Harper, R. and others",
    title          = {Evolution of silicon microstrip detector currents during proton irradiation at the CERN PS},
    journal        = "Nucl. Instrum. Meth. A",
    volume         = "479",
    year           = "2002",
    pages          = "548-554",
    doi            = "https://doi.org/10.1016/S0168-9002(01)00945-7",
%    url            = "https://www.sciencedirect.com/science/article/pii/S0168900201009457",
}

@misc{Harper,
    author        = "Harper, R.",
    title         = "{Radiation Damage studies of the silicon microstrip detectors and finding an intermediate mass Higgs Boson at ATLAS}",
    note          = "{PhD thesis: Sheffield Univesity, U.K.}",
    year          = "2001",
    url           = "https://cds.cern.ch/record/2759942",
}

@Article{IDET-2020-01,
    author         = "{ATLAS Collaboration}",
    title          = "{Measurements of sensor radiation damage in the ATLAS inner detector using leakage currents}",
    journal        = "JINST",
    volume         = "16",
    year           = "2021",
    pages          = "P08025",
    doi            = "10.1088/1748-0221/16/08/P08025",
%    reportNumber   = "CERN-EP-2021-055",
%    eprint         = "2106.09287",
%    archivePrefix  = "arXiv",
%    primaryClass   = "hep-ex",
%    note           = "Accepted by JINST",
%    url            = "https://cds.cern.ch/record/2773132",
}

@Book{CERN-2021-001,
      author        = "Dawson, I and others",
      title         = "{Radiation effects in the LHC experiments: Impact on detector performance and operation}",
      publisher     = "CERN",
      address       = "Geneva",
      series        = "CERN Yellow Reports: Monographs",
      reportNumber  = "CERN-2021-001",
      year          = "2021",
      url           = "https://cds.cern.ch/record/2764325",
      doi           = "10.23731/CYRM-2021-001",
}


%% file: ATLAS-useful.bib
@Article{Evans:2008zzb,
      author         = "Evans, Lyndon and Bryant, Philip",
      title          = "{LHC Machine}",
      journal        = "JINST",
      volume         = "3",
      pages          = "S08001",
      doi            = "10.1088/1748-0221/3/08/S08001",
      year           = "2008",
      SLACcitation   = "%%CITATION = JINST,3,S08001;%%",
}

@article{Martin:2009iq,
      author         = "Martin, A. D. and Stirling, W. J. and Thorne, R. S. and Watt, G.",
      title          = "{Parton distributions for the LHC}",
      journal        = "Eur. Phys. J. C",
      volume         = "63",
      year           = "2009",
      pages          = "189",
      doi            = "10.1140/epjc/s10052-009-1072-5",
      eprint         = "0901.0002",
      archivePrefix  = "arXiv",
      primaryClass   = "hep-ph",
      reportNumber   = "IPPP-08-95, DCPT-08-190, CAVENDISH-HEP-08-16",
      SLACcitation   = "%%CITATION = ARXIV:0901.0002;%%"
}

@Article{Sjostrand:2006za,
     author    = "Sj{\"o}strand, Torbjorn and Mrenna, Stephen and Skands, Peter Z.",
     title     = "{PYTHIA 6.4 physics and manual}",
     journal   = "JHEP",
     volume    = "05",
     year      = "2006",
     pages     = "026",
     eprint    = "hep-ph/0603175",
     archivePrefix = "arXiv",
     doi       = "10.1088/1126-6708/2006/05/026",
     SLACcitation  = "%%CITATION = HEP-PH/0603175;%%"
}

@article{Sjostrand:2014zea,
      author         = "Sj{\"o}strand, Torbj{\"o}rn and Ask, Stefan and Christiansen,
                        Jesper R. and Corke, Richard and Desai, Nishita and Ilten,
                        Philip and Mrenna, Stephen and Prestel, Stefan and
                        Rasmussen, Christine O. and Skands, Peter Z.",
      title          = "{An introduction to PYTHIA 8.2}",
      journal        = "Comput. Phys. Commun.",
      volume         = "191",
      year           = "2015",
      pages          = "159",
      doi            = "10.1016/j.cpc.2015.01.024",
      eprint         = "1410.3012",
      archivePrefix  = "arXiv",
      primaryClass   = "hep-ph",
      reportNumber   = "LU-TP-14-36, MCNET-14-22, CERN-PH-TH-2014-190,
                        FERMILAB-PUB-14-316-CD, DESY-14-178, SLAC-PUB-16122,
                        --FERMILAB-PUB-14-316-CD",
      SLACcitation   = "%%CITATION = ARXIV:1410.3012;%%"
}

@article{Agostinelli:2002hh,
      author         = "{GEANT4 Collaboration} and Agostinelli, S. and others",
      title          = "{\textsc{Geant4} -- a simulation toolkit}",
      journal        = "Nucl. Instrum. Meth. A",
      volume         = "506",
      year           = "2003",
      pages          = "250",
      doi            = "10.1016/S0168-9002(03)01368-8",
      reportNumber   = "SLAC-PUB-9350, FERMILAB-PUB-03-339",
      SLACcitation   = "%%CITATION = NUIMA,A506,250;%%"
}


%% file: ATLAS.bib
@Article{PERF-2007-01,
    author         = "{ATLAS Collaboration}",
    title          = "{The ATLAS Experiment at the CERN Large Hadron Collider}",
    journal        = "JINST",
    volume         = "3",
    year           = "2008",
    pages          = "S08003",
    doi            = "10.1088/1748-0221/3/08/S08003",
    primaryClass   = "hep-ex",
}

@Article{IDET-2010-01,
    author         = "{ATLAS Collaboration}",
    title          = "{The ATLAS Inner Detector commissioning and calibration}",
    journal        = "Eur. Phys. J. C",
    volume         = "70",
    year           = "2010",
    pages          = "787",
    doi            = "10.1140/epjc/s10052-010-1366-7",
    eprint         = "1004.5293",
    archivePrefix  = "arXiv",
    primaryClass   = "hep-ex",
}

@Article{SOFT-2010-01,
    author         = "{ATLAS Collaboration}",
    title          = "{The ATLAS Simulation Infrastructure}",
    journal        = "Eur. Phys. J. C",
    volume         = "70",
    year           = "2010",
    pages          = "823",
    doi            = "10.1140/epjc/s10052-010-1429-9",
    eprint         = "1005.4568",
    archivePrefix  = "arXiv",
    primaryClass   = "physics.ins-det",
}

@Article{IDET-2013-01,
    author         = "{ATLAS Collaboration}",
    title          = "{Operation and performance of the ATLAS semiconductor tracker}",
    journal        = "JINST",
    volume         = "9",
    year           = "2014",
    pages          = "P08009",
    doi            = "10.1088/1748-0221/9/08/P08009",
    reportNumber   = "CERN-PH-EP-2014-049",
    eprint         = "1404.7473",
    archivePrefix  = "arXiv",
    primaryClass   = "hep-ex",
}

@Article{PIX-2018-001,
    author         = "Abbott, B. and others",
    title          = "{Production and integration of the ATLAS Insertable B-Layer}",
    journal        = "JINST",
    volume         = "13",
    year           = "2018",
    pages          = "T05008",
    doi            = "10.1088/1748-0221/13/05/T05008",
    eprint         = "1803.00844",
    archivePrefix  = "arXiv",
    primaryClass   = "physics.ins-det",
}

@Article{IDTR-2019-05,
    author         = "{ATLAS Collaboration}",
    title          = "{Alignment of the ATLAS Inner Detector in Run 2}",
    journal        = "Eur. Phys. J. C",
    volume         = "80",
    year           = "2020",
    pages          = "1194",
    doi            = "10.1140/epjc/s10052-020-08700-6",
    reportNumber   = "CERN-EP-2020-108",
    eprint         = "2007.07624",
    archivePrefix  = "arXiv",
    primaryClass   = "hep-ex",
}

@Booklet{ATL-SOFT-PUB-2021-003,
    author         = "{ATLAS Collaboration}",
    title          = "{ATLAS Computing Acknowledgements}",
    howpublished   = "{ATL-SOFT-PUB-2021-003}",
    url            = "https://cds.cern.ch/record/2776662",
}

@Report{ATLAS-TDR-05,
    author         = "{ATLAS Collaboration}",
    title          = "{ATLAS Inner Detector: Technical Design Report, Volume 2}",
    type           = "ATLAS-TDR-5, CERN-LHCC-97-017",
    year           = "1997",
    url            = "https://cds.cern.ch/record/331064",
}

@Report{ATLAS-TDR-19,
    author         = "{ATLAS Collaboration}",
    title          = "{ATLAS Insertable B-Layer: Technical Design Report}",
    type           = "ATLAS-TDR-19; CERN-LHCC-2010-013",
    year           = "2010",
    url            = "https://cds.cern.ch/record/1291633",
    related        = "ATLAS-TDR-19-addm",
    relatedstring  = "Addendum:",
}


%% file: PubNotes.bib
@Booklet{ATL-PHYS-PUB-2016-017,
    author         = "{ATLAS Collaboration}",
    title          = "{The Pythia~8 A3 tune description of ATLAS minimum bias and inelastic measurements incorporating the Donnachie--Landshoff diffractive model}",
    howpublished   = "{ATL-PHYS-PUB-2016-017}",
    url            = "https://cds.cern.ch/record/2206965",
    year           = "2016",
}
